\documentclass[reqno]{amsart}
\usepackage{epsf}
\def\be{\begin{equation}}
\def\ee{\end{equation}}
\def\ba{\begin{array}}
\def\ea{\end{array}}
\def\bea{\begin{eqnarray}}
\def\eea{\end{eqnarray}}

\def\noi{\noindent}

\def\ldd{\\}

\begin{document}

{}\hfill{DSF$-$1/2008}%

%
%

\title[Fermi and the nuclear pile: the retrieval of novel
documents]{Enrico Fermi and the Physics and Engineering of a
nuclear pile: the retrieval of novel documents}

\author{S. Esposito and O. Pisanti}
\address{{\it S. Esposito}: Dipartimento di Scienze Fisiche,
Universit\`a di Napoli ``Federico II'' \& I.N.F.N. Sezione di
Napoli, Complesso Universitario di M. S. Angelo, Via Cinthia,
80126 Napoli ({\rm Salvatore.Esposito@na.infn.it})}%
\address{{\it O. Pisanti}: Dipartimento di Scienze Fisiche,
Universit\`a di Napoli ``Federico II'' \& I.N.F.N. Sezione di
Napoli, Complesso Universitario di M. S. Angelo, Via Cinthia,
80126 Napoli ({\rm Ofelia.Pisanti@na.infn.it})}%

\begin{abstract}
We give a detailed account of the recent retrieval of a consistent
amount (about 600 pages) of documents written by Enrico Fermi
and/or his collaborators, coming from different sources previously
unexplored. These documents include articles, patents, reports,
notes on scientific and technical meetings and other papers,
mainly testifying Fermi's activity in the 1940s about nuclear pile
physics and engineering. All of them have been carefully
described, pointing out the relevance of the given papers for
their scientific or even historical content. From the analysis of
these papers, a number of important scientific and technical
points comes out, putting a truly new light on the Fermi's (and
others') scientific activity about nuclear piles and their
applications. Quite unexpectedly intriguing historical remarks,
such as those regarding the relationships between U.S. and
Britain, just after the end of the war, about nuclear power for
pacific and/or military use, or even regarding long term physics
research and post-war research policy, emerge as well.
\end{abstract}

\maketitle


\section{Introduction}

\noi The name of Enrico Fermi is universally associated to the
successful development of the self-sustaining nuclear chain
reaction culminated, at first, in the operation of the first
controlled nuclear pile in Chicago on December 2, 1942 and, then,
in the realization of the first nuclear explosion in the Trinity
test of the Los Alamos laboratory on July 16, 1945. The
acknowledgment of the fundamental role played by the Italian
scientist in this project comes directly from the papers written
by Fermi himself, and collected in the 1960s by his former
collaborators \cite{FNM}, but especially from the numerous  and
detailed accounts by the people who worked with him in that
project. In fact, as it is natural to expect, the work on nuclear
fission carried out during the war years was classified, so that
only essential reports were written on that work, and only part of
them was declassified  after some time, and made publicly
available. These reports testifying Fermi's activity, as available
in the first 1960s, were published in the {\it Collected Papers}
of Ref. \cite{FNM}. However, by giving an accurate look at this
material, it results quite evident that what published, though
amounting to a large figure, is not the complete story. This
conclusion comes out not only by assuming the existence of
possible documents, earlier classified, but also by analyzing and
cross-checking different testimonies (among the others, we quote
for example the book written by Emilio Segr\`e \cite{FermiFisico}
and the recollections by Herbert L. Anderson and Albert Wattenberg
in Ref. \cite{Sachs}).

A careful work devoted to investigate towards such a direction has
now been performed by one of us (S.E.) and has lead to the
retrieval of a consistent amount of papers, reports or other
documents written by Fermi himself and/or his collaborators,
directly pointing out the peculiar activity of the Italian
scientist about different aspects of Nuclear Physics, but {\it
not} limited to the scientific point of view (in a strict sense).

We have completed the analysis of all this novel material,
amounting to about 600 pages, and present here the results of this
study.

The major sources of the documents are the following:

\begin{enumerate}
\item[1.] the Albert Wattenberg Papers at the University Library
of the University of Illinois at Urbana-Champaign;

\item[2.] the United States Patent and Trademark Office;

\item[3.] the Papers of Sir James Chadwick at the Churchill
Archives Centre in Cambridge (U.K.).
\end{enumerate}

\noi In the archives of the University of Illinois at
Urbana-Champaign a number of papers are deposited donated by
Albert Wattenberg (1917-2007), a former collaborator of Fermi, who
collected documents pertaining to the Manhattan Project. Part of
them were retrieved by Wattenberg as joint editor of the {\it
Collected Papers} by Fermi, and then published in Ref. \cite{FNM}.
However, among these documents we found many unpublished notes and
reports, all dealing with the activity by Fermi and others on
nuclear fission topics, ranging from 1942 to 1944. In particular
we have found 23 notes on meetings of different Councils, where
explicit interventions by Fermi were annotated, 2
scientific/technical reports written by Fermi and collaborators, 5
periodic reports edited by Fermi and others, 1 anonymous
scientific/technical report classified by Wattenberg among the
Fermi papers.

In the U.S. Patent Office, instead, we have found the most
important papers, from a strictly scientific viewpoint, that is
the papers for 15 patents filed up to 1952 (the vast majority
ranging from 1944 to 1946), all but two directly dealing with
nuclear reactors. In practice, all these patents were issued many
years after their application to the competent office, some of
them being even posthumous, and were never published (in Ref.
\cite{FNM}, for example, except for one case, they hadn't  even
been mentioned).

Finally, among the papers deposited by Sir James Chadwick at the
Churchill Archives Centre in Cambridge (U.K.), the most relevant
one related to Enrico Fermi is a complete version of the set of
lectures given by Fermi at Los Alamos in 1945 on Neutron Physics,
containing some material {\it not present} in the known
``American'' version (published on page 440ff of Volume II of Ref.
\cite{FNM}).

Other ``minor'' documents have been recovered as well; all the
material not published or mentioned in the Fermi's {\it Collected
Papers} and now retrieved will be discussed in some detail in the
following.

\section{The path to the exploitation of nuclear energy}

\noi In order to put in the right context what is the object of
the present analysis, it will be preceded by a rapid summary of
what was known about the activity of Fermi on nuclear fission and
related topics before our recent retrieval.

\subsection{The discovery of the fission of uranium and the
possibility to produce a chain reaction}

\begin{quote}
It has been known for many years that vast amounts of energy are
stored in the nuclei of many atomic species and that their release
is non in contradiction with the principle of the conservation of
the energy, nor with any other of the accepted basic laws of
physics. In spite of this recognized fact, it was the general
opinion among physicists until recently that a large scale release
of the nuclear energy would not be possible without the discovery
of some new phenomenon. \cite{FNM223}
\end{quote}

\noi Such a new phenomenon, as mentioned by Fermi in one of his
reviews of 1946, was that observed by Otto Hahn and Fritz
Strassmann in the Fall of 1938 at the Kaiser Wilhelm Institute in
Berlin, when bombarding the uranium nucleus with neutrons from a
radium-beryllium source. The correct explanation of the Hahn and
Strassmann experiments was soon given by Lise Meitner and Otto R.
Frisch who interpreted the observed phenomenon as due to the
splitting of uranium, from which two elements formed, each of
approximately half of its original mass. The mass which
``disappeared'' was assumed to be converted into energy, according
to Einstein's theory of relativity.

The news of the novel phenomenon reached the other side of the
Atlantic Ocean just after Fermi and his family arrived in America,
after receiving the Nobel Prize in Stockholm.

\begin{quote}
Niels Bohr, who had come for a stay at Princeton, was on his way
to attend a conference in Washington. [...] By the time he was
ready to leave Princeton, Bohr had heard the results of Frisch's
experiments. It was a most exciting development.
\cite{AndersonSachs}
\end{quote}

\noi Willis Lamb was in Princeton at that time and, after heard
from Bohr of this breaking news, he went to Columbia University
and communicated it to Fermi \cite{FermiFisico}. Quite
independently, according to Anderson's recollections,

\begin{quote}
on his way to Washington, Bohr thought it would be a good idea to
drop by and see Fermi to tell him about the exciting new physics.
He came to the Pupin Physics Laboratory looking for Fermi. [...]
He didn't find Fermi; he found me instead. I was the only person
around. He hadn't see me before but that didn't stop him. He
grabbed me by the shoulder and said, ``Young man, let me tell you
about fission.'' [...] I had heard enough to catch the excitement.
[...] When Bohr left I felt I had something to tell Fermi. [...]
``Professor Fermi, I've come to tell you that I have just seen
Professor Bohr. He was looking for you and he told me some very
interesting things.'' Fermi interrupted me. A smile broke out and
he said, ``Let ME tell you about fission.'' Then I heard again,
but this time much more graphically, how the energy would appear
when the uranium was split and the pieces flew apart by Coulomb
repulsion. \cite{AndersonSachs}
\end{quote}

\noi After the news spread out, many physicists (including Fermi
and collaborators) confirmed the results by Hahn and Strassmann
and proved true the interpretation and suggestions by Frisch,
working rapidly for a better understanding of the phenomenon.

\begin{quote}
In the spring of 1939 it was generally known that a fission that
can be produced by the collision of a single neutron with a
uranium atom was capable of producing more than one new neutron,
probably something of the order of two or three. It was felt at
that time by many physicists that a chain reaction based on the
uranium fission was a possibility well worth investigating.
\cite{FNM223}
\end{quote}

The idea of a nuclear chain reaction able to liberate energy on a
large scale came to Leo Szilard as early as in 1933-4, when it was
believed that beryllium (instead of uranium) was unstable and that
neutrons would split off when this element disintegrated. This
proved soon incorrect, but the possibility to create a process
that would emit more neutrons than were absorbed (or, in other
words, with a multiplication or reproduction factor greater than
one) came back into the picture when the fission of uranium was
discovered. This was promptly recognized by Szilard who, according
to Anderson \cite{AndersonSachs}, ``was very anxious to work with
Fermi, or at least to have discussion with him'' in order to
achieve effectively a chain reaction.

\subsection{Natural uranium and graphite}

In 1939 a number of experiments were performed to put the problem
of fission on a quantitative basis. The first important fact to be
realized was that the cross section for neutron fission was higher
for low energy neutrons, while the second one was that the key
isotope of uranium involved in  the fission induced by slow
neutrons was the rare one of mass 235, instead of the most
abundant $^{238}$U. The problem was. however, complicated by the
fact that, besides producing fission, slow neutrons can also give
rise to the production of the radioactive isotope $^{239}$U by
simple capture. In particular the capture of neutrons with thermal
energies (thermal neutrons) was proved to be due  to a strong
resonance absorption at somewhat higher energies \cite{FNM131}.
Such a process competes with fission in taking up the neutrons
which are needed to sustain a chain reaction, so that a major
problem in making the chain reaction to be effective was to avoid
losses due to this absorption.

In any case, the first basic point to be cleared up was the choice
of the fissile material to be used and, in this respect, two
alternatives were opened at the end of 1939. The first one was the
separation of $^{235}$U from the natural uranium, thus eliminating
the absorption by the most abundant isotope $^{238}$U. Obviously,
for this method to work, the major difficulty for that time was to
produce large quantities of the isotope needed. The alternative
choice was, instead, to use directly natural uranium, with the
evident drawback caused by the undesirable absorption of neutrons
by the most abundant isotope, which may lower significatively the
multiplication factor for the self-sustaining reaction to be
achieved. The problem with both the alternative methods were
serious, and Fermi chose to work out the one where more physical
effects should be understood and kept under control, i.e. he
decided to study the possibility of a chain reaction with natural
uranium. It is quite interesting to observe that Fermi was very
confident that such a way was the right one:

\begin{quote}
``Herbert,'' he said, ``if you stick with me we'll get the chain
reaction first. The other guys will have to separate those
isotopes first, but we'll make it work with ordinary uranium.''
\cite{AndersonSachs}
\end{quote}

\noi Such an attitude, as usual for him, came from the appropriate
quantitative results he and his collaborators obtained from an
extensive experimental work. Here, as already mentioned, the
discriminating factor was the slowing down of the incident
neutrons, which makes more effective the cross section for fission
with respect to that for absorption.

The problem of  the slowing down of neutrons and its effect on the
development of neutron-induced nuclear reactions (and, in
particular, the production of radioactive elements) had been the
subject of intense and fruitful researches by Fermi and his group
in Rome as early as 1934 \cite{DeGSlow}, and lead to several
important papers, collected in Ref. \cite{FNM}. A patent for the
practical applications of the results obtained was as well issued;
the interesting subsequent anecdotes related to this patent have
been narrated in Ref. \cite{FermiFisico}. It was recognized that
the most efficient way to slow down neutrons was to pass them
through hydrogen, the lightest chemical elements present in water,
paraffin, etc., so that the obvious conclusion for getting a
reproduction factor high enough for a chain reaction was to
disseminate uranium powder in water. However, measurements
revealed \cite{FNM132} that thermal neutron absorption by hydrogen
was too large for water to make it a usable medium for slowing
down neutrons in a chain reaction, since that absorption (leading
to formation of deuterium) would lower substantially the
multiplication factor. Thus, other light elements should be taken
into consideration.

\begin{quote}
Out of Szilard's thinking came the idea of using graphite instead
of water to slow down the neutron. [...] Fermi had also been
thinking about graphite. \cite{AndersonSachs}
\end{quote}

Measurements showed \cite{FNM136} that the absorption of neutrons
on graphite was small enough to make it the obvious choice for a
material for slowing down the neutrons, so that Fermi set forth
also the basic theoretical techniques for describing the behaviour
of neutrons in such substances. It was also shown that, after the
neutrons reached thermal energies, a second diffusion process
began in which the neutrons continued to diffuse through the
material until they either escaped or were absorbed. The
advantages of graphite against water, as a moderator for neutrons,
came out from experiments with a pile of graphite aimed at
measuring the absorption of carbon \cite{FNM138}. In such a pile
the neutrons were slowed down more slowly than in water, but once
they reached thermal energies the neutrons would diffuse longer
and reach greater distances from the source. As a consequence, a
physical separation of the thermal neutrons from higher energy
ones could be obtained, and this property was later used by Fermi
in many different ways.

\subsection{Experimental piles}

At this point of the story, the next step was to design a chain
reacting pile that would work, and, to this end, a number of
experimental piles were built, early at Columbia University in New
York and then at Chicago, to study directly the properties of
uranium and graphite (or other moderators) in a pile.

The key ingredient was, of course, to work with sufficiently pure
materials; these were obtained from different factories (with
quite different degree of purity), and always were tested by Fermi
and his collaborators. A chemical method, involving ether
separation, was used to purify uranium \cite{FNM137} while the
absorption of neutrons by graphite was especially measured.

Graphite bricks were stacked into the so-called ``sigma pile''
(denoted with the Greek letter ``sigma''), designed to measure the
absorption cross section. A neutron source was placed near the
bottom of the pile and indium foils were exposed at various points
on the vertical axis above the source; from the radioactivity
induced in these foils the absorption cross section of graphite
was deduced. To this regard, standard procedures were introduced
\cite{FNM140} by which indium (and rhodium) foils could be
calibrated in order that the measurement of their radioactivity
could be used to give either the slow neutron density or the
slowing down density in absolute units. The graphite column
erected at Columbia was also used as a source of thermal neutrons
in the measurement of the absorption cross section of boron. This
element, in fact, had importance in absolute neutron measurement,
because of its high neutron absorption cross section and its
dependence on the inverse of the velocity of neutrons
\cite{FNM148}.

For uranium, apart from its purification, an important problem was
that of resonance absorption, as mentioned before. The idea then
came out of using uranium in lumps, just to reduce the resonance
absorption. Also, Fermi measured the resonance absorption for
uranium oxide compressed into spheres and, in particular, when
these spheres were embedded in graphite \cite{FNM139}. Evidently,
he was already thinking about experiments to test a ``complete''
uranium-graphite reactor.

Meanwhile, the fission of uranium induced by fast (rather than
slow) neutrons was as well investigated to some extent, not only
for the possibility of obtaining a fast neutron chain reaction,
but also for measuring the contribution of fast neutron-induced
reactions to the slow neutron chain reaction \cite{FNM145}.

Fermi and Szilard had the very important idea of placing the
uranium oxide in a lattice in the graphite, instead of spreading
it out uniformly \cite{WattenSachs}. Here the problem was ``to
ascertain whether a given lattice of uranium oxide lumps embedded
in graphite could give a divergent chain reaction if its
dimensions were made sufficiently large'' \cite{FNM150}, by
exercising the greatest care in keeping under control possible
losses of neutrons.

In order to test with a smaller structure whether a larger one
would work, Fermi invented the ``exponential experiment''. Uranium
was placed among the graphite bricks in a cubic lattice array,
with a radium-beryllium neutron source near the bottom and indium
foils exposed at various distances from it on the vertical axis.
The arrangement is, thus, similar to that of the sigma pile, but
the exponential pile was much larger than the sigma pile. The
exponential decrease in the neutron density along the axis is
greater or less than that expected due to leakage according to
whether the reproduction factor is less or greater than one.

Such exponential piles were developed at Columbia in Summer-Fall
of 1941 \cite{FNM223}; they produced results indicating that even
an infinite amount of material would not lead to a self-sustaining
structure, this being due mainly to the impurities in the
graphite. The situation changed when, during the following Spring
(1942), some new graphite was available. The last two experiments
performed at Columbia, before the move to Chicago, gave
encouraging results \cite{FNM151, FNM152}, and definitively
demonstrated an understanding of the physical effects being
involved.

\subsection{Achieving the first nuclear chain reaction}

The National Academy of Science Committee, whose chairman was
Arthur H. Compton of the University of Chicago, was charged to
review the uranium projects of the United States and to judge
their military importance. At the end of 1941 the Committee
decided that the work made by the Fermi group using natural
uranium was important and, one day before the Pear Harbor attack
on December 7, 1941, the Metallurgical Laboratory was established
with Compton as its scientific head in Chicago. For people working
on a chain reaction using natural uranium, Chicago became the only
game going and, finally, Fermi and his group at Columbia
definitively moved to Chicago in April 1942.

\begin{quote}
Under Compton leadership a large number of people came too. Among
them there was Szilard who worked hard getting the graphite free
from neutron absorbing impurities, and Norman Hilberry, who did a
marvellous job procuring what was needed. Soon large quantities of
graphite began to appear for us to test. Equally strenuous efforts
were expended getting uranium in forms sufficiently pure. First we
worked with uranium oxide. Then various people worked to produce
uranium metal. Outstanding among those was Frank Spedding from
Iowa State University. [...] Spedding's uranium was an important
component of the first chain reaction. \cite{AndersonSachs}
\end{quote}

\noi A number of engineers then came into the project to produce
an appropriate and feasible design of a chain reacting system, so
that a first practical problem was to ``translate'' the known
physical achievements into a form suitable to them who had little
knowledge in a field completely new. To this end, Fermi invented
the notion of ``danger coefficient'' \cite{FNM153} for identifying
the impurities which were dangerous for the realization of the
chain reaction, due to their high neutron absorption cross
section. The effect of such impurities was, then, taken into
account directly on the evaluation of the multiplication factor
through those danger coefficients. For example, it was determined
the effect of gases in the interstices of the graphite, mainly
concerning with the appreciable amount of nitrogen impurity in the
porous graphite, or even the effect of the undesirable impurity of
water in graphite or uranium \cite{WattenSachs}.

Another problem studied was the stability of the pile against
temperature changes, since the heat production in the reactor
would have altered the reactivity of the pile \cite{FNM154163}.

The study of the uranium-graphite reactor was not the sole work
carried out at the Metallurgical Laboratory; other possible
systems were as well considered and some measurements made. This
is the case, for example, of the so-called ``water boiler'', that
is a reactor system made of a central uranium core enriched with
$^{235}$U and water around it serving as a moderator
\cite{FNM155}. Also, the multiplication factor of a uranium oxide
system with a beryllium metal as neutron moderator was measured
\cite{WattenSachs}.

Turning back to the study of the main uranium-graphite reactor,
the first important result was obtained in August 1942, when very
pure uranium oxide was delivered to the Laboratory, making the
reproduction factor $k$ greater than one for the first time
\cite{FNM166}. The 4\% excess available ($k=1.04$) effectively
opened the road to the building of the first self-sustaining pile,
the Chicago Pile No. 1 (CP-1).

The major engineering problem with it was the choice of an
adeguate cooling system with sufficiently low neutron absorption,
since the ``official'' motivation for the project was to produce
plutonium, another fissile material (other than $^{235}$U) to be
used also for military purposes. Indeed, ``a large effort was
underway for planning the pilot and production reactors, on the
assumption that CP-1 would succeed'' \cite{WattenSachs}.
Alternative choices \cite{FNM176} were proposed to cool the system
by gas (preferably helium), water, or even liquid bismuth, this
ingenious proposal by Szilard being later set aside because of the
lack of engineering experience with this material. The Chicago
group definitively worked on the design for a helium cooled plant
submitted by the engineers T.V. Moore and M. Leverett.

\begin{quote}
So it happened that on 15th of November [1942] we started to build
the pile in the West Stands [of the Stagg Field, in Chicago.]
[...] Fermi wanted to build the pile with a shape as close  to
spherical as possible. This would minimize the surface/volume
ratio and make the best use of the material which would became
available. [...] A major change in design came when we had news
that Spedding would be sending some of his high purity uranium
metal. The best place for this was as close to the center as
possible. As a result, the shape of the pile was changed as we
went along. The spherical shape we started with got squashed
somewhat as we went along because the purity of the material we
were getting was better than we had anticipated.
\cite{AndersonSachs}
\end{quote}

\noi The delivery of the Spedding's metal avoided the use of
another ingenious trick proposed by Anderson, i.e. to build the
pile inside an envelope made of ballon cloth to remove the air
(and replace it with Carbon dioxide), in order to minimize the
absorption of neutrons by the nitrogen in the air within the pile,
with a gain of about 1\% in the reproduction factor \cite{FNM168,
AndersonSachs}.

To initiate the chain reaction, it was not necessary (as in
experimental piles) to introduce in the pile a separate neutron
source since, as already experimentally measured, the uranium also
had a non-vanishing probability for spontaneous fission, so that
it emits a few neutrons of its own. However, when the pile was
building, to keep it from becoming too reactive once it began to
approach the critical size, some neutron absorber was needed to
control the reactivity of the chain reaction. Control rods were,
then, inserted within the pile, made simply of strips of cadmium,
since such element was known to be a strong neutron absorber. The
pile was controlled and prevented from burning itself to complete
destruction just by these cadmium rods, which absorb neutrons and
stop the bombardment process of uranium. Further safety
arrangements were as well conceived and set up by Fermi for the
first reactor (see Ref. \cite{AndersonSachs}, for example), whose
construction resulted to be completed about a week earlier than
the director of the Metallurgical Laboratory had officially
anticipated. In the afternoon of December 2, 1942, in fact, the
Chicago Pile No. 1 finally got critical and a chain reaction
successfully started for the first time.

\begin{quote}
We had built the pile, and Fermi had established that we could get
a self-sustaining nuclear reaction that we could control in a very
predictable manner. \cite{WattenSachs}
\end{quote}

\subsection{Further studies on nuclear piles during the war years}

The further development in the studies on nuclear pile, during the
three years 1943-1945 was of course focused on the main objective
of producing weapons, so that it is natural to expect very few
detailed information on these classifies topics. Indeed, none of
these appeared in the {\it Collected Papers} by Fermi \cite{FNM},
and our source of information is only composed of eyewitnesses
(see, for example, \cite{FermiFisico}). However, quite
fortunately, some reports exist that testify on part of Fermi's
activity during these years, not strictly and directly related to
military applications, though those reports had been classified
for some time (see Ref. \cite{FNM}). In the following we will
briefly discuss only such known activity.

First of all, the pile was used as a suitable device for checking
directly the purity of the uranium and for studying a number of
features of the uranium-graphite lattice, unaccessible before
\cite{FNM182183}. However, after about three months of operation,
the original CP-1 pile was explored sufficiently to learn how to
rebuild it with many improvements. A second pile, CP-2, was
effectively built at the Argonne site, near Chicago, in March of
1943, and several studies started to be done. These were mainly
aimed at designing an efficient pilot plant for producing
plutonium or for isotope separation. Such plants were actually
erected (at the end of 1943 and later on) at Oak Ridge, Tennessee
(known as ``Site X'') and at Hanford, Washington (known as ``Site
W''). An example is the designing and test of a radiation shield
for the production piles to be built at Hanford, mentioned in Ref.
\cite{FNM187}.

The pile was also used as a tool to measure neutron absorption
cross sections by several elements. Samples of these elements were
put in the pile, and the compensating changes in control rod
position were determined \cite{FNM210}. This method also became a
routine tool for checking for neutron absorbing impurities in the
materials used in reactors.

Some explicit ``physics works'' was, furthermore, carried out when
the so-called ``thermal column'' was devised by Fermi and
incorporated in experimental piles \cite{FNM189}. A graphite
column was, in fact, set up on the top of a pile, where thermal
neutrons could be found with substantial intensity and essentially
free from those of higher energy. This lead to the discovery of a
novel phenomenon, that is the diffraction of thermal neutrons by
graphite lattice \cite{FNM191}, which opened the road to
investigate the wave properties of neutrons \cite{FNM217}. The
increased neutron intensity available from a pile also allowed to
obtain truly monochromatic beams of neutrons for different
experiments (such as, for example, the measurement of the boron
cross section at a well definite neutron velocity). This was made
possible by a thermal neutron velocity selector designed by Fermi
at the Los Alamos Laboratory (known as ``Site Y'') \cite{FNM200}
and then built at Argonne.

The fission spectrum of uranium was also measured accurately by
exploiting the slow neutrons provided by a pile, which were then
absorbed by a layer of uranium. Other physical properties of
$^{235}$U and $^{239}$Pu were as well determined \cite{FNM212},
and these measurements, performed at Los Alamos in 1944 with the
active collaboration of Chicago's people, revealed somewhat
unexpected properties of plutonium. In the same period some
interesting work was also done on the theoretically possible
phenomenon of ``breeding'' \cite{FNM211}, namely of producing more
fissionable material in a reactor than was consumed, clearly
depending on the effective number of neutrons available in the
chain reaction.

The increased production of heavy water in 1943 made possible to
take seriously into account a proposal by H.C. Urey of April 1942
to use heavy water as neutron moderator. This lead to the
construction of an experimental reactor, known as P-9 and later
becoming CP-3 pile, which would have much more power than CP-2,
thus extending the experimental possibilities \cite{FNM194}.

Finally, other effects were studied during 1944, ranging from the
dissociation pressure of water due to fission \cite{FNM214} to the
measurement of the amount of nitrogen in the first production pile
at Hanford. An unexpected problem with the Hanford pile was also
studied, and independently solved by Fermi and J. Wheeler, on the
xenon poisoning, which caused the full stop of the chain reaction
\cite{FNM218}.

Further works by Fermi until the end of the Second World War
concerned mainly the realization of the atomic bomb at the Los
Alamos Laboratory, so that the corresponding written reports were
strictly classified and not available for the {\it Collected
Papers}. A relevant exception are the lecture notes \cite{FNM222}
for a course that Fermi gave at Los Alamos just after the end of
the war, in the fall of 1945. Here he summarized the results
achieved on neutron physics, with particular reference to nuclear
piles. These lectures are, however, an example of the didactic
ability of Fermi rather than a source of information about his
research work.

\section{Novel documents}

\noi We have indulged above on the works carried out by Fermi and
his collaborators about pile physics, to give an as complete as
possible overview of all the related topics tackled in the first
1940s. This has been done not only for giving an appropriate
context for the documents recently retrieved, but also for a
better comprehension of the novel material present in it, which we
now prepare to discuss in some detail.

In order to identify the documents retrieved in different places,
we have used a simple coding for them made of three or four
letters determining the source archive: USP for U.S. Patents, WAT
for the papers in the Wattenberg Archive and CHAD for the
documents in the Chadwick Archive. The number following the USP
code enumerates the Fermi patents, in chronological order.
Instead, for the WAT code we have used the same cataloguing number
of the Library of the University of Illinois at Urbana-Champaign;
however, since in several cases this cataloguing number refers to
more than one document, we have also added an additional (lower
case) letter differentiating the diverse documents (the
alphabetical order corresponding to the chronological order).
Minor documents coming from other sources do not seemingly
necessitate of additional codes.

Some of the papers just retrieved were not directly written by
Fermi, but are directly related to works performed by him, such as
lecture notes, reports or notes on meetings, and so on. In order
to let the reader to recognize promptly these papers, we have used
special symbols. In particular, we have denoted with a
$\diamondsuit$ those where the contribution by Fermi is explicitly
recognizable (typically, notes on meetings), and with a
* those where such contribution can be deduced only indirectly
(lecture notes or edited reports).

\subsection{Patents}

From the strictly scientific point of view, the most important
part of the present work concerns with the retrieval of the
patents authored (or co-authored) by Fermi on pile physics and
engineering. Except USP1 and USP8, all the patents deal with the
technical and operative construction of nuclear reactors. The
activity by Fermi on this subject was early well recognized from
the accounts given by the living testimonies (see, for example,
Ref. \cite{FermiFisico}) and partially documented by several
papers appeared in the Fermi's {\it Collected Papers}, as
discussed above. Nevertheless, from the newly retrieved papers, a
number of important scientific and technical points comes out,
putting some new bright light on the Fermi's activity in the
project. In practice, what Fermi {\it effectively} did for the
success of the project is here technically documented, and very
clearly emerges from these papers. It is quite impressive the fact
that, just from the accurate reading of the patents, anyone who
has at his own disposal the necessary materials could effectively
build a working reactor, with a number of possible alternatives.

A detailed account of these patents then follows.
\\

\begin{itemize}
\item[{\tiny USP1}] {\it Process for the production of radioactive
substances} (7 pages + 2 figures),
\\
by E. Fermi, E. Amaldi, B. Pontecorvo, F. Rasetti and E. Segr\`e,
\\
filed Oct. 3, 1935 (Patent No. 2,206,634; July 2, 1940); \\
original patent application filed Oct. 26, 1934 in Italy (Patent
No. 324,458).
\\

\noi ``This invention relates to the production of isotopes of
elements from other isotopes of the same or different elements by
reaction with neutrons, and especially to the production of
artificial radioactivity by formation of unstable isotopes. [...]
It is an object of the present invention to provide a method and
apparatus by which nuclear reactions can be carried on with high
efficiency and with the heavier as well as the lighter elements. A
more specific object of the invention is to provide a method and
apparatus for artificially producing radio-active substances with
efficiency such that their cost may be brought below that of
natural radio-active materials. Our invention is based upon the
use of neutron instead of charged particles for the bombardment
and transformation of the isotopes.'' \\
\end{itemize}

\noi Indeed, in this paper, a very detailed description of the
experimental results obtained by studying the radioactivity
induced in a number of chemical elements by irradiation of slow
neutrons is reported, along with a corresponding theoretical
interpretation.

The original patent application, {\it Metodo per accrescere il
rendimento dei procedi\-menti per la produzione di radioattivit\`a
artificiali mediante il bombardamento con neutroni} (Method for
increasing the efficiency of the processes for the production of
artificial radioactivities by neutron bombardment), was submitted
in Italy just after the achievement (on October 22, 1934) of the
first experimental results, and later extended to U.S.A. and other
countries. The intriguing story about this patent (seemingly
without reference to its content), which resulted to be of
fundamental relevance for the subsequent development of the atomic
energy, is well described in the literature (see, for example,
Ref. \cite{FermiFisico}).

Particularly interesting is the mention of the possible discovery
of ``transuranic'' elements given in the present patent. Even
here, some caution was adopted about its interpretation, as well
as the theoretical interpretation of the effects induced by slow
neutrons considered in the paper: ``The theoretical statements and
explanations are, of course, not conclusive and our invention is
in no way dependent upon their correctness. We have found them
helpful and give them for the aid of others, but our invention
will be equally useful if it should prove that our theoretical
conclusions are not altogether correct.''\footnote{Similar
sentences appear also in other patents for evident legal reasons,
but here the dubious ``theoretical correctness'' is particularly
pointed out.}

The reference article for the material here contained is Ref.
\cite{FNM107} of February 15, 1935 to which we refer the reader
for further details. However, at least in part, specific results
discussed here are somewhat different from those in Ref.
\cite{FNM107}.
\\

\begin{itemize}
\item[{\tiny USP2}] {\it Test exponential pile} (11 pages + 11
figures),
\\
by E. Fermi,
\\
filed May 4, 1944 (Patent No. 2,780,595; Feb. 5, 1957).
\\

\noi ``My invention relates to the general subject of nuclear
fission and more particularly to a means and method to creating
and measuring a chain reaction obtained by nuclear fission of
natural uranium having a $^{235}$U isotope content of
approximately $1/139$.'' \\
\end{itemize}

\noi The paper contains an extremely detailed description of an
atomic pile employing natural uranium as fissile material and
graphite as moderator. Apart from the discussion of the theory of
the intervening phenomena, a report on the very construction of
such a pile (with many detailed drawings) and on the experimental
test of the pile (discussing experimental data, their
interpretation and possible improvements) is given. Particularly
relevant is the reported ``invention'' of the exponential
experiment, aimed at ascertaining if the pile under construction
would be divergent (i.e. with a neutron multiplication factor $k$
greater than 1) by making measurements on a smaller pile. The idea
is to measure the exponential decrease of the neutron density
along the length of a column of uranium-graphite lattice, where a
neutron source is placed near its base. Such an exponential
decrease is greater or less than that expected due to leakage,
according to whether the $k$ factor is less or greater than 1, so
that this experiment is able to test the criticality of the pile,
its accuracy increasing with the size of the column.

For the present paper, there is no ``reference'' published
article, although some material appears also in the important Ref.
\cite{FNM150} of March 26, 1942. More in general, some results are
as well present in several papers of Volume II of the Fermi {\it
Collected Papers} \cite{FNM}, but many details (including several
figures) are reported only in the present patent.
\\

\begin{itemize}
\item[{\tiny USP3}] {\it Neutronic reactor} (30 pages + 42
figures),
\\
by E. Fermi and L. Szilard,
\\
filed Dec. 19, 1944 (Patent No. 2,708,656; May 17, 1955).
\\

\noi ``The present invention relates to the general subject of
nuclear fission and particularly to the establishment of
self-sustaining neutron chain fission reaction in systems
embodying uranium having a natural isotopic content.'' \\
\end{itemize}

\noi As emphasized in {\it The New York Times} of May 19, 1955,
this ``historic patent, covering the first nuclear reactor'', is
the first one issued by the U.S. Patent Office, and served as a
reference for the subsequent patents on the same subject. In this
long paper, the theory, experimental data and principles of
construction and operation of ``any'' type of nuclear reactor
known at that time are discussed in an extremely detailed way.
Various possible fission fragments produced by the reactor,
several forms of the uranium employed (metal, oxide and so on,
grouped in different geometrical forms), various materials adopted
as moderators, several cooling systems, different geometries of
the reactors, etc. are considered accurately.

The theoretical description, centered around the achievement of a
self-sustaining chain reaction, is exhaustive, and great attention
is devoted to any possible cause of neutron loss, to the resonance
capture of neutron and to the effect of the presence of relevant
impurities in the reactor. The production chain of neutrons in the
pile is described in great detail, along with the theoretical
arguments underlying the exponential experiment.

The problem of the variation of the multiplication factor due to
the production of radioactive elements, such as xenon, is
discussed extensively. In particular it is pointed out that,
although the initial production of xenon lowers the multiplication
factor $k$ due to its relevant neutron absorption, it subsequently
increases again due to the decay of xenon into another isotope
which absorbs fewer neutrons.

The building up of reactors with solid (graphite) or liquid (heavy
water) moderators is discussed, as well as other possible
moderators such as light water or beryllium. In particular, the
ratio is given of the absorption cross section to the scattering
cross section for several moderators.

Procedures for the purification of uranium are described as well.
Several methods (i.e., the exponential pile or the ``shotgun''
method; see Patent No. 2.969,307) are reported for testing the
purity against neutron absorption of different materials. The
effect of the boron and vanadium impurities in the graphite and
light water in the heavy water are considered.

Different cooling systems for the reactors are considered and
compared in the paper, based on the circulation of a gas
(typically, air) or a liquid (light or heavy water, diphenyl,
etc.).

The principles and practice for the construction, functioning and
control of several kinds of reactors are reported in detail.

One reactor considered in the present paper is a low power
uranium-graphite one without cooling system, where the active part
consists in (small) cylinders of metallic uranium or
pseudo-spheres of uranium oxide (or cylinders of $U_3O_8$). The
control rods are made of steel with boron inserts, while
limitation and safety rods are made of cadmium.

In addition, an uranium-graphite pile cooled by air or even by
water or diphenyl is considered. It is pointed out that dyphenil
should usually be preferred with respect to water, due to its
lower absorption of neutrons and to its higher boiling
temperature, but the disadvantage related to its use is mainly due
to the closed pumping system required and to the possible
occurrence of polymerization which makes the fluid viscous.

Another kind of reactor described in detail is made of uranium
(vertical) bars immersed in heavy water. When, during the
operation, the heavy water is dissociated into $D_2$ and $O_2$,
these two gaseous elements are carried by an inert gas (helium)
into a recombination device. The control and safety rods are made
of cadmium.

Hybrid reactors composed of different lattices in the same
neutronic reactor, in order to increase the multiplication factor
$k$, are considered as well.

A description of the possible uses of nuclear reactors, other than
as power supplies, including the production of collimated beams of
fast neutrons, the production of plutonium (a fissionable material
usable in other reactors) or several other radioactive isotopes
(for possible utilization in medicine) is as well given.

As it results clear, no published reference article behind the
present paper exists. Some partial results may be found in several
papers\footnote{Just to cite some of them, we mention Ref.
\cite{FNM139} for the use of uranium spheres or in lumps, Ref.
\cite{FNM140} for the use of indium foils to measure slow neutron
density, Ref. \cite{FNM153} for the introduction of danger
coefficients, Ref. \cite{FNM176} for the methods of cooling, Ref.s
\cite{FNM180} and \cite{FNM181} for the discussion about the
location of uranium and control rods in the pile, Ref.
\cite{FNM194} for the use of heavy water as moderator, and so on.}
of Volume II of Ref. \cite{FNM} (see, for example, \cite{FNM181}),
but here very many technical data and some information of historic
interest (mainly on the experiments performed in order to obtain
the data reported) are given.
\\

\begin{itemize}
\item[{\tiny USP4}] {\it Chain reacting system} (13 pages + 23
figures),
\\
by E. Fermi and M.C. Leverett,
\\
filed Feb. 16, 1945 (Patent No. 2,837,477; June 3, 1958).
\\

\noi ``The present invention relates to the subject of nuclear
fission and more particularly to a plant wherein the heat
generated as a result of the fission process can be removed at a
rapid rate and preferably in such a manner that it can be utilized
for the production of power. In addition, the products resulting
from the fission process in the plant can readily be removed
without requiring complete dismantling of the plant.'' \\
\end{itemize}

\noi This paper focuses mainly on an automatic system for the
control rods in a nuclear reactor (in the present case made of
natural uranium and graphite) reporting, aside from several
related theoretical points (already considered in previous
patents), a detailed description of it. The purpose of the control
circuit, ruling the position of boron or cadmium rods within the
reactor, is just that of achieving a suitable neutron density to
produce the desired temperature in the system.

The cooling medium is gaseous helium circulating in the active
regions of the reactor, i.e. directly in contact with the uranium,
where approximately the 92\% of the heat is produced. The choice
of such noble gas is made in order to minimize the possible
corrosion of the fissile material and the absorption of neutrons,
which are crucial to self-sustain the fission reaction. However,
other possible choices for the coolant gas (such as air, oxigen or
water vapor) are discussed as well in terms of their ``danger
coefficients'' affecting the determination of the multiplication
factor \cite{FNM153}.

The discussion of some methods of cooling chain reacting piles was
initiated in Ref. \cite{FNM176}, but no reference published paper
exists of the material presented here.
\\

\begin{itemize}
\item[{\tiny USP5}] {\it Neutronic reactor} (8 pages + 12
figures),
\\
by E. Fermi,
\\
filed May 12, 1945 (Patent No. 2,931,762; Apr. 5, 1960).
\\

\noi ``My invention relates to the general subject of nuclear
fission and particularly to the establishment of self-sustaining
neutron chain reactions, compositions of matter and methods of
producing such compositions suitable for use in creating a
self-sustaining chain reaction by nuclear fission of uranium by
slow neutrons in a neutronic reactor.'' \\
\end{itemize}

\noi Particular attention is paid, in this paper, to the problem
of removing heat from a chain reacting device. The system proposed
(and carried into effect) is to cool the moderator (and not
directly the uranium) with a liquid circulating in tubes of
aluminium or some other material.

This paper is, in practice, an ``evolution'' of the previous
patents (especially Patent No. 2,708,656) where, apart from the
presentation of the novel kind of reactor mentioned above, several
new physical data are presented. In particular, some details about
the construction and operation of the system, including
interesting tricks, are reported.

The main subject of this patent does not appear in any other
published paper.
\\

\begin{itemize}
\item[{\tiny USP6}] {\it Air cooled neutronic reactor} (11 pages +
12 figures),
\\
by E. Fermi and L. Szilard,
\\
filed May 29, 1945 (Patent No. 2,836,554; May 27, 1958).
\\

\noi ``The present invention relates to a neutronic reactor which
is capable of numerous uses but is particularly adapted to use for
the production of the transuranic element\footnote{That is
plutonium, $^{239}$Pu.} $^{239}$94 and/or radioactive fission
products by neutrons released during a self-sustaining nuclear
chain reaction through fission of uranium with slow neutrons. More
particularly, our invention relates to the removal of the heat of
the neutronic reaction to such an extent that the reaction may be
conducted at a more rapid rate and the production of element
$^{239}$94 and/or fission products may be accelerated. Natural
uranium may be used in the reaction and contain the isotopes
$^{238}$92 and $^{235}$92 in the ratio of approximately 139 to
1.'' \\
\end{itemize}

\noi The specific reactor considered in this paper is an
uranium-graphite one cooled by air, circulating within the porous
graphite, and with control and safety rods made of cadmium or
boron. The air serving as coolant passes only once through the
reactor, so that it is not too much enriched in radioactive
$^{41}$Ar. Furthermore, it is exhausted at a substantial distance
above ground (from the top of a stack), in order that the
radioactive argon in the cooling air is sufficiently dispersed in
and diluted by fresh atmospheric air before reaching any person on
the ground.

Since the main object of this patent is to produce plutonium, some
constructional details aimed at removing plutonium for the
reactor, when a certain concentration of it is achieved, are
illustrated. In particular, the mechanism for the loading or
unloading of the uranium slugs is made of iron or lead in order to
shield it from the radioactive bars in case they are loaded. It is
also interesting to note with the authors that even after the
uranium slugs have been extracted, they are so exceedingly
radioactive that the produced heat would melt themselves if not
immersed in water.

The production of plutonium was considered by Fermi in some
previously issued reports (see, for example, Ref. \cite{FNM} on
pages 391 and 411), but what discussed here in so great detail
(including the basic air cooling) is present in no other published
paper.
\\

\begin{itemize}
\item[{\tiny USP7}] {\it Testing material in a neutronic reactor}
(8 pages + 9 figures),
\\
by E. Fermi and H.L. Anderson,
\\
filed Aug. 28, 1945 (Patent No. 2,768,134; Oct. 23, 1956).
\\

\noi ``Our invention relates to the general subject of nuclear
fission and more particularly to a means and method for testing
materials by means of a self-sustaining nuclear chain reaction
system. Such a chain reaction system may be created by the nuclear
fission of uranium by thermal neutrons, utilizing natural uranium
having a $^{235}$U isotope content of as low as the natural ratio
of approximately 1/139 of $^{238}$U or an
enriched uranium having a higher $^{235}$U content.'' \\
\end{itemize}

\noi The main object of this paper is to give a suitable method
for determining neutron absorption by different materials, when
they are irradiated with the neutrons coming from a nuclear pile.
Such a reactor has deliberately a low reproduction factor, due to
the apertures made in the system, from which neutrons are lost
(the use of a coolant, in particular, is not provided).

The test is carried out by means of a comparison of the effects
produced by the testing material with respect to those of the
standard active material in the reactor. If the equilibrium
position of the control rod with the test material in the reactor
is further out of the reactor than it was with the standard lump
in the reactor, then the test material absorbs more neutrons than
the standard metal did. The opposite conclusion is, instead,
reached if the control rod must be pushed further into the reactor
to achieve equilibrium with the test material into the system.

Several details about the calibration of the control rod with
different units are given, together with a discussion of the
corrective effects due to a pressure change.

The main subject of this patent does not appear in any other
published paper.
\\

\begin{itemize}
\item[{\tiny USP8}] {\it Neutron velocity selector} (6 pages + 6
figures),
\\
by E. Fermi,
\\
filed Sept. 18, 1945 (Patent No. 2,524,379; Oct. 3, 1950).
\\

\noi ``The present invention relates to neutron velocity selector
apparatus and particularly to apparatus of this type
which utilizes a rotating shutter.'' \\
\end{itemize}

\noi This paper presents a detailed description of the
construction and operation of a velocity selector for neutrons
with velocities up to $6000 \div 7000$ m/s. This apparatus employs
a rotating shutter designed in such a way that neutrons are passed
during a portion of each rotation of the shutter, the shutter
blocking all neutron radiation at other times.

The selector is built up with alternate laminations of a material
with high neutron capture cross section (such as, for example,
cadmium, boron or gadolinium), and parallel laminations of a
material with low capture probability (such as, for example,
aluminium, magnesium or beryllium). This is required in order to
provide a path through the shutter to the neutrons, which then
pass into a ionization chamber.

The timing mechanism, adopted to activate or deactivate the
neutron detection and measuring means at given times following
each opening or closing of the shutter, is electronic (not
mechanic), controlled by a photocell unit.

The reference published article for the main topic of the present
patent is Ref. \cite{FNM200}.
\\

\begin{itemize}
\item[{\tiny USP9}]  {\it Neutronic reactor} (5 pages + 8
figures),
\\
by E. Fermi and L. Szilard,
\\
filed Oct. 11, 1945 (Patent No. 2,807,581; Sept. 24, 1957).
\\

\noi ``The present invention relates to the general subject of
nuclear fission and particularly to the establishment of
self-sustaining neutron chain fission reactions in systems
embodying uranium having a natural isotopic content.'' \\
\end{itemize}

\noi This paper gives, indeed, a detailed description of a variant
of the reactor presented in the previous Patent No. 2,708,656 by
the same authors; it makes use of uranium arranged in plates,
instead of spheres or rods. Such a different geometry is
particularly efficient when a liquid moderator (for example heavy
water) is used; in this case the moderator itself serves also as a
coolant. In the paper, however, the use of solid moderators (like
graphite or beryllium) is discussed as well.

The adoption of the given geometry leads to greater neutron losses
in the reactor (due to resonant capture in uranium), but they are
compensated by the mentioned use of a liquid moderator/coolant.

The main subject of this patent does not appear in any other
published paper.
\\

\begin{itemize}
\item[{\tiny USP10}] {\it Neutronic reactor} (3 pages + 4
figures),
\\
by E. Fermi, W.H. Zinn and H.L. Anderson,
\\
filed Oct. 11, 1945 (Patent No. 2,852,461; Sept. 16, 1958).
\\

\noi ``This invention relates to the general subject of nuclear
fission, and more particularly to a novel means for improving the
establishment of self-su\-sta\-in\-ing nuclear fission
chain reaction.''\\
\end{itemize}

\noi An improvement of the reactors described in the previous
patents, aimed at increasing the reproduction factor, is reported
here, obtained by diminishing the neutron loss due to impurities
within the reactor. This is achieved by encasing the reactor in a
rubberized balloon cloth housing (or something like this) in order
to eliminate the atmospheric air therefrom, thus eliminating both
the effect of the danger coefficient of nitrogen (70\% of the
atmospheric air) and that of the argon present in the air, that
can become radioactive. Since the removal of the air from the
reactor may result in structural problems, caused by the forces
brought into play by that evacuation, the reactor is then filled
by a non-reactive (from a chemical and nuclear standpoint) gas
such as helium or carbon dioxide.

It is also interesting to point out that the authors consider also
the possibility to control (a little) the reproduction ratio of
the reactor by varying the air content of it.

Just a rapid mention of the main idea of the present patent (i.e.
the encasing of the pile in a balloon cloth) appeared in Ref.
\cite{FNM168}, but no detailed description of the system
considered here is reported in any other published paper.
\\

\begin{itemize}
\item[{\tiny USP11}] {\it Neutronic reactor} (3 pages + 5
figures),
\\
by E. Fermi and W.H. Zinn,
\\
filed Nov. 2, 1945 (Patent No. 2,714,577; Aug. 2, 1955).
\\

\noi ``The present invention relates generally to neutronic
reactors and, more particularly, to novel articles of manufacture
used in and in combination with such reactors, and to the
combination of such novel articles of manufacture with neutronic
reactors. [...] More specifically, an object of the present
invention is to provide novel shielding means for the active
portion of a neutronic reactor adapted to be used in combination
therewith. Another object is to provide in a neutronic reactor a
novel cooled shield. Another object is to provide a novel
composite rod adapted to be used as part of the active portion of
a neutronic reactor. Another object is to provide a novel rod for
use as part of the active portion of a neutronic reactor which is
constructed with fissionable material in a portion thereof only.
Another object is to provide in a neutronic reactor novel means
for introducing foreign subject matter into the active portion of
the neutronic reactor for bombardment by neutrons. Another object
is to provide in a neutronic reactor a novel collimated beam for
utilizing the active effects of the neutronic reactor upon objects
exposed exteriorly of the reactor.''\\
\end{itemize}

\noi Indeed, this paper describes a series of technical
improvements of a chain reacting pile, as reported above. Some
attention is paid, for instance, to the shielding of the active
part of the reactor, the design of the uranium-containing rods and
to the recombination of $D_2$ and $O_2$ in $D_2O$ (since heavy
water is expensive).

Particularly interesting, from the scientific point of view, is
the opportunity to have a well inside the active part of the
reactor (which is the part most rich in neutrons and gamma rays),
where objects to be bombarded with $n, \gamma$ may be placed, and
from which collimated beams of such particles to be used outside
the reactor may be formed.

The description of the mentioned technical improvements is not
reported in any other published paper (see, however, Ref.
\cite{FNM186} for the radiation shield, Ref. \cite{FNM214} for the
dissociation of (light) water and Ref. \cite{FNM217} for the
collimation of a neutron beam).
\\

\begin{itemize}
\item[{\tiny USP12}] {\it Method of testing thermal neutron
fissionable material for purity} (4 pages + 4 figures),
\\
by E. Fermi and H.L. Anderson,
\\
filed Nov. 21, 1945 (Patent No. 2,969,307; Jan. 24, 1961).
\\

\noi ``This invention relates to a novel method of testing the
neutronic purity of uranium or other material to be used in a
neutronic reactor.''\\
\end{itemize}

\noi The main aim of this paper is, in fact, to outline a method
for determining the ``neutronic purity'' (i.e., with respect to
elements with an high cross section for neutron capture) of given
materials to be used in a pile.

The ``shotgun'' test is conducted by placing an indium foil (as a
neutron detector) near a neutron source, and measuring its induced
radioactivity with a Geiger-Muller counter. The same measure is
performed when a given quantity of boron (a standard neutron
absorbing pellet) is placed near the detector foil and,
subsequently, by replacing the boron with the material containing
impurities. A direct comparison between the absorption caused by
the unknown composition and the standard boron absorber gives the
desired result for the sum of the danger coefficients of the
impurities (in terms of boron equivalent).

Some theoretical developments show, as well, that the fractional
absorption of the impurities with respect to uranium is
approximately equal to the variation of the reproduction factor in
the pile, induced by the presence of the impurities themselves.

Neutron absorption by impurities is considered in Ref.
\cite{FNM210} (published in 1947 but referring to work made in
1943-4), but the method adopted is completely different from that
described in the present patent, which is not reported in any
other published paper.
\\

\begin{itemize}
\item[{\tiny USP13}] {\it Method of sustaining a neutronic chain
reacting system} (9 pages + 16 fi\-gu\-res),
\\
by E. Fermi and M.C. Leverett,
\\
filed Nov. 28, 1945 (Patent No. 2,813,070; Nov. 12, 1957).
\\

\noi ``The present invention relates to devices of primary use for
the production of neutrons by virtue of a self-sustaining chain
reaction through fission of uranium or other fissionable
isotopes with slow neutrons, known as neutronic reactors.''\\
\end{itemize}

\noi This paper gives a general discussion of a reactor with
variable critical dimensions. The pile considered is an
uranium-graphite one, cooled by air and with control rods of
cadmium or boron (the uranium rods are placed in aluminium
jackets).

Of particular interest is the discussion of the variation of the
reproduction factor $k$ due to long and short term effects. Long
term effects are, for instance, the increase of $k$ due to the
production of plutonium and its decrease due to the production of
fission impurities. Instead, among the short term effects
considered are the production of xenon, which absorbs neutrons,
and the effect of retarded neutrons.

It is also of some relevance the pointing out that a moderator
with a thickness of 1-2 feet around the uranium in the reactor
acts as a reflecting screen for neutrons, with the same efficiency
of an infinite thickness one. From this it follows that by using a
moderator of 10 feet, for instance, the uranium content of the
pile may be increased, with no relevant consequence on the
efficacy of the screen.

A peculiar curiosity is the suggestion that the presence of
nitrogen (as an impurity) in the reactor, which may change due to
changes in the atmospheric pressure, could be used to obtain a
sensitive barometer.

Some partial results may be found already in other patents and/or
in several papers of Volume II of Ref. \cite{FNM} (in particular,
the realization of xenon poisoning is narrated on pages 428-429 of
this reference). No published reference article behind the present
paper exists.
\\

\begin{itemize}
\item[{\tiny USP14}] {\it Neutronic reactor shield} (2 pages + 6
figures),
\\
by E. Fermi and W.H. Zinn,
\\
filed Jan. 16, 1946 (Patent No. 2,807,727; Sept. 24, 1957).
\\

\noi ``This invention relates to radiation shielding devices and
more particularly to a radiation shield that is suitable for
protection of personnel from both gamma rays and neutrons.''\\
\end{itemize}

\noi The mentioned shield from dangerous radiations is achieved to
the best by the combined action of a neutron slowing material (a
moderator) and a neutron absorbing material. Hydrogen is
particularly effective for such a shield since it is a good
absorber of slow neutrons and a good moderator of fast neutrons.
The neutrons slowed down by hydrogen may, then, be absorbed by
other materials such as boron, cadmium, gadolinium, samarium or
steel. Steel is particularly convenient for the purpose, given its
effectiveness in absorbing also the gamma rays from the reactor
(both primary gamma rays and secondary ones produced by the
moderation of neutrons).

In particular, in the present patent a shield is described, made
of alternate layers of steel and masonite (an hydrolized
ligno-cellulose material).

The object of the present paper is not discussed in any other
published paper.
\\

\begin{itemize}
\item[{\tiny USP15}] {\it Method of operating a neutronic reactor}
(30 pages + 42 figures),
\\
by E. Fermi and L. Szilard,
\\
filed Dec. 1, 1952 (Patent No. 2,798,847; July 9, 1957).
\\

\noi ``The present invention relates to the general subject of
nuclear fission and particularly to the establishment of
self-sustaining neutron chain fission reaction in systems
embodying uranium having a natural isotopic content.''\\
\end{itemize}

\noi This paper is a later\footnote{Note that the application for
the present patent was filed on the tenth anniversary of the
operation of the first chain reacting pile at Chicago, on December
2, 1942.} almost faithful copy of Patent No. 2,708,656, already
described above. It was probably prepared (by the authors) in
order to correct several misprints of the previous version. The
most ``relevant'' change is the replacement of the 8 claims of the
original mentioned patent by the following only one claim, which
well summarizes the work done:

``A method of operating a neutronic reactor including an active
portion having a neutron reproduction ratio substantially in
excess of unity in the absence of high neutron absorbing bodies,
said method comprising the steps of inserting in the active
portion a shim member consisting essentially of a high neutron
absorbing body in an amount to reduce the neutron reproduction
ratio to a value slightly higher than unit to prevent a dangerous
reactivity level, controlling the reaction by moving  a control
member consisting essentially of a second high neutron absorbing
body inwardly and outwardly in response to variations in neutron
density, to maintain the neutron reproduction ratio substantially
at unity, and withdrawing successive portions of the shim member
to the extent necessary to enable the reactor to be controlled by
movement of the control member after the neutron reproduction
value has been lowered to the point where the outward movement of
the control member is insufficient to maintain the neutron
reproduction ratio at the desired point, and thus to maintain the
range of control effected by such movement of the control member
substantially constant despite diminution of neutron reproduction
ratio caused by operation of the reactor, the active portion being
substantially free of high neutron absorber other than the control
member and the shim member.''
\\

\subsection{Scientific reports}

Here we give an account on three scientific reports from the
Wattenberg archive, different from the patent papers, not
comprised in the {\it Collected Papers} \cite{FNM} (the first of
these reports not showing the list of authors).
\\

\begin{itemize} \item[{\tiny WAT1043v}]
{\it * The Fourth Intermediate Pile} (15 pages),
\\
Metallurgical Project,
\\
Report C-102.
\\

\noi ``The experiments reported here belong to a set of
experiments designed to establish the arrangement of $3^{\prime
\prime}$ cubes of alloy oxide in a graphite moderating medium
which will produce the highest multiplication factor. Previous
experiments were made on a simple cubic lattice of these alloy
oxide cubes of such dimensions that the ratio between the volume
of the graphite and the volume of the oxide was roughly 20/1. This
gave a multiplication factor of 0.94. Half of the oxide cubes were
then removed, leaving a face-centered lattice in which the volume
ratio of graphite to oxide was 40/1. In this arrangement the
multiplication factor fell to 0.86. In the present report, a body
centered structure was assembled in which the volume ratio of
graphite to oxide was about 10/1. It is found that the
multiplication factor is again close to 0.86. From the results of
these structures it is concluded that the simple cubic lattice in
which the volume ratio was 20/1 represents closely the optimum
conditions. Theoretical calculations
support this experimental result.''\\
\end{itemize}

\noi This paper did not report explicitly the list of the authors,
but it was classified by Wattenberg among the Fermi papers. A
careful analysis has shown that, apart from indirect information
on Fermi's activity, it was to some extent effectively written
(or, at least, ``inspired'') by Fermi. The date of writing was, as
well, not given but, according to the material presented in the
paper, it was likely written in 1942.

The results presented here, very well summarized in the abstract
above, were not reported in any other published paper.
\\

\begin{itemize}
\item[{\tiny WAT1043t}] {\it Report of the Committee for the
Examination of the Moore-Leverett Design of a He-Cooled Plant} (18
pages),
\\
by E. Fermi, S.K. Allison, C. Cooper and E.P. Wigner,
\\
Report CE-324 (1942).\footnote{This report was likely written
around October 29, 1942.}
\\
\end{itemize}

\noi This report was likely written by Fermi (the chairman of the
Committee) with memoranda by Allison, Cooper, Wigner, and a letter
from Szilard.

In it a pile of dimensions considerably larger than that
originally planned in the Moore-Leverett design is considered,
this urging for a re-design of the lattice, for a reduction of the
amount of uranium metal, and the consideration of the possibility
to use a non cubic cell (as stated in previous conferences).

The employment of centrifugal (turbo) compressors (for the
coolant) is considered, instead of reciprocating compressors, with
high purity helium to avoid corrosion of uranium.

A number of technical problems, such as that of an adequate
radiation shielding, the production of radioactive materials in
the reactor which can be collected by helium during the shut-down
of operations, or precaution on helium released in atmosphere are
discussed. Problems of emergency measures for serious loss of
helium and to prevent the activated uranium from melting (if the
the cooling system with helium is switched off) are as well
considered.

It is here pointed out that the operation of the control rods
takes place by looking at the neutron density, rather than at the
temperature of the reactor. Attention is also paid to possible
displacements in the arrangement of the graphite due to the
thermal expansion, that can cause damages to the structure and
interfere with the operation of the control rods. Wigner, in
particular, proposes a cylindrical arrangement instead of the
spherical one.

As a conclusion, the Moore-Leverett design of a He-cooled power
plant can work satisfactorily, although several details have still
to be worked out.

The object of the present report is not discussed in any other
published paper.
\\

\begin{itemize}
\item[{\tiny WAT149}] {\it Measurement of the Cross Section of
Boron for Thermal Neutrons} (4 pages),
\\
by E. Bragdon, E. Fermi, J. Marshall and L. Marshall,
\\
Report CP-1098 (January 11, 1944).
\\

\noi ``Measurements of the boron cross section have been made for
slow neutrons from different sources. The cross section of boron
for neutrons of velocity = 2kT/m = 2200 meters/second at
293$^{\mathrm o}$ K is found to be $705 \times 10^{-24}$
cm$^2$/atom. The cross section varies widely with different
moderators, due to the fact that the temperature of the thermal
neutrons depends on the nature of the moderator.''\\
\end{itemize}

\noi As stated in the abstract, this papers deals with the
measurement of the cross section of thermal neutrons on boron for
different velocities of the neutrons. Velocities ranging from 1700
to 5000 m/s were obtained with a velocity selector, not described
in this paper (see, however, USP8). The relevant measurements are
done by varying also the pressure.

The results of the present paper converged later in the published
article in Ref. \cite{FNM200} (see also the comment to this paper
in Ref. \cite{FNM}, noting the different number of authors), where
the velocity selector was described as well.
\\

\subsection{Notes on Meetings}

A substantial part of the documents testifying for Fermi's
activity and retrieved in the Wattenberg archive consists of many
notes on meetings about nuclear piles and related matters,
attended by Fermi mostly in 1942. Some of these notes were already
published in Ref. \cite{FNM}, but many of them were not included
in the {\it Collected Papers}, probably because the corresponding
material does not present itself as reports, but largely as
minutes of discussions. However, these documents are of great
importance both from a purely scientific point of view and for
historical reasons, since all the notes but the last one
(accounting for a meeting of April 1944) directly reported on the
activity that lead to the achievement of the first chain reaction,
ranging from May to November, 1942. In fact, although the final
scientific results obtained in Chicago were later collected and
discussed in subsequent papers (patents or, in few cases,
published articles), the present notes testify on {\it how} those
results were obtained and, in some cases, also give detailed
information on further, practically unknown, achievements, not
reported in published papers.

Just to quote few examples, we mention an interesting trick
suggested by Fermi for lowering the temperature in the pile,
inspired by what happens in wind tunnels; or the control of the
multiplication factor by means of the pressure of nitrogen in a
liquid cooled pile. Much attention was, indeed, paid to the
problem of heat transfer in the planned power and production
plants, and to that of an effective and easy control of the chain
reaction. Some discussions were also carried out on chain reacting
piles working with fast (instead of slow) neutrons, and on
different schemes for the uranium-graphite pile.

From an history of science viewpoint, these notes also present
very interesting information, not available from other sources, on
the Metallurgical Project, its formation and development, social
and political implications (interventions of General Groves to one
of the Meetings considered are registered in the corresponding
notes), and so on. Several interesting and annoying discussions
reveal, in fact, the urgency of the production of plutonium or
other fissile material for military rather than civil applications
already in 1942, the position of the problem of the moral effect
of the operation and that of the relations with Army, including
the issue of security and the distribution of information.
However, different matters related to the physiological effect of
the radiations developed in a pile (a problem raised more than
once by Fermi), were considered as well, along with discussions
about power utilization and long term research after the
conclusion of the war.

A number of other interesting topics treated in the Meetings may
be found in the detailed account of any document which follows,
including (at the end) the notes on the Meeting of April 1944.
\\

\begin{itemize} \item[{\tiny WAT1043a}]
{\it $\diamondsuit$ Meeting of Engineering Council} (4 pages),
\\
Present: Moore, Allison, Fermi, Leverett, Wheeler, Compton,
Hilberry and Doan,
\\
Report CE-106 (May 28, 1942).
\\

\end{itemize}
\noi The main discussion is on the cooling of the uranium-graphite
pile by water, helium or both; some discussion is present on
problems related to the pumping of the coolant. An interesting
remark by Fermi is the following: since the temperature in wind
tunnels is controlled by changing the cross section of the tubes,
this trick can be used as well in piles for obtaining lower
temperatures.
\\
Minor discussions are on the design of a pilot power plant and a
pilot extraction plant, with a remark by Fermi on the possibility
of long lived activity induced in iron. Fermi also suggested a way
for avoiding non uniform production of power in piles, just by
blocking part of channels by graphite.
\\
Other minor discussion is on leakage, where Fermi suggested an
external graphite layer of 1 feet.

\

\begin{itemize} \item[{\tiny WAT1043b}]
{\it $\diamondsuit$ Meeting of the Planning Board} (3 pages),
\\
Present: Hilberry, Spedding, Allison, Wigner, Doan, Szilard,
Wheeler, Fermi, Moore and Compton,
\\
Reports CS-112, CS-185 (June 6, 1942).
\\

\end{itemize}
\noi Discussion on the status and organization of the activities:
first self-sustaining pile, open pile working at a low rate of
operation, helium cooling, other best cooling agents.
\\
Re-organization of the work at Chicago (concentration of physics
under Fermi), with discussion of problems on finding location,
which should be chosen according to facilities and personnel at
disposal (the decision, however, lies with Washington).
\\
Discussion on the steps to undertake to protect Government's
position about patents and on the understanding with British (a
common patent pool).
\\
Other scientific discussions are on the purity of the graphite
supplied from various factories, with a remark by Fermi on the
properties of different samples. Wigner discussed the results by
Creutz on resonance absorption.

\

\begin{itemize} \item[{\tiny WAT1043c}]
{\it $\diamondsuit$ Meeting of the Engineering Council} (9 pages),
\\
Present: Moore, Wilson, Seaborg, Wheeler, Leverett, Fermi,
Hilberry, Compton, Spedding and Allison,
\\
Report CS-131 (June 11, 1942).
\\

\end{itemize}
\noi The choice of the site location of the first pile is
discussed in detail, with reference to: water power for the
cooling system; alternative schemes for supplying the power needs
of the plant (about 65000 kW obtained either directly from the
electric lines or from {boilers} by using engines producing
mechanical power); schedule of plants (100 W, 100 kW, 100-100000
kW, 10$^6$ kW); housing and list of the personnel; advantages and
disadvantages of the possible association of the {separation}
project to the atomic power project. Alternatives on the site
location are Chicago or Tennessee Valley: the majority of the
presents is for the first one.

\

\begin{itemize} \item[{\tiny WAT1043d}]
{\it $\diamondsuit$ Meeting of the Engineering Council} (7 pages),
\\
Present: Moore, Leverett, Fermi, Wheeler, Seaborg, Doan, Wilson
and Spedding,
\\
Report CS-135 (June 18, 1942).
\\

\end{itemize}
\noi Discussions on: optimum lattice constants, neutrons losses
due to ducts and channels, {batch} versus continuous operation of
the pile and relation with possible ``chemical'' experiments to
plan (especially on the transformations undergone by U and Pu).
\\
Fermi made some estimates on optimum lattice constants and gave
relations among the multiplication factor $k$, the critical size
and the neutron density. He favored the possibility to have more
small rectangular ducts instead of less large squared ones.
\\
Moreover, Fermi also suggests to use nitrogen as well for
controlling a pile with liquid coolant, since $k$ depends on
pressure and would control up to 4\% in $k$.

\

\begin{itemize} \item[{\tiny WAT1043e}]
{\it $\diamondsuit$ Meeting of the Engineering Council} (4 pages),
\\
Present: Moore, Leverett, Doan, Szilard, Allison, Teller, Seaborg,
Wheeler, Fermi and Wilson,
\\
Report CS-147 (June 25, 1942).
\\

\end{itemize}
\noi Fermi describes the advantages of the possibility to work
with a pile {\it not} operating at optimum $k$.
\\
Discussions on cooling by gas or liquid (but people later focused
on the former): problems with hydrogen that reacts with uranium
metal, with graphite (producing methane), etc.; problems with
helium about leakage, large {sound velocity} (in relation to the
use of blowers or compressors). The general agreement is to use
helium for the {100000 kW pile}.
\\
Further discussions are on the diffusion of the fission products
(in the circulating helium, the walls of the pile, etc.), with a
suggestion by Fermi to not use water spray, and about {blowers and
compressors}.

\

\begin{itemize} \item[{\tiny WAT1043f}]
{\it $\diamondsuit$ Meeting of the Engineering Council} (6 pages),
\\
Present: Fermi, Allison, Seaborg, Whitaker, Doan, Wilson, Moore,
Leverett, Wheeler, Szilard, Compton, Spedding, Hilberry and
Wollan,
\\
Report CS-163 (July 2, 1942).
\\

\end{itemize}
\noi Discussion on the construction, installation (at Chicago) and
operation of pile I, pile II and pilot plant. During the
discussion on the necessary facilities, Fermi recommends to avoid
limitations on water and electric supply in order not to ``cut the
wings''.
\\
Several studies and possible experiments for the piles {I and II}
are considered. Fermi proposes to: study the thermal stability in
pile I or intermediate pile with heat supplied from external
sources; test different cooling mechanisms on the various piles;
have experiments with the piles with the precaution that the
produced radioactivity does not influence them. Fermi also
suggests to: measure $k$ by using the theory of anisotropic pile
or, otherwise, preferably depend altogether on theoretical
calculations; design pile I for evacuation (that is: for
extracting uranium oxide from the pile); etc. (gas tight, sheet
metal, balloon cloth).

\

\begin{itemize} \item[{\tiny WAT1043g}]
{\it $\diamondsuit$ Meeting of the Engineering Council} (4 pages),
\\
Present: Moore, Leverett, Fermi, Wigner, Allison, Wollan, Wheeler,
Seaborg, Spedding, Szilard, Steams and Wilson,
\\
Report CS-174 (July 9, 1942).
\\

\end{itemize}
\noi Fermi reports on a test of chemical stability of uranium, its
reaction with graphite, etc.. He also discusses a number of other
topics as follows.
\\\
About the problem of heat transfer in an energy producing pile,
Fermi proposes to study the behavior of the pile both when its
temperature is large and when it is small, noting that the
Reynolds number depends on temperature.
\\
About the control mechanisms, he suggests an alternative scheme of
control by putting in an absorbing gas, as well as to regulate the
pile by raising or lowering the water level in it .
\\
Finally, about cooling, Fermi observes that calculations are
sufficient for helium but not for water, for which the
intermediate experiment is required. Also, he points out that the
exponential experiment is not suitable for measuring the effect of
the {helium} diffusion through ducts.

\

\begin{itemize} \item[{\tiny WAT1043h}]
{\it $\diamondsuit$ Meeting of the Technical Council} (7 pages),
\\
Present: Fermi, Compton, Allison, Moore, Szilard, Wigner and
Wheeler,
\\
Report CS-184 (July 14, 1942).
\\

\end{itemize}
\noi Fermi discusses the use of beryllium (giving also some data)
as moderator and neutron reflector, pointing out that it is not
convenient to have an all-beryllium structure, but rather a pile
with 2 cm of beryllium around the uranium metal, since the thermal
absorption is not compensated by the $n\rightarrow 2\, n$
reaction.
\\
A discussion on the shielding of experimental plant and the
handling of materials (with respect to protection problems)
follows, with remarks by Fermi.
\\
About the problem of cooling, a general discussion is made on the
temperature dependence of conductivity of uranium oxide and
graphite. In particular, Fermi considers the possible behavior of
U$_3$O$_8$ in a pile working at 100 W (corresponding to a
temperature of about $1000^\circ C$), and alternative choices of
cooling. For the last point, Szilard proposes the use of bismuth
as coolant.
\\
About the control system, Fermi also observes that, for proper
consideration of the problem of removing oxide, it is important to
decide if control columns have to take out from the pile
horizontally or vertically, such a choice coming out from
practical attempts (build different structures and try out).
\\
A minor discussion on the necessity of more people involved in
operations is also present.

\

\begin{itemize} \item[{\tiny WAT1043i}]
{\it $\diamondsuit$ Meeting of the Engineering Council} (3 pages),
\\
Present: Moore, Leverett, Lewis, Fermi, Wollan, Hilberry,
Whitaker, Wilson, Wheeler, Allison, Wigner, Seaborg and Doan,
\\
Report CE-194 (July 21, 1942).
\\

\end{itemize}
\noi General discussions on power plant, extraction plant (by
using fluorination, precipitation with a carrier), and pile
problems (radiation, recharging, breakdown, etc.) are present.
\\
Fermi observes that it would be useful to make some of the
discussed problems clear to a radiologist.

\

\begin{itemize} \item[{\tiny WAT1043j}]
{\it $\diamondsuit$ Meeting of the Technical Council} (5 pages),
\\
Present: Fermi, Szilard, Wigner, Compton, Whitaker, Allison,
Moore, Wheeler and Doan,
\\
Report CS-202 (July 25, 1942).
\\

\end{itemize}
\noi Fermi discusses the value for the multiplication factor $k$
(equal to 1.06) obtained by using several technical solutions and
with UO$_2$ compressed in pseudo-spheres.
\\
A general discussion on the providers of uranium metal and the
product provided then follows.
\\
Fermi raises also the question of the physiological effects of
radiations, by claiming the need for a physician. On this point,
Compton displays data from several medical institutions (National
Cancer Institute and  Chicago Tumor Institute).

\

\begin{itemize} \item[{\tiny WAT1043k}]
{\it $\diamondsuit$ Meeting of the Planning Board} (4 pages),
\\
Present: Spedding, Fermi, Szilard, Wigner, Doan, Moore, Wheeler,
Compton and Hilberry,
\\
Report CS-213 (August 1, 1942).
\\

\end{itemize}
\noi About the problem of heat transfer and cooling, Fermi reports
some values for the conductivity of the materials involved in
different conditions.
\\
A proposal comes out to study the chemistry of the pile under
radiation, with Fermi observing that two tubes leading into the
pile were planned for insertion of samples. Other technical topics
under discussion regards the use of carbide in pilot plant, but
not in experimental plant, the recommendation to use graphite
around the pile (for reflecting escaping neutrons), and the
problem of carbide production.
\\
Also interesting are the discussions about the responsibility for
clearance and security (leaved to Army), and the authorization for
giving information about the pile to ``anyone'', Conant being
recognized as the final authority on the distribution of
information. A mention is present about the use of the pile for
power but not for production of explosives.

\

\begin{itemize} \item[{\tiny WAT1043l}]
{\it $\diamondsuit$ Meeting of the Engineering Council} (2 pages),
\\
Present: Moore, Leverett, Steinbach, Fermi, Spedding, Wheeler,
Wigner, Seaborg and Wilson,
\\
Report CE-229 (August 8, 1942).
\\

\end{itemize}
\noi Minor discussions are present about safety and control rods,
the use of a covering layer of graphite, possible external (rather
than inside the pile) cooling with {oil}, etc..
\\
Fermi estimates a 90\% probability for achieving thermal stability
in the pile.

\

\begin{itemize} \item[{\tiny WAT1043m}]
{\it $\diamondsuit$ Meeting of the Technical Council} (2 pages),
\\
Present: Nichols, Hilberry, Spedding, Doan, Fermi, Steinbach,
Grafton, Boyd, Moore and Wheeler,
\\
Report CS-251 (September 4, 1942).
\\

\end{itemize}
\noi Fermi reports on the status of the values obtained in the
multiplication factor $k$. Other interesting discussions regard
the possible combination of water cooling and bismuth cooling
(with bismuth circulating in the uranium-containing channels,
between the aluminium jacket and the uranium rod), the status of
production of uranium (and, in particular, about material provided
by Alexander), and problems with radiation protection.
\\
Fermi proposes to consider the possibility to test uranium metal
with an exponential pile, with a remark that ``the metal won't be
better than the oxide''.

\

\begin{itemize} \item[{\tiny WAT1006}]
{\it $\diamondsuit$ Discussion of Helium Cooled Power Plant} (3
pages),
\\
Present: Leverett, Cooper, Moore, Wigner, Steinbach, Fermi,
Szilard and Wheeler,
\\
Report CS-267 (September 16, 1942).
\\

\end{itemize}
\noi Several problems related to He cooled power plant (steel and
compressors, control and safety rods, oil unaffected by radiation,
etc.) are discussed in this meeting. A list of topics to be
studied (and tasks to assign) is reported.
\\
It is also interesting to point out a remark by Szilard, which
argued that $10^5$ kW of power were necessary to win the war.

\

\begin{itemize} \item[{\tiny WAT1043n}]
{\it $\diamondsuit$ Meeting of the Technical Council} (11 pages),
\\
Present: Compton, Wheeler, Moore, Allison, Szilard, Fermi and
Spedding,
\\
Report CS-274 (September 18, 1942).
\\

\end{itemize}
\noi General discussions on technical details, including a water
cooled plant, are made.
\\
Quite interesting is the discussion of several points regarding
the policy for the site X. These points included: triple
extraction plant; water supply; association with the production
unit.
\\
It is pointed out that, while the work by {Lawrence and Urey} was
carried out under the only control of OSRD, that performed by the
Metallurgical Laboratory was under the control of OSRD through the
Army. Possible alternatives were considered to propose to work not
under the direct supervision of the Army, Szilard suggesting to
take no action until the pile was effectively operating.
Discussions about the ``future'' of the people working in the
project are made, with particular reference to the research policy
of the next 20 years.
\\
After other discussions about alternative locations for production
piles (Argonne, Tennessee, {Palos} Park), the attention is turned
on the committee proposed by Bush for the decision about the
destination of the fissile material: this had to be employed for
the design and production of bombs.
\\
Finally, the problem of further man power is discussed. Szilard
recommends to associate Auger, Rasetti, Goldhaber and Rossi,
taking care of an advice by Conant that the inclusion of such
people in the project was possible only if the work is made in a
restricted area. Another name proposed by Szilard is that of
Lewis, although it is recognized that he had already too many
duties, among which the development of methods for explosives in
China.

\

\begin{itemize} \item[{\tiny WAT1043o}]
{\it $\diamondsuit$ Meeting of the Technical Council} (8 pages),
\\
Present: Allison, Fermi, Moore, Wigner, Compton, Wheeler and
Oppenheimer,
\\
Report CS-281 (September 29, 1942).
\\

\end{itemize}
\noi The main discussions deals mainly with the work related to
the pile operating with fast neutrons, and the question of the
move of the work to Site X (logistic problems with the Site X,
etc.). To this regard, it has to be pointed out that Fermi
dissented about the shift of the work on fast neutrons to Site X.
\\
The choice of the Site X as the new main basis of the
Metallurgical Project, had by now been definitive (a preliminary,
rough map was also included), this having been favored to a
certain extent by Army, who preferred to have all the relevant
work in only one enclosure.
\\
Other non physics arguments, touched in the meeting, are about the
patent rights assignment to the American government and the
collaboration of British engineers. Instead, minor topics of
scientific interest are about helium and bismuth cooled plants and
water cooling.

\

\begin{itemize} \item[{\tiny WAT1043p}]
{\it $\diamondsuit$ Meeting of the Technical Council} (10 pages),
\\
Present: Allison, Fermi, Wigner, Compton, Whitaker, Moore, Cooper,
Szilard, Manley, McMillan, Wheeler and Doan,
\\
Report CS-284 (October 1, 1942).
\\

\end{itemize}
\noi As in the meeting of two days before, the arguments of the
discussion deals with work on fast neutrons and, especially, with
the move to Site X (with evaluation of alternative, possible
options). The majority of the presents prefer Chicago as a
suitable site in order not to waste time and, moreover, they
express their preference to work not under the supervision of Army
and independently of industry.
\\
Other discussions concern the work by Fermi on slow neutrons. The
development work for piles 1 and 2 was assumed to be out of way by
March 1 (1943); after this was completed, the schedule comprised
to work on other piles, including fast neutron pile and heavy
water pile.
\\
Mentions are made to bismuth cooling and to other technical
issues.

\

\begin{itemize} \item[{\tiny WAT1043q}]
{\it $\diamondsuit$ Meeting of the Technical Council} (9 pages),
\\
Present: Allison, Fermi, Szilard, Moore, Wigner, Whitaker,
Wheeler, Steinbach, Compton and Groves,
\\
Report CS-286 (October 5, 1942).
\\

\end{itemize}
\noi Detailed discussions regard results about different schemes
for the pile: 1) (Wigner) uranium rods, water cooling with pipes,
graphite as moderator; 2) (Fermi) uranium lumps {imbedded} in
graphite, cooling by occasional water pipes; 3) (Cooper) metal
pipes, {shot}, and graphite as moderator, with removal of uranium
metal and recharging; 4) external cooling (which gives about 300
kW power only). No definitive decision is adopted on this
argument.
\\
At a certain point of the meeting, Compton and General Groves came
in reporting, quite interestingly, that War Department considered
the Metallurgical Project important. The discussion then changed a
little, with Allison's claims that one couldn't win the war with
an externally cooled plant and Fermi's remark that the program
will be delayed by several months due to change plan at the
Argonne site.

\

\begin{itemize} \item[{\tiny WAT1043r}]
{\it $\diamondsuit$ Meeting of the Technical Council} (9 pages),
\\
Present: Allison, Wigner, Compton, Whitaker, Szilard, Wheeler,
Fermi, Moore, Cooper, Steinbach and Kirkpatrick,
\\
Report CS-290 (October 7, 1942).
\\

\end{itemize}
\noi Various arguments are treated, all related (directly or
indirectly) to the move to the Site X.
\\
Groves is said to be eager to have explosives, by June 15, 1943,
several plants being assumed to be operating after June 15 (dates
by Groves have not been reported in the minutes). Some
discouragement in Army is reported, that the project had not
achieved more so far. {Moral} effect of operations is discussed as
well.
\\
Technical council recommends the construction at site X of a 300
kW pile (Pile 2) by March 15 and, to this end, Fermi notes that
there was no need that such a pile be made with uranium only in
the form of metal.
\\
The decision by Groves about production and extraction of fissile
material at Site X is discussed, as well as concentration of the
work on fast neutrons at the same site (Groves' decision urged by
Oppenheimer). Topics related to power utilization and long term
research (not) at the same Site X, with possible
``countermeasures'' to undertake regarding this last point, are
also considered.
\\
The time schedule for Pile 1 is committed to Fermi, that for Pile
2 to Whitaker, etc.. In connection to possible changes in the
plans due to move of the work, Fermi notes that the original plan
(for Pile 1) was to prove that chain reaction goes and to flash
pile for a limited time.
\\
A minor discussion on the question of plating uranium (and
possible reactions of it) is also made.

\

\begin{itemize} \item[{\tiny WAT1043s}]
{\it $\diamondsuit$ Meeting of the Technical Council} (6 pages),
\\
Present: Allison, Szilard, Wigner, Moore, Wheeler and Fermi,
\\
Report CS-294 (October 12, 1942)
\\

\end{itemize}
\noi The questions of machining graphite and sintered uranium
metal are discussed in some detail.
\\
Several options for an externally cooled pile are considered by
using: 1) copper pipes; 2a) copper shell cooled by air; 2b) copper
shell cooled by water spray. Fermi's preference is for choice 2a),
assuming that copper didn't leak; however he himself points out
that it would be difficult to find leak (then, the rubber cap
method was considered). A general agreement is expressed to leave
this problem to Fermi to look into.
\\
A peculiar remark by Fermi is about his feeling that time
estimates for pile to work were not certain, since the amounts of
uranium metal required (to prevent loss in $k$) were probably
underestimated. Nevertheless, he considered a mistake to wait any
time at all for producing the chain reaction in this way, while he
favored to put the uranium oxide in form of spheres.
\\
Other interesting suggestions by Fermi are about the change in
picture of $k$, and the use of a ``sandwich'' experiment with 4
layers of uranium metal (about 1 ton); U$_3$O$_8$ pile was
suggested to be used as a standard, and a gain of 0.8\% in $k$ was
estimated just by removing nitrogen.
\\
More discussions are again about the move to Site X: people felt
not to transfer until the working situation was not clear. Fermi
observes that all matter about Site X appeared to arise from a
mistaken impression that experimental work was practically
finished.

\

\begin{itemize} \item[{\tiny WAT1008}]
{\it $\diamondsuit$ Conference on Lattice Spacing} (2 pages),
\\
Present: Steinbach, Leverett, Fermi, Wigner and Wheeler,
\\
Memo \#15 (October 21, 1942).
\\

\end{itemize}
\noi Estimates about optimum lattice spacing in the pile and C/U
ratio for the helium cooled plant are presented, by taking into
account the request of a minimum total amount of uranium metal.
Fermi suggests to save the amount of metal by diminishing the
proportion of U to C toward the outside of the pile.

\

\begin{itemize} \item[{\tiny WAT1043u}]
{\it $\diamondsuit$ Meeting of the Technical Council} (4 pages),
\\
Present: Allison, Compton, Fermi, Moore, Spedding, Szilard and
Wigner,
\\
Report CS-356 (November 19, 1942).
\\

\end{itemize}
\noi Preliminary agreements are discussed in this meeting about
the preparation of a report for a committee (which included
industrial specialists on production problems) formed for
examining the Metallurgical project in Chicago (alternative to the
project based at Berkeley), that will come on November 26. The
topics discussed regard the purity of the final product
(plutonium), its radioactivity, spontaneous heating, etc., by
pointing out that very little was known about the metallurgy of
plutonium, and the processes proposed for producing it seemed very
far from industrial possibilities.
\\
Remarks are present about the availability in U.S.A. of 1000 tons
of uranium, and that no slowing down of neutrons would be required
for Piles 1 and 2, although Wigner pointed out that studies on
fast neutron reactions were still preliminary.
\\
About non strictly scientific issues, it is quite interesting to
note the invitation from Washington authorities to go ahead with
the production of plutonium, and the remark on General Groves who
was interested in all possible military uses of what studied in
Chicago, rather than applications to power production.
\\
In these notes it is pointed out that a report on the chain
reaction of about 10 pages should have been prepared by Fermi
(along with the contributions of other scientists on other
subjects), this report being unknown.

\

\begin{itemize} \item[{\tiny WAT}]
{\it $\diamondsuit$ Notes on Meeting of April 26-28, 1944} (5+8
pages),
\\
Present: Fermi, Allison, Wigner, Smyth, Szilard, Morrison, Watson,
Feld, Hogness, Young, Weinberg, Creutz, Cooper, Vernon and
Ohlinger,
\\
Report N-1729, Eck-209.
\\

\end{itemize}
\noi This report is composed of two papers accounting for a
two-day meeting (April 26-28, 1944).
\\
A large part of the first paper, with much of the Fermi's
intervention at the two-day meeting, was already published in the
{\it Collected Papers} \cite{FNM211}. There, the discussion
focused on chain reaction for the production of a power of about
$10^6$ kW. A large {mother} plant was conceived for producing
plutonium to be used as fissile material in smaller plants; Fermi
noted that this arrangement could be useful for the heating of
towns. Then, after a brief theoretical discussion, with numerical
estimates and data, for the full metal utilization, Fermi focused
on four different types of piles, both operating with slow
neutrons and fast neutrons, and depending on the percentage of the
enrichment of the fissile material and the moderator employed. The
remaining part of the first paper, not published in \cite{FNM211},
dealt with alternatives to what discussed by Fermi proposed by
Szilard, coming out by ``assuming more optimistic values of the
constants so as to indicate other potentialities''.
\\
The second paper dealt mainly with a discussion by Morrison on
several scientific, economic, and social issues related to pile
producing power, resulting quite interesting from an historical
point of view. Among the remarks to the Morrison's relation, we
mention that ``Fermi questioned the estimated value of [the number
of neutrons produced/number of fissionable atoms used up] = 2.5 on
the ground that it might be too optimistic and pointed out that
there is a long range future in developing the full utilization of
[$^{238}$U] and thorium''.
\\

\subsection{Edited reports}

In the Wattenberg archive several reports are present, describing
some of the work performed in 1943 and 1944 by the division of the
Metallurgical Laboratory headed by Fermi. As it is already known
(see, for example, Ref. \cite{FermiFisico} or the introduction to
the papers of Volume II of Ref. \cite{FNM}), Fermi played a very
active role in the work officially assigned to others, so that
although none of these reports describes (theoretical or
experimental) activities directly performed by him, they
nevertheless reveal precious information about part of the work
done under the supervision of Fermi. Of course, these few reports
account only for activities not directly related to military
applications, although they were finalized primarily to the
production of fissile material for explosives.

The detailed description of the four reports available follows
below.
\\

\begin{itemize} \item[{\tiny WAT1023}]
{\it * Report for Month Ending September 25, 1943} (55 pages),
\\
edited by A.H. Compton, S.E. Allison and E. Fermi,
\\
Report CP-964.
\\

\end{itemize}
\noi This report does not contain the description of the work
performed directly by Fermi, but rather it describes some of the
activities performed in the month of September 1943 by different
people under Fermi's supervision.\footnote{In this report, the
summary of these activities was written (on September 27, 1943) by
Wigner ``in the absence of Mr. Fermi''.} The most interesting ones
are summarized below.
\\
Much work, both experimental and theoretical, was devoted to the
study of a so called P-9 pile, that is a chain reacting system
with heavy water as a coolant. The Zinn group was involved in
making plans for an experimental P-9 plant at Argonne, while the
Young group worked on the design of a P-9 pile, both for a
heterogeneous and a homogeneous pile. In a P-9 pile more fissile
material had to be used for several technical reasons (related to
pumps and heat exchangers employed), but this was compensated by a
higher value of the effective multiplication factor $k$. Other
problems to be solved were that of the separation of the uranium
oxide from the circulating heavy water and the method to choose
for separating the heavy water from the cooling liquid (in order
to use it again after a given cycle), the determination of the
critical size for a P-9 pile, etc.. A sketch showing one possible
arrangement for a (near) homogeneous P-9 slurry pile was presented
as well.
\\
Another research conducted by some people of the Zinn group
regarded the ``cell saturation'' effect, induced by increasing the
absorption cross section of a single cell (to be used in the
lattice of a pile) to such an extent that the change in reactivity
of the pile was no longer proportional to the amount of impurity
added, but rather to its square.
\\
The group headed by Anderson studied, among the other things, the
residual radioactivity of control rods made of different
materials, the effect of fast fission on the multiplication
factor, and remeasured the ratio of the absorption cross sections
of boron and hydrogen.
\\
Marshall, instead, studied and prepared a velocity selector
consisting of a sandwich of aluminium and cadmium sheets for
obtaining measurements on neutrons with definite energy (see
previous section), while Morrison performed an experimental study
on the boundary conditions for the neutron density between
paraffin and graphite for a study on a neutron reflector, with the
determination of the temperature effect on the diffusion length in
graphite.
\\
The Feld group got involved in the investigation on inelastic
cross sections of several heavy elements (lead, bismuth, iron and
uranium), which were relevant for fast neutron chain reaction,
while started novel measurements on the (n,2n) and ($\gamma$,n)
reactions on beryllium with the paraffin pile technique.
\\
In this report, a large account is also given to some theoretical
activity performed by several people, as emphasized by Wigner
himself: ``the Theoretical Physics Section's report for this month
is in considerably more detail than was the custom in previous
months''. Then Wigner interestingly continues: ``It is not
expected to report in similar detail in the future, as a good part
of the work done by us is principally for our own use. However, it
was intended to give a more adequate picture of the work that we
are doing''.
\\
As already mentioned, much of the theoretical work was devoted to
the P-9 pile, but another interesting investigation (by some
people of the Weinberg group) dealt instead with the fast chain
reacting pile, having found that the measured cross section for
fast fission was smaller than previously assumed (the ratio of the
fast fission neutrons to the thermal fission ones was previously
measured incorrectly).
\\
Other researches regarded the study of possible danger situations
in which control rods could not be governed due to a pressure
damping in the cooling circuit, or the studies on the preferred
geometrical form of a pile, its dimensions or other technical
details (pumps, valves, heat exchangers, etc.)

\

\begin{itemize} \item[{\tiny WAT1027}]
{\it * Report for Month Ending December 25, 1943 - Part II} (6
pages),
\\
edited by A.H. Compton, S.E. Allison and E. Fermi,
\\
Report CN-1190.
\\

\end{itemize}
\noi This report does not contain the description of the work
performed directly by Fermi, but rather it describes some of the
activities performed in the month of December 1943 by different
people under Fermi's supervision. The summary of the first part of
this report is in Ref. \cite{FNM199}, while the second part
dealing with the work performed by the Anderson group is
considered here.
\\
The Anderson group determined the yield of plutonium per kWh for
the Argonne pile, and predicted that the Hanford 250 kW operations
should produce about 230 grams of plutonium per day. They also
studied neutron yields from polonium by irradiating different
samples (lithium sulphur, chlorine and argon) with alpha
particles.

\

\begin{itemize} \item[{\tiny WAT1037}]
{\it * Report for Month Ending August 26, 1944} (38 pages),
\\
edited by A.H. Compton, E. Fermi and W.H. Zinn,
\\
Report CP-2081.
\\

\end{itemize}
\noi This report does not contain the description of the work
performed directly by Fermi, but rather it describes some of the
activities performed in the month of August 1944 by different
people under Fermi's supervision. The most interesting ones are
summarized below. It was here noted that the Chicago Pile N.3
completed its second {month} of operation, while a silver {tape}
recorder was completed and installed on CP-2.
\\
The Seren group studied the properties of the thermal column (see
the previous section), while Zinn performed measurements on the
Bragg reflection of a highly collimated beam of thermal neutrons
and, more in general, on neutron spectroscopy.
\\
Licthenberger studied, instead, the scattering from strong
absorbers, Morrison and Teller having identified the isotope
$^{112}$Cd as that responsible for the strong capture of thermal
neutrons, while Wattenberg prepared photo-neutron sources made by
activation of several nuclides.
\\
Finally, Anderson focused on the neutrons from the reaction
$^3$H+$^2$D$\rightarrow^4$He+n and, in general, studied possible
transformations of thermal into fast neutrons, while the Nagle's
group measured the yield of delayed neutrons.

\

\begin{itemize} \item[{\tiny WAT1039}]
{\it * Report for Month Ending October 28, 1944} (27 pages),
\\
edited by A.H. Compton, E. Fermi and W.H. Zinn,
\\
Report CP-2301.
\\

\end{itemize}
\noi This report does not contain the description of the work
performed directly by Fermi, but rather it describes some of the
activities performed in the month of October 1944 by different
people under Fermi's supervision. The most interesting ones are
summarized below.
\\
The group guided by Zinn studied the poisoning of the chain
reaction by $^{135}$Xe in CP-3 (see the previous section) and
related arguments, and May and Anderson measured the nuclear
constants of $^{233}$U, whose behavior they found similar to that
of $^{235}$U but giving larger values for $k$, so that the use of
$^{233}$U was suggested to be more favorable than that of Pu.
\\
Langsdorf studied, instead, the resonance scattering of neutrons,
while the Seren group measured the activation cross section of
columbium Cb, which resulted to be a useful information for
producing stainless alloy with uranium.

\

\begin{itemize} \item[{\tiny WAT1040}]
{\it * Report for Month Ending November 25, 1944} (24 pages),
\\
edited by A.H. Compton, E. Fermi and W.H. Zinn,
\\
Report CP-2436.
\\

\end{itemize}
\noi This report does not contain the description of the work
performed directly by Fermi, but rather it describes some of the
activities performed by different people under Fermi's
supervision. The most interesting ones are summarized below.
\\
The Zinn group continued their studies on the Bragg reflection of
thermal neutrons from a crystal (considered as a neutron
spectrometer), and observed also the total reflection by Cu, Al,
Be, glass and graphite mirrors.
\\
The Lichtenberger group, instead, made boron absorption
measurements in order to study the variation with energy of the
resonance absorption of $^{238}$U, while the Wattenberg group
mainly focused on photo-neutrons from U+Be sources.
\\
Finally, the Hill group studied the tuning of coincidences in
$\alpha$-chambers and made an analysis of a number of pictures
from cloud chambers searching for ternary fissions (and possible
appearance of three-particle fissions).
\\

\subsection{Lecture Notes}

Once the pile program of the Metallurgical Project in Chicago was
sufficiently advanced not to need a continuous attention by Fermi,
he definitively moved to Los Alamos (in September 1944) to join
the Manhattan Project. Here Fermi began to give isolated lectures
on many different subjects \cite{AndersonSachs,FermiFisico},
related to that project, for the benefit of the people who worked
at Los Alamos, many of them being just students or graduated guys.
Then, after the end of the war, in the Fall of 1945 he taught a
regular course on neutron physics to about thirty students: this
was the first time that such a complete course was given, ranging
over more than ten years of important discoveries, and also the
first occasion for the scientists who contributed in those
achievements to pause and reason a bit more on the results
obtained.

We know about the content of this course from the notes taken down
in class by one of the attending students, I. Halpern, who
assembled them into a (classified) typescript on February 5, 1946.
A first part of the Fermi lectures at Los Alamos, containing
neutron physics without reference to chain reactions, was
declassified on September 5, 1946, while the remaining part has
been declassified only in 1962. Both parts have been later
published in the {\it Collected Papers} by Fermi \cite{FNM222}.
Leaving aside the pregnant didactic style by Fermi, the main
relevance of such notes is, as we have already mentioned, that
they present for the first time a complete and accurate treatment
of neutron physics from its beginning, including a detailed study
of the physics of the atomic piles. In this respect it is not
surprising that especially the second part of the notes, dealing
just with chain reactions and pile physics, was considered as
``confidential'' material by governmental offices.

However, we have recently recovered a {\it different} version of
the Fermi lectures at Los Alamos, formerly belonged to James
Chadwick and now deposited at the Chur\-chill Archive Centre in
Cambridge (U.K.). The folders relevant to us are essentially two.
The first one (CHAD I 17/3) contains a letter from R.T. Batson of
the Atomic Energy Commission (A.E.C.), a copy of the paper {\it
Elementary Theory of the Pile} by Fermi\footnote{This paper is
reproduced in Ref. \cite{FNM}; in particular see page 538 of
Volume II.} and a copy of only the {\it first part} of the Halpern
notes of the Fermi lectures. The second folder (CHAD I 4/1)
contains a version of the {\it complete} set of lectures made by
A.P. French, dated June 23, 1947.

It is apparently not strange that the material of the first folder
belonged to Chadwick, since he was the respected (also by
Americans) leader of the British Mission in the United states. The
biggest part of the British contingent was, in fact, at Los
Alamos, and Chadwick himself was present at the world's first
nuclear test at Alamogordo on July 16, 1945. Several scientists of
the British Mission were very young and, among the others, it was
Anthony P. French who graduated in Physics at the Cambridge
University just in 1942. In the same year he joined the atomic
bomb project (``Tube Alloys'') at the Cavendish Laboratory, and
was later sent to Los Alamos in October 1944 as a member of the
British Mission. Here he worked with E. Bretscher, O.R. Frisch, J.
Hughes, D.G. Marshall, P.B. Moon, M.J. Poole, J. Rotblat, E.W.
Titterton and J.L. Tuck in the field of experimental nuclear
physics \cite{Szasz},\footnote{The remaining part of the British
Mission was composed by B. Davison, K. Fuchs, D.J. Littler, W.G.
Marley, R.E. Peierls, W.G. Penney, G. Placzek, H. Sheard and
T.H.R. Skyrmes.} and returned to the United Kingdom in 1946,
working for two years at the just newly formed Atomic Energy
Research Establishment (A.E.R.E.). The second folder of the
Chadwick papers mentioned above contains just the notes of Fermi
course on neutron physics taken by French on his own, when he was
at Los Alamos, and later (1947) re-organized into a final version
when he came back to England.

The present recovery thus shows a clear historical and scientific
relevance. However, while the historical interest is the main
subject of a different paper \cite{FermiFrench}, we here focus
only on the scientific relevance of that recovery, which is
clearly centered about the fact that our previous knowledge of the
Fermi course was incomplete and, to some extent (limited to the
Halpern notes) misleading. As his usual, Fermi was very accurate
in the choice of the topics, that he developed in detail and in a
very clear manner, a peculiarity which does not often emerge from
the notes taken directly down in class by students, and later
arranged into the Halpern version.

We have performed a careful analysis of the mentioned documents,
whose main results are summarized below. First of all, our study
has shown that the French notes do {\it not} depend on the Halpern
ones, but French probably saw them (the organization of the
introduction is similar). The topics covered are exactly the same,
although to a certain (minor) extent the material is organized in
a little different manner. The detailed table of contents
(including sections and subsections) follows.
\\

\begin{itemize} \item[{\tiny CHAD}]
{\it * Neutron Physics. A Course of Lectures by E. Fermi} (113+iii
pages),
\\
Notes by A.P. French (June 23, 1947).
\\
\end{itemize}
\noi

\begin{itemize}
\item[1.] Sources of neutrons\\
- Natural Sources\\
${}$ \quad (a) Alpha particle sources \\
${}$ \quad (b) Photo-neutron sources\\
- Artificial Sources
\item[2.] The Isotopic Chart: Nuclear Masses and Energies\\
- The Isotopic Chart\\
- Energy Balance of Reactions\\
- The Binding Energies of Nuclei\\
- The Packing Fraction Curve
\item[3.] The Scattering of Neutrons (Part 1)\\
- General Considerations\\
- Elementary Theoretical Treatment\\
${}$ \quad 1) Elastic Scattering \\
${}$ \quad 2) Inelastic $(n,n)$ scattering\\
${}$ \quad 3) Inelastic $(n,m)$ scattering\\
${}$ \quad 4) Inelastic $(n,\gamma)$ scattering
\item[4.] Resonance: Models of the Nucleus\\
- Resonance in Nuclear Reactions\\
- Two Models of the Nucleus
\item[5.] The Scattering of Neutrons (Part 2)\\
- The Solution of Schrodinger's Equation\\
- The Scattering Cross Section\\
- Neutron-Proton Scattering\\
- The Breit-Wigner Formula\\
- The Effect of Chemical Binding on Scattering\\
- Scattering Cross Sections for Other Elements
\item[6.] Slow Neutrons as Waves\\
- Introduction\\
- Isotopic Effects\\
- Penetration of Thermal Neutrons\\
- The Production of Very Slow Neutrons\\
- Reflection and Refraction
\item[7.] The Slowing Down of Neutrons\\
- Introduction\\
- The Energy Loss in One Collision\\
- Many Collisions\\
- The Spatial Distribution of Slowed Neutrons\\
- Theory of the Spatial Distribution
\item[8.] The Age Equation\\
- Derivation of the Age Equation\\
${}$ \quad 1) Diffusion\\
${}$ \quad 2) Energy Drift\\
- The Problem of Point Source
\item[9.] Thermal Neutron Distribution\\
- The Basic Equation\\
- Point Sources of Slow Neutrons\\
- Point Sources of Fast Neutrons\\
- Bounded Media\\
- The Measurement of Diffusion Length\\
- Diffusion in Graphite\\
- Some Useful Quantities and Relationships
\item[10.] The Reflection of Neutrons\\
- Introductions\\
- Approximate Solution of the Escape Problem\\
- Exact Solution by the Integral Equation Method\\
- The Albedo\\
- Measurement of the Albedo
\item[11.] The Stability of Nuclei\\
- The Binding Energy of a Nucleus\\
${}$ \quad 1) The Liquid Drop Model\\
${}$ \quad 2) Nuclear Composition\\
${}$ \quad 3) Coulomb Forces\\
${}$ \quad 4) The Odd-Even Effect\\
- Determination of Coefficients\\
- The Binding of Neutrons in Nuclei
\item[12.] Nuclear Fission\\
- The Possibility of Fission\\
- Limitations on the Occurrence of Fission\\
- The Liquid Drop Model in Fission\\
- The Particles of Fission\\
- Cross Section for Fission and Other Processes
\item[13.] The Possibility of a Chain Reaction\\
- The Properties of Natural Uranium\\
${}$ \quad 1) The High Energy Region\\
${}$ \quad 2) The Thermal Region\\
- Moderators\\
- Homogeneous and Lumped Graphite Piles\\
- The Possibility of a Homogeneous Pile
\item[14.] The Heterogeneous Pile\\
- The Design of a Lumped Pile\\
- The Determination of Pile Constants\\
${}$ \quad 1) The Magnitude of $\epsilon$\\
${}$ \quad 2) $(1-f_R)$\\
${}$ \quad 3) Calculation of $f_T$\\
- Reproduction Factor and Critical Size
\item[15.] The Time Dependence of a Pile\\
- The Time-Dependent Equation\\
- Evaluation of the Period
\item[16.] Practical Aspects of Pile Physics\\
- The Determination of $k$\\
- The Study of Pile Materials\\
- Energy and Radiation Production\\
- Shielding\\
- Other Types of Pile
\item[17.] Fast Reactors\\
- Elementary Considerations\\
- The Integral Equation to the Neutron Distribution\\
- The Critical Size for a Fast Reactor\\
- Supercritical Reactors\\
${}$ \\
Problems and Exercises\\
\end{itemize}

Almost all the topics listed above were expounded by Fermi;
according to French, when Fermi was absent, R.F. Christy and E.
Segr\`e treated the scattering of neutrons and the albedo in the
reflection of neutrons, respectively.

The text of the notes is different in the French and Halpern
versions; in few cases, however, similar or even identical words
or sentences are present in both versions, likely denoting quotes
from an original wording by Fermi. In general, the French notes
are much more detailed and accurate (as may be roughly deduced
even looking at the table of contents reported above), with a
great number of shorter or larger peculiar additions\footnote{The
case is completely different, for example, from that of the
revision of the (first part of the) Halpern notes made by J.G.
Beckerley in 1951 (document AECD 2664 of the Atomic Energy
Commission). Here the author {\it re-wrote} the Fermi lectures by
including several additions from {\it other} sources, ``where
clarity demanded more information and where the addition of recent
data made the text more complete.'' Contrarily to the present case
(as it is evident from the text of the notes), Beckerley ``was not
privileged to attend the course'' by Fermi.} (explanations,
calculations, data or other, and 5 more exercises) not present in
the Halpern notes. It is quite interesting that the greater detail
already present in the French notes increases even more in quality
(especially figures and data) in the last part, directly related
to chain reactions and their applications, and, moreover, explicit
references to bomb applications are made (see below). By limiting
ourselves to significative scientific remarks or discussions, the
French version of the Fermi lecture notes contains about 100
additions, 18 of them being quite relevant while the remaining
part accounts for minor remarks, calculation details or figures.
Instead the peculiar additions present in the Halpern version but
not in the French one are only about 30 (and 3 more exercises),
only one of them being relevant. Also, the French paper contains
the six questions which were set as a final examination at the end
of the lecture course.

The most relevant additions deals with the following (the page
number refers to that present in the French manuscript):
\\

- (French, page 6) the entire section {\it The Binding Energies of
Nuclei}, where the definition of the binding energy and an example
for calculating it in a specific case is reported;

- (French, page 19) the introduction of the first section of
chapter 5: ``In this section we consider the solution by wave
mechanics of a simple problem in nuclear scattering. The nucleus
is considered as a centre of force, the force being of short
range, so that it ceases to exist beyond a certain distance $r_0$
from the origin. The actual shape of the nuclear potential then
approximates to a square well, as shown in Fig. 14. The potential
$U$ is negative and constant over most of the nucleus. This
corresponds to the facts, as far as we know them, of the
interaction between a neutron and a nucleus. The depth of the
nuclear potential well is equal to the binding energy, that is
about 8 MeV'';

- (French, page 27) some details about the Bragg scattering of
slow neutrons by an element with different isotopic composition,
ending with the following remark: ``the total scattering intensity
is thus given by $I_{\rm sc} = {\rm const.} \, ( \sigma_1 +
\sigma_2 \pm \sqrt{\sigma_a \sigma_2})$, and may be seen to
consist of coherent and incoherent contributions, the latter not
being subject to interference'';

- (French, pages 35-36) a long discussion, with detailed
calculations about the spatial distribution of slowed neutrons,
aimed at calculating the source strength in neutrons per second
both for a thermal detector and for a resonance detector (final
explicit expressions are reported);

- (French, page 42) calculation details about the neutron
scattering in a medium (with the determination of the mean free
path), ending with a prediction for the neutron-proton scattering
cross section (in water) of $\sigma \simeq 20$ barns  which
``agrees very closely with the accepted value'';

- (French, pages 52-53) discussion on calculation details aimed at
solving the so-called (Fermi) age equation for the diffusion of
neutrons from a Ra-Be source in a column of graphite of square
section (with length of side $a$) and infinite length; the
effective length of a side of the column, $a=a_{\rm geometrical} +
2 \cdot (0.67 \, \lambda)$ ($\lambda$ being the mean free path),
and the range of the neutrons, $r_0=\sqrt{4 \tau}$ ($\tau$ being
the age parameter), are introduced; the numerical values of $r_0$
(instead of only $\tau$ as in the Halpern notes) for three
(instead of two) typical neutron energies are given; the addition
in the French notes ends with the peculiar observation that ``we
have the somewhat paradoxical result that the system can be made
infinite for fast neutrons being slowed down, but not for the same
neutrons when they have become thermal'';

- (French, pages 59-60) after some calculations, the section on
the measurement of the albedo ends with the observation that ``a
thermal neutron in these media [paraffin and water] makes about
100 collisions before being captured. The distance it travels,
measured along the path, is about 80 cm on the average, and the
time it takes to do this, which is its lifetime as a thermal
neutron, is something less than a millisecond'';

- (French, page 61) introductory remarks on the binding energy of
a nucleus, with the theoretical expression for the measured mass
of an atom in terms of $A,Z$ and the said binding energy;

- (French, pages 63-65) several important additions related to the
stability of nuclei (according to the even- and odd-ness of $Z$,
$A$ or both) and the accurate determination of the expression of
the binding energy of nuclei in terms of $Z$ and $A$, with several
numerical data (and a graph);

- (French, pages 72-73) the inclusion of three graphs for the
cross section of $(n,\gamma)$ and $(n,{\rm fission})$ processes on
uranium as function of the incident neutron energy;

- (French, pages 79-80) relevant additions about homogeneous and
lumped graphi\-te piles: explicit calculations of the neutron
absorption volume by uranium spheres of 3 cm radius (and of other
quantities) lead to the conclusion that ``no homogeneous pile of
this type will work, and we must therefore devote our attention
(if we are considering only the U-graphite combination) to
heterogeneous piles'';

- (French, page 81) small introduction (with key comments) to the
design of a lumped pile, with figures of lattice structures with
spherical lumps or rods of uranium;

- (French, page 84) two relevant figures (and related discussion)
about efficient cooling systems (by blast of air or water flowing)
for piles; introductory remarks to the section dealing with the
reproduction factor and critical size of a pile, with a graph of
the actual neutron density in a finite pile as a function of the
distance from the center of the pile;

- (French, pages 86-87) explicit expressions and related comments
on the reproduction factor $k$ as a function of geometrical and
other parameters of the pile;

- (French, pages 99-100) ``We have discussed the mechanism of
thermal neutron chain reactions. The question now arises how to
produce a nuclear explosion''; introductory remarks about fast
reactors starting from the calculated expression for the growth of
neutron density in a reactor;

- (French, pages 100-101) definition, calculations and related
discussion on the transport cross section and transport mean free
path for neutrons in a fast reactor with a core of $^{235}$U and a
tamper (neutron reflector);

- (French, pages 103-104) discussion of equilibrium conditions
(with explicit expressions) for a fast reactor and mathematical
expressions for some quantities describing neutron losses;

- (French, page 110) the end of the chapter on fast reactors (and,
then, of the lecture notes) is: ``in this way one can calculate
the e-folding time for a fast reactor, and its value thus found
will be valid until mechanical effects set in  -- these having to
be known before the efficiency etc. of the bomb can be
estimated''.
\\

The only relevant addition in the Halpern version is, instead:
\\

- the description of a fast neutron detector, based on the
scattering of a neutron flux by a paraffin layer (see page 471 of
Volume II of Ref. \cite{FNM}).
\\

\subsection{Other contributions}

Few other minor documents, of different nature and relevance, have
come to light during our research.

Although not properly a paper or a report, the first document we
point out here is a letter written by Fermi to Lord Rutherford as
early as in 1934, when he and his group in Rome started to study
the radioactivity induced by neutron bombardment. As recalled
above, these studies led, in October 1934, to the discovery of the
important properties of slow neutrons (see Patent USP1). This
document is presently conserved among the Rutherford Papers at the
Cambridge University Library (U.K.). The text of the letter is as
follows:
\\
\begin{quote}
{}\hfill{Rome, June 15th, 1934}\\

My dear Prof. Rutherford, \\
I enclose a reprint of a paper on the present status of our
researches on the activation of uranium. The same results shall
appear shortly in Nature.

We have been forced to publish these results of a research which
is actually not yet finished by the fact that the newspapers have
published so many phantastic [sic!] statements about our work that
we found it necessary to state clearly our point of view.

We are now engaged in trying to understand the influence of the
neutron energy on the activation of elements. We try to do this
using neutrons from a source of Em+B.

We are interested in this problem not only as it can throw some
light on the processes involved, but also because we plan to
construct a neutron tube similar to that of the Cavendish
laboratory.

In this connection I would be much obliged if in case you have
tested your tube for activating elements, you would let me know
some data on the intensities of the activations.

The construction of this tube would be much facilitated for us if
it were possible for some assistants of our laboratory (Drs.
Amaldi and Segr\'e) to come this summer to the Cavendish
Laboratory in order to see the apparatus and possibly be
instructed about its use.

I would be very grateful to you if you will give me an answer on
this point.

With kindest regards \\

{}\hfill{\begin{tabular}{c} yours very truly \\ \\ Enrico Fermi
\end{tabular}\qquad}
\end{quote}

Here, the relevant information, mainly from an historical point of
view, is the reference to the construction of ``a neutron tube
similar to that of the Cavendish laboratory'' and, particularly,
the request by Fermi of ``some data on the intensities of the
activations'' ought to be obtained by the Rutherford group at the
Cavendish Laboratory. Indeed, this testify for an attempt made by
Fermi to set up some collaboration among the two groups even {\it
before} that Amaldi and Segr\`e came to Cambridge in the summer of
1934. Although four letters by Rutherford to Fermi were known
(they are conserved at the Domus Galilaeana in Pisa, Italy), such
collaboration at a distance had not been addressed previously. The
reply to the letter by Fermi is, indeed, known\footnote{It is
reported by E. Amaldi in his not very known paper {\it Neutron
Work in Rome in 1934-36 and the Discovery of Uranium Fission},
Riv. Stor. Sci. {\bf 1}, 1-24 (1984).}, but this fact cannot be
deduced from it. The answer by Rutherford to the specific request
by Fermi was negative, denoting the advantage of the Rome group
over that of the Cavendish Laboratory on this point (probably
unexpected by Fermi): ``I cannot at the moment give you definite
statement as to the output of the neutrons from our tube but it
should be of the same order as from an Em+Be tube containing 100
millicurie and may be pushed much higher.''
\\

The second document we consider here is the following:
\\

\begin{itemize}
\item[{\tiny }] {\it Total Collision Cross Section of Negative
Pions on Protons},
\\
by D.E. Nagle, H.L. Anderson, E. Fermi, E.A. Long and R.L. Martin,
\\
Phys. Rev. {\bf 86}, 603 (1952).
\\

\noi ``The transmission of negative pions in liquid hydrogen has
been measured using the pion beams of the Chicago
synchrocyclotron. Pion beams with energies from 60 to 230 MeV were
used. The transmissions were measured using scintillation counting
techniques. The total collision cross section increases with
energy starting from small values at 30 MeV and rising to the
``geometrical'' value of about $60 \times 10^{-27}$ cm$^2$ at
about 160 MeV. Thereafter up to 220 MeV, the cross section remains
close to this value. The steep energy dependence at low energies
is consistent with interpretation that the pion
is pseudoscalar with a pseudovector interaction.'' \\
\end{itemize}

As the companion paper in Ref. \cite{FNM249a}, it was presented at
the 1952 Annual Meeting of the American Physical Society held at
New York on January 31 - February 2, 1952; in the mentioned
journal, only the abstract of both papers were reported, as custom
for the proceedings of that meeting. It testifies for some of the
work performed at Chicago by Fermi and his collaborator on pion
physics \cite{FNMpion}; the results are summarized in the abstract
reported entirely above. Strange enough, the paper considered does
not appear among the {\it Collected Papers} \cite{FNM}, contrary
to what happen for the paper in \cite{FNM249a}, although both
abstracts were published in the journal on the same page.

The last document is a popular article written by Fermi for a
newspaper\footnote{Although this paper was effectively published,
again it was not included among the {\it Collected Papers}, so
that it is practically unknown.}, in the occasion of the tenth
anniversary of the operation of the first chain reacting pile at
Chicago, on December 2, 1942:
\\

\begin{itemize}
\item[{\tiny }]  {\it Fermi's own story},
\\
by E. Fermi,
\\
Chicago Sun-Times, November 23, 1952.
\\
\end{itemize}
\noi Here Fermi gave a personal description of that event,
preceded by a short story of the main stepping-stones that leaded
to the realization of the first chain reaction, starting from the
discovery of radioactivity by H.A. Becquerel. It is particularly
interesting the conclusion of this article, where Fermi stated his
view (and hope) about science and possible military applications
of it:
\begin{quote}
The further development of atomic energy during the next three
years of the war was, of course, focused on the main objective of
producing an effective weapon.

At the same time we all hoped that with the end of the war
emphasis would be shifted decidedly from the weapon to the
peaceful aspects of atomic energy.

We hoped that perhaps the building of power plants, production of
radioactive elements for science and medicine would become the
paramount objectives.

Unfortunately, the end of the war did not bring brotherly love
among nations. The fabrication of weapons still is and must be the
primary concern of the Atomic Energy Commission.

Secrecy that we thought was un unwelcome necessity of the war
still appears to be an unwelcome necessity. The peaceful
objectives must come second, although very considerable progress
has been made also along those lines.

The problems posed by the world situation are not for the
scientist alone but for all people to resolve. Perhaps a time will
come when all scientific and technical progress will be hailed for
the advantages that it may bring to man, and never feared on
account of its destructive possibilities.
\\
\end{quote}

\section{Conclusions}

\noi In the present paper we have given a detailed account of the
many documents recently retrieved, and mainly testifying Fermi's
activity in the 1940s about pile physics and engineering. These
documents include patents, reports, notes on scientific and
technical meetings and other papers; all of them have been
carefully described, pointing out the relevance of the given paper
for its scientific or even historical content.

From a purely scientific point of view, the patents on nuclear
reactors, some reports or notes, and the complete set of lecture
notes for a course on neutron physics are the most important
documents. Quite intriguing are the papers written for the patents
issued at the U.S. Patent Office, since they directly deal with
the technical and operative construction of the nuclear reactors.
Although the activity by Fermi on this was early well recognized
from the accounts given by the living testimonies and partially
documented by several papers appeared in the Fermi's {\it
Collected Papers} (published in the 1960s), as it is evident from
what discussed above, from the newly retrieved papers a number of
important scientific and technical points come out, putting a
truly new light on the Fermi's activity in the project. In few
words, at last we can recognize exactly what Fermi {\it
effectively} did for the success of the pile project, since it is
now well documented.

The other papers, especially the notes on meeting, are also of
particular relevance for the history of the achievement of the
knowledge on chain reactions (with particular reference to the
construction of the first chain reacting pile in Chicago at the
end of 1942) and its application in the Manhattan Project. In some
documents, explicit references to weapons, their use during the
Second World War, and related matters appear. Quite a persistent
``obsession'', even as early as in 1942, for the production of
fissile material (mainly plutonium) for military uses emerges from
many documents, a feature which was not at all considered in
previous historical reconstructions. The attitude of Fermi on this
point comes out very clear: he is not ``obsessed'' at all by
military applications (like, instead, several other colleagues),
but rather by civil use of nuclear energy (for ``the heating of
towns'') and, quite unexpectedly, by the physiological effects of
radiations. Quite important (and, again, unexpected) are, as well,
the discussions at several meeting of long term physics research
and post-war research policy, and those regarding the
relationship, about nuclear power for pacific and/or military use,
between U.S. and Britain just after the end of the war.

Although we have discussed in some a great detail all of the novel
documents, given their obvious relevance, we can expect that
accurate studies on them, to be performed in order to explore the
full implications of them, have still to come, probably committing
the people interested in scientific and historical matters for
some time in the near future.



\subsection*{Acknowledgments}
The active and valuable cooperation of the staff of the
Chur\-chill Archive Centre, Cambridge (U.K.), the Department of
Manuscripts of the Cambridge University Library, Cambridge (U.K.),
the University Library of the University of Illinois at
Urbana-Champaign, and of the Information Resource Center of the
U.S. Embassy in Rome is here gratefully acknowledged. The authors
are also indebted to A. De Gregorio for valuable discussions and
to G. Miele for his kind support and encouragement.


\appendix

\section{Enrico Fermi's papers}

\noi In the following we report the complete list of works by
Enrico Fermi, as recognized on January 2008. This list is probably
incomplete; a number of papers issued in U.S.A during the Second
World War, for example, have been treated as restricted data for
some time by the competent governmental authorities.

In this list we have included also papers not directly written by
Fermi, but directly related to works performed by him, such as
lecture notes, reports or notes on meetings, and so on. These
papers have been pointed out by special symbols. In particular, we
have denoted with a $\diamondsuit$ those where the contribution by
Fermi is explicitly recognizable (typically, notes on meetings),
and with a * those where such contribution can be deduced only
indirectly (lecture notes or edited reports).

The following list has been compiled by adopting a full
chronological criterion. We have ordered the papers according to:
1) the explicit date reported in it; 2) the date of reception of
the paper by the publishing house; 3) the publication date. If
none of this applies, we have made recourse to some internal
analysis of the given paper and to the comparison with other
papers, also taking into account (if possible) the ordering of the
Enrico Fermi's {\it Collected Papers} (FNM) prepared by E. Segr\`e
et al. (which, however, is not strictly chronological).

The papers already present in the {\it Collected Papers} have been
pointed out by the code FNM followed (if applicable) by the
related list number. Those unpublished present only (for what we
know) in the Wattenberg Archive have been denoted by the code WAT
followed (if applicable) by the list number of that archive.
Finally, the patents registered at the U.S. Patent Office have
been pointed out by the code USP followed by the chronological
list number.

For books, we have reported only the first, original edition of
them, omitting the subsequent translations.
\\

\footnotesize

\begin{itemize}
\item[] \noi {\bf Books}
\bigskip
\end{itemize}

\begin{enumerate}

      \item[Book1]
      {\it Lezioni di elettrodinamica}, pubblicate a cura dello studente Adelino
      Morelli, pp. 95, Stabilimento tipo-litografico del Genio Civile, Roma,
      [s.d.].\\

      \item[Book2]
      {\it Lezioni di Fisica teorica}, dettate dal Prof. E. Fermi, raccolte dai
      Dott. Dei e Martinozzi, pp. 60, [s.n.], Roma, 1927.\\

      \item[Book3]
      {\it Introduzione alla fisica atomica}, pp. 330 Zanichelli, Bologna, 1928. \\

      \item[Book4]
      {\it Fisica ad uso dei Licei}, vol. I, pp. 239 and vol. II, pp. 243,
      Zanichelli, Bologna, 1929. \\

      \item[Book5]
      {\it Molecole e cristalli}, pp. 303 Zanichelli, Bologna, 1934.
      \\

      \item[Book6]
      {\it Thermodynamics}, pp. VII-160, Prentice-Hall, New York, 1937. \\

      \item[Book7]
      E. Fermi and E. Persico, {\it Fisica per le Scuole Medie Superiori},
      pp. 314 Zanichelli, Bologna, 1938. \\

      \item[Book8]
      {\it * Nuclear Physics. A Course given at the University of Chicago},
      Notes compiled by J. Orear, A.H. Rosenfeld, and R.H. Shulter, pp. VII+246,
      The University of Chicago Press, Chicago, 1949. \\

      \item[Book9]
      {\it Elementary Particles}, pp. XII+110, Yale university Press, New Haven,
      1951. \\

      \item[Book10]
      {\it Notes on Quantum Mechanics}, pp. VII+171,
      The university of Chicago Press, Chicago, 1961. \\
\end{enumerate}

\


\begin{itemize}
\item[] \noi {\bf Papers}
\bigskip
\end{itemize}

\begin{itemize}
\item[] \noi \underline{\bf 1921} \bigskip
\end{itemize}

\renewcommand{\labelenumi}{P\arabic{enumi}.}

\begin{enumerate}

      \item{}[FNM1]
 {\it Sulla dinamica di un sistema rigido di cariche elettriche}, \\
      Nuovo Cimento {\bf 22}, 199-207 (1921). \\ 

      \item{}[FNM2]
 {\it Sull'elettrostatica di un campo gravitazionale uniforme e sul peso delle masse elettromagnetiche}, \\
      Nuovo Cimento {\bf 22}, 176-188 (1921). \\ 

\noi \underline{\bf 1922} \\

      \item{}[FNM3]
 {\it Sopra i fenomeni che avvengono in vicinanza di una linea oraria}, \\
      Rend. Lincei {\bf 31}(1), 21-23, 51-52, 101-103, (1922). \\ 

      \item{}[FNM4b]
 {\it Correzione di una grave discrepanza tra la teoria delle masse elettromagnetiche e la teoria della relativit\`a.
      Inerzia e peso dell'elettricit\`a}, \\
      Rend. Lincei {\bf 31}(1), 184-187 (1922). \\ 

      \item{}[FNM4b]
 {\it Correzione di una grave discrepanza tra la teoria elettrodinamica e quella relativistica delle masse
      elettromagnetiche. Inerzia e peso dell'elettricit\`a}, \\
      Rend. Lincei {\bf 31}(1), 306-309 (1922). \\ 

      \item{}[FNM38b]
 {\it Un teorema di calcolo delle probabilit\`a ed alcune sue applicazioni}, \\
      Tesi di abilitazione della Scuola Normale Superiore. Pisa, 1922.
      \\ 

      \item{}[FNM4a]
 {\it \"Uber einen Winderspruch zwischen der elektrodynamishen und der relativistishen Theorie der
      elektromagnetischen Masse}, \\
      Phys. Zeits. {\bf 23}, 340-344 (1922). \\ 

      \item{}[FNM6]
 {\it I raggi Rontgen}, \\
      Nuovo Cimento {\bf 24}, 133-163, (1922). \\ 

      \item{}[FNM7]
 {\it Formazione di immagini coi raggi Rontgen}, \\
      Nuovo Cimento {\bf 25}, 63-68, (1923). \\

      \item{}[FNM14]
 {\it Sulla teoria statistica di Richardson dell'effetto fotoelettrico}, \\
      Nuovo Cimento {\bf 26}, 97-104, (1923). \\ 

\noi \underline{\bf 1923} \\

      \item{}[FNM5]
 {\it Le masse nella teoria della relativit\`a.} Da A. Kopff, {\it I fondamenti della relativit\`a Einsteiniana} Edizione
      italiana a cura di R. Contu e T. Bembo. Hoepli, Milano 1923, 342-344 \\

      \item{}[FNM8]
 {\it Sul peso dei corpi elastici}, \\
      Mem. Lincei {\bf 14}, 114-124, (1923). \\

      \item{}[FNM9]
 {\it Sul trascinamento del piano di polarizzazione da parte di un mezzo rotante}, \\
      Rend. Lincei {\bf 32}(1), 115-118, (1923). \\ 

      \item{}[FNM10]
 {\it Sulla massa della radiazione in uno spazio vuoto}, \\
      E. Fermi and A. Pontremoli, \\
      Rend. Lincei {\bf 32}(1), 162-164, (1923).
      \\ 

      \item{}[FNM4c]
 {\it Correzione di una contraddizione tra la teoria elettrodinamica e quella relativistica delle masse
      elettromagnetiche}, \\
      Nuovo Cimento {\bf 25}, 159-170, (1923). \\ 

      \item{}[FNM12]
 {\it Il principio delle adiabatiche ed i sistemi che non ammettono coordinate angolari}, \\
      Nuovo Cimento {\bf 25}, 171-175, (1923). \\ 

      \item{}[FNM11b]
 {\it Dimostrazione che in generale un sistema meccanico normale \`e quasi ergodico}, \\
      Nuovo Cimento {\bf 25}, 267-269, (1923). \\ 

      \item{}[FNM13]
 {\it Alcuni teoremi di meccanica analitica importanti per la teoria dei quanti}, \\
      Nuovo Cimento {\bf 25}, 271-285, (1923). \\ 

      \item{}[FNM11a]
 {\it Beweis dass ein mechanisches Normalsystem im allgemeinen \\ quasi-ergodisch ist}, \\
      Phys. Zeits {\bf 24}, 261-265, (1923). \\ 

      \item{}[FNM15]
 {\it Generalizzazione del teorema di Poincar\`e sopra la non esistenza di integrali uniformi di un sistema di
      equazioni canoniche normali}, \\
      Nuovo Cimento {\bf 26}, 105-115, (1923). \\ 

      \item{}[FNM16]
 {\it Sopra la teoria di Stern della costante assoluta dell'entropia di un gas perfetto monoatomico}, \\
      Rend. Lincei {\bf 32}(2), 395-398, (1923). \\ 

      \item{}[FNM17a]
 {\it Sulla probabilit\`a degli stati quantici}, \\
      Rend. Lincei {\bf 32}(2), 493-495, (1923). \\ 

\noi \underline{\bf 1924} \\

      \item{}[FNM18]
 {\it Sopra la riflessione e la diffusione di risonanza}, \\
      Rend. Lincei {\bf 33}(1), 90-93, (1924). \\ 

      \item{}[FNM19]
 {\it Considerazioni sulla quantizzazione dei sistemi ch contengono degli elementi identici}, \\
      Nuovo Cimento {\bf 1}, 145-152, (1924). \\ 

      \item{}[FNM20]
 {\it Sull'equilibrio termico di ionizzazione}, \\
      Nuovo Cimento {\bf 1}, 153-158, (1924). \\ 

      \item{}[FNM11a]
 {\it \"Uber die existenz quasi-ergodisher Systeme}, \\
      Phys. Zeits. {\bf 25}, 166-167, (1924) \\ 

      \item{}[FNM17b]
 {\it \"Uber die Wahrscheinlichkeit der Quantenzustande}, \\
      Z. Physik {\bf 26}, 54-56, (1924). \\ 

      \item{}[FNM21a]
 {\it Berekeningen over de intensiteiten van spektraallijnen}, \\
      Physica {\bf 4}, 340-343, (1924). \\

      \item{}[FNM23b]
 {\it \"Uber die Theorie des Stosses zwischen Atomen und elektrisch geladenen Teilchen}, \\
      Z. Physik {\bf 29}, 315-327, (1924). \\

\noi \underline{\bf 1925} \\

      \item{}[FNM22]
 {\it Sui principi della teoria dei quanti}, \\
      Rend. Seminario matematico Universit\`a di Roma {\bf 8}, 7-12, (1925). \\

      \item{}[FNM24]
 {\it Sopra l'urto tra atomi e nuclei di idrogeno}, \\
      Rend. Lincei {\bf 1}, 77-80, (1925). \\ 

      \item{}[FNM21b]
 {\it Sopra l'intensit\`a delle righe multiple}, \\
      Rend. Lincei {\bf 1}, 102-124, (1925). \\ 

      \item{}[FNM25]
 {\it Una relazione tra le costanti delle bande infrarosse delle molecole triatomiche}, \\
      Rend. Lincei {\bf 1}, 386-387, (1925). \\ 

      \item{}[FNM23a]
 {\it Sulla teoria dell'urto tra atomi e corpuscoli elettrici}, \\
      Nuovo Cimento {\bf 2}, 143-158, (1925). \\ 

      \item{}[FNM26]
 {\it Effect of an Alternating Magnetic Field on the Polarization of the Resonance Radiation of Mercury Vapour}, \\
      E. Fermi and F. Rasetti, \\
      Nature (London),  {\bf 115}, 764, (1925).
      \\ 

      \item{}[FNM28]
 {\it Effetto di un campo magnetico alternato sopra la polarizzazione della luce di risonanza}, \\
      E. Fermi and F. Rasetti, \\
      Rend. Lincei {\bf 1}, 716-722, (1925).
      \\ 

      \item{}[FNM28]
 {\it Ancora dell'effetto di un campo magnetico alternato sopra la polarizzazione della luce di risonanza}, \\
      E. Fermi and F. Rasetti, \\
      Rend. Lincei {\bf 2}, 117-120, (1925).
      \\ 

      \item{}[FNM29]
 {\it Sopra la teoria dei corpi solidi}, \\
      Periodico di Matematiche {\bf 5}, 264-274, (1925). \\

      \item{}[FNM34]
 {\it Problemi di chimica, nella fisica dell'atomo}, \\
      Periodico di Matematiche {\bf 6}, 19-26, (1926). \\ 

      \item{}[FNM27]
 {\it \"Uber den Einfluss eines wechselnden magnetischen Feldes auf die Polarization der Resonanzstrahlung}, \\
      Z. Physik {\bf 33}, 246-250, (1925). \\ 

\noi \underline{\bf 1926} \\

      \item{}[FNM30]
 {\it Sulla quantizzazione del gas perfetto monoatomico}, \\
      Rend. Lincei {\bf 3}, 145-149, (1926). \\ 

      \item{}[FNM35]
 {\it Sopra l'elettrone rotante}, \\
      F. Rasetti and E. Fermi, \\
      Nuovo Cimento {\bf 3}, 226-235, (1926). \\ 

      \item{}[FNM32]
 {\it Sopra l'intensit\`a delle righe proibite nei campi magnetici intensi}, \\
      Rend. Lincei {\bf 3}, 478-483, (1926). \\ 

      \item{}[FNM38a]
 {\it Sopra una formula di calcolo delle probabilit\`a}, \\
      Nuovo Cimento {\bf 3}, 313-318, (1926). \\ 

      \item{}[FNM37]
 {\it Il principio delle adiabatiche e la nozione di forza viva nella nuova meccanica ondulatoria}, \\
      E. Fermi and E. Persico, \\
      Rend. Lincei {\bf 4}(II), 452-457, (1926).
      \\

      \item{}[FNM31]
 {\it Zur Quantelung des idealen einatomigen Gases}, \\
      Z. Physik {\bf 36}, 902-912, (1926). \\ 

      \item{}[FNM39]
 {\it Quantum Mechanics and the Magnetic Moment of Atoms}, \\
      Nature (London) {\bf 118}, 876, (1926). \\ 

      \item{}[FNM33]
 {\it Argomenti pro e contro la ipotesi dei quanti di luce}, \\
      Nuovo Cimento {\bf 3}, XLVII-LIV, (1926). \\

\noi \underline{\bf 1927} \\

      \item{}[FNM40b]
 {\it Una misura del rapporto h/k per mezzo della dispersione anomala del tallio}, \\
      E. Fermi and F. Rasetti, \\
      Rend. Lincei {\bf 5}, 566-570, (1927).
      \\ 

      \item{}[FNM42]
 {\it Sul meccanismo dell'emissione nella meccanica ondulatoria}, \\
      Rend. Lincei {\bf 5}, 795-800, (1927). \\ 

      \item{}[FNM36]
 {\it Zur Wellenmechanik des Stossvorganges}, \\
      Z. Physik {\bf 40}, 399-402, (1927). \\ 

      \item{}[FNM40a]
 {\it Eine Messung des Verhaltnisses h/k durch die anomale Dispersion des Thalliumdampfes}, \\
      E. Fermi and F. Rasetti, \\
      Z. Physik {\bf 43}, 379-383, (1927).
      \\ 

      \item{}[FNM41]
 {\it Gli effetti elettro e magnetoottici e le loro interpretazioni}, \\
      Fascicolo speciale dell'Energia Elettrica nel $1^{\mathrm o}$
      centenario della morte di A. Volta, Uniel, Roma (1927). \\

      \item{}[FNM43]
 {\it Un metodo statistico per la determinazione di alcune propriet\`a dell'atomo}, \\
      Rend. Lincei {\bf 6}, 602-607, (1927). \\ 

\noi \underline{\bf 1928} \\

      \item{}[FNM44]
 {\it Sulla deduzione statistica di alcune propriet\`a dell'atomo. Applicazione alla teoria del sistema periodico
      degli elementi}, \\
      Rend. Lincei {\bf 6}, 602-607, (1927). \\ 

      \item{}[FNM47]
 {\it Eine statistische Methode zur Bestimmung Einiger Eigenschaften des Atoms und ihre Anwendung auf die Theorie des
      periodischen System der Elemente}, \\
      Z. Physik {\bf 48}, 73-79, (1928). \\ 

      \item{}[FNM46]
 {\it Anomalous Group in the Periodic System of Elements}, \\
      Nature (London) {\bf 121}, 502, (1928). \\ 

      \item{}[FNM45]
 {\it Sulla deduzione statistica di alcune propriet\`a dell'atomo.
 Calcolo della correzione di Rydberg per i termini s}, \\
      Rend. Lincei {\bf 7}, 726-730, (1928). \\ 

      \item{}[FNM48]
 {\it Statistische Berechnung der Rydbergkorrektionen der s-Terme}, \\
      Z. Physik {\bf 49}, 550-554, (1928). \\ 

      \item{}[FNM49]
 {\it \"Uber die Anwendung der statistischen Methode auf die Probleme des Atombaues}, \\
      Falkenhagen, Quantentheorie und Chemie. Leipziger Vortrage, 95-111, (1928). Hirzel, Leipzig, 1928 \\

\noi \underline{\bf 1929} \\

      \item{}[FNM50]
 {\it Sopra l'elettrodinamica quantistica}, \\
      Rend. Lincei {\bf 9}, 881-887, (1929). \\ 

      \item{}[FNM51]
 {\it Sul moto di un corpo di massa variabile}, \\
      Rend. Lincei {\bf 9}, 984-986, (1929). \\ 

      \item{}[FNM52]
 {\it Sulla teoria quantistica delle frange di interferenza}, \\
      Rend. Lincei {\bf 10}, 72-77, (1929); \\
      Nuovo Cimento {\bf 7}, 153-158, (1930).
      \\ 

      \item{}[FNM53]
 {\it Sul complesso 4d della molecola di elio}, \\
      Rend. Lincei {\bf 10}, 515-517, (1929); \\
      Nuovo Cimento {\bf 7}, 159-161, (1930).
      \\ 

      \item{}[FNM55]
 {\it Magnetic Moments of Atomic Nuclei}, \\
      Nature (London) {\bf 125}, 16, (1930). \\ 

      \item{}[FNM56]
 {\it I fondamenti sperimentali delle nuove teorie fisiche}, \\
      Atti Soc. It. Progr. Sci., $18^{\mathrm a}$ Riunione, vol. 1, 365-371, (1929). \\

      \item{}[FNM58]
 {\it Problemi attuali della fisica}, \\
      Annali dell'istruzione media {\bf 5}, 424-428, (1929). \\

\noi \underline{\bf 1930} \\

      \item{}[FNM57a]
 {\it Sui momenti magnetici deli nuclei atomici}, \\
      Mem. Accad. d'Italia {\bf 1} (Fis.), 139-148, (1930). \\ 

      \item{}[FNM63]
 {\it Sul calcolo degli spettri degli ioni}, \\
      Mem. Accad. d'Italia {\bf 1} (Fis.), 149-156 (1930); \\
      Nuovo Cimento, {\bf 8}, 7-14, (1931).
      \\ 

      \item{}[FNM54b]
 {\it Sul rapporto delle intensit\`a nei doppietti dei metalli alcalini}, \\
      Nuovo Cimento {\bf 7}, 201-207, (1930). \\ 

      \item{}[FNM66]
 {\it La th\`eorie du rayonnement}, \\
      Annales de l'Inst. H. Poincar\`e {\bf 1}, 53-74, (1930). \\

      \item{}[FNM57b]
 {\it \"Uber die magnetischen Momente der Atomkerme}, \\
      Z. Physik {\bf 60}, 320-333, (1930). \\ 

      \item{}[FNM59]
 {\it L'interpretazione del principio di causalit\`a nella meccanica quantistica}, \\
      Rend. Lincei {\bf 11}, 980-985, (1930). \\ 

      \item{}[FNM54a]
 {\it \"Uber das Intensitatsverhaltnis der Dublettkomponenten der Alkalien}, \\
      Z. Physik {\bf 59}, 680-686, (1930). \\ 

      \item{}[FNM64]
 {\it Sopra l'elettrodinamica quantistica}, \\
      Rend. Lincei {\bf 12}, 431-435, (1930). \\ 

      \item{}[FNM60]
 {\it Atomi e stelle}, \\
      Atti. Soc. It. Progr. Sci., $19^{\mathrm a}$ Riunione, vol. 1, 228-235, (1930). \\

      \item{}[FNM61]
 {\it I fondamenti sperimentali della nuova meccanica atomica}, \\
      Periodico di matematiche {\bf 10}, 71-84, (1930). \\

      \item{}[FNM62]
 {\it La fisica moderna}, \\
      Nuova Antologia {\bf 65}, 137-145, (1930). \\

\noi \underline{\bf 1931} \\

      \item{}[FNM68]
 {\it \"Uber den Ramaneffekt des Kohlendioxyds}, \\
      Z. Physik {\bf 71}, 250-259, (1931). \\ 

      \item{}[FNM65]
 {\it Le masse elettromagnetiche nella elettrodinamica quantistica}, \\
      Nuovo Cimento {\bf 8}, 121-132, (1931). \\ 

      \item{}[FNM69]
 {\it \"Uber den Ramaneffekt des Steinsalzes}, \\
      E. Fermi and F. Rasetti, \\
      Z. Physik {\bf 71}, 689-695, (1931).
      \\ 

\noi \underline{\bf 1932} \\

      \item{}[FNM67]
 {\it Quantum Theory of Radiation}, \\
      Rev. Mod. Phys. {\bf 4}, 87-132, (1932). \\ 

      \item{}[FNM71]
 {\it L'effetto Raman nelle molecole e nei cristalli.}, \\
      Mem. Accad. d'Italia {\bf 3} (Fis.), 239-256, (1932). \\ 

      \item{}[FNM70]
 {\it \"Uber die Wechselwirkung: von zwei Elektronen}, \\
      H. Bethe and E. Fermi, \\
      Z. Physik {\bf 77}, 296-306, (1932).
      \\ 

      \item{}[FNM72a]
 {\it La physique du noyau atomique.}, \\
      Congr\`es International d'\'electricit\'e. Paris (1932), C.R., 1 Sect., Rep. 22, 789-807.
      \\ 

      \item{}[FNM72b]
 {\it Lo stato attuale della fisica del nucleo atomico}, \\
      Ricerca Scientifica {\bf 3}(2), 101-113, (1932). \\ 

      \item{}[FNM73]
 {\it Sulle bande di oscillazione e rotazione dell'ammoniaca.}, \\
      Rend. Lincei {\bf 16}, 179-185, (1932); \\
      Nuovo Cimento {\bf 9}, 277-283, (1932).
      \\ 

\noi \underline{\bf 1933} \\

      \item{}[FNM74]
 {\it Azione del campo magnetico terrestre sulla radiazione penetrante.}, \\
      E. Fermi and B. Rossi, \\
      Rend. Lincei {\bf 17}, 346-350, (1933).
      \\ 

      \item{}[FNM75b]
 {\it Sulla teoria delle strutture iperfini}, \\
      E. Fermi and E. Segr\`e, \\
      Mem. Accad. d'Italia {\bf 4}(Fis), 131-158, (1933).
      \\ 

      \item{}[FNM77b]
 {\it On the Recombination of Electrons and Positrons}, \\
      E. Fermi and G. Uhlenbeck, \\
      Phys. Rev. {\bf 44}, 510-511, (1933).
      \\ 

      \item{}[FNM75a]
 {\it Zur Theorie der Hyperfeinstrukturen}, \\
      E. Fermi and E. Segr\`e, \\
      Z. Physik {\bf 82}, 11-12, 729-749 (1933).
      \\ 

      \item{}[FNM77a]
 {\it Sulla ricombinazione di elettroni e positroni}, \\
      E. Fermi and G. Uhlenbeck, \\
      Ricerca Scientifica {\bf 4}(2), 157-160, (1933). \\

      \item{}[FNM78]
 {\it Uno spettrografo per raggi $\ll$gamma$\gg$ a cristallo di bismuto}, \\
      E. Fermi and F. Rasetti, \\
      Ricerca Scientifica {\bf 4}(2), 299-302, (1933). \\

      \item{}[FNM76]
 {\it Tentativo di una teoria di emissione dei raggi $\ll$beta$\gg$}, \\
      Ricerca Scientifica {\bf 4}(2), 491-495, (1933). \\

      \item{}[FNM79]
 {\it Le ultime particelle costitutive della materia}, \\
      Atti Soc. It. Progr. Sci., $22^{\mathrm a}$ Riunione, vol. 2, 7-14,
      (1933); \\
      Scientia {\bf 55}, 21-28 (1934). \\

\noi \underline{\bf 1934} \\

      \item{}[FNM80a]
 {\it Tentativo di una teoria dei raggi $\beta$}, \\
      Nuovo Cimento {\bf 11}, 1-19, (1934). \\ 

      \item{}[FNM80b]
 {\it Versuch einer Theorie der $\beta$-Strahlen. I}, \\
      Z. Physik {\bf 88}, 161-171, (1934). \\ 

      \item{}[FNM95]
 {\it Sopra lo spostamento per pressione delle righe elevate delle serie spettrali}, \\
      Nuovo Cimento {\bf 11}, 157-166 (1934). \\ 

      \item{}[FNM84]
 {\it Radioattivit\`a provocata da bombardamento di neutroni. I}, \\
      Ricerca Scientifica, {\bf 5}(1), 283, (1934). \\ 

      \item{}[FNM85]
 {\it Radioattivit\`a provocata da bombardamento di neutroni. II}, \\
      Ricerca Scientifica, {\bf 5}(1), 330-331, (1934). \\

      \item{}[FNM93]
 {\it Radioactivity Induced by Neutron Bombardment}, \\
      Nature (London) {\bf 133}, 757 (1934). \\ 

      \item{}[FNM86]
 {\it Radioattivit\`a provocata da bombardamento di neutroni. III}, \\
      E. Amaldi, O. D'Agostino, E. Fermi, F. Rasetti and E. Segr\`e, \\
      Ricerca Scientifica, {\bf 5}(1), 452-453, (1934).
      \\ 

      \item{}[FNM82]
 {\it Le orbite $\infty$ s degli elementi}, \\
      Mem. Accad. d'Italia {\bf 6}(1) (Fis.), 119-149, (1934). \\ 

      \item{}[FNM94]
 {\it Sulla possibilit\`a di produrre elementi di numero atomico maggiore di 92}, \\
      E. Fermi, F. Rasetti, O. D'Agostino, \\
      Ricerca Scientifica {\bf 5}(1), 536-537 (1934).
      \\ 

      \item{}[FNM99]
 {\it Possible Production of Elements of Atomic Number Higher than 92}, \\
      Nature (London) {\bf 133}, 898-899 (1934). \\ 

      \item{}[FNM87]
 {\it Radioattivit\`a provocata da bombardamento di neutroni. IV}, \\
      E. Amaldi, O. D'Agostino, E. Fermi, F. Rasetti and E. Segr\`e, \\
      Ricerca Scientifica, {\bf 5}(1), 652-653, (1934).
      \\ 

      \item{}[FNM96]
 {\it Radioattivit\`a prodotta da bombardamento di neutroni}, \\
      Nuovo Cimento {\bf 11}, 429-441 (1934). \\ 

      \item{}[FNM97]
 {\it Nuovi radioelementi prodotti con bombardamenti di neutroni}, \\
      E. Amaldi, E. Fermi, F. Rasetti and E. Segr\`e, \\
      Nuovo Cimento {\bf 11}, 442-447 (1934). \\ 

      \item{}[FNM88]
 {\it Radioattivit\`a provocata da bombardamento di neutroni. V}, \\
      E. Amaldi, O. D'Agostino, E. Fermi, F. Rasetti and E. Segr\`e, \\
      Ricerca Scientifica, {\bf 5}(1), 21-22, (1934).
      \\ 

      \item{}[FNM98]
 {\it Artificial Radioactivity Produced by Neutron Bombardment}, \\
      E. Fermi, E. Amaldi, O. D'Agostino, F. Rasetti and E. Segr\`e, \\
      Proc. Roy. Soc. (London) Series A
 {\it 146}, 483-500 (1934). \\ 

      \item{}[FNM81]
 {\it Zur Bemerkung von G. Beck und K. Sitte}, \\
      Z. Physik {\bf 89}, 522, (1934). \\ 

      \item{}[FNM101]
 {\it Conferencias}, Faculdad de Ciencias Exactas Fisicas y Naturales, \\
      Publicacion {\bf 15}, Buenos Aires (1934). \\

      \item{}[FNM102]
 {\it Natural Beta Decay}, \\
      International Conference on Physics,
      London 1934. Vol. 1. Nuclear Physics, 66-71, Physical Society (London). \\

      \item{}[FNM103]
 {\it Artificial Radioactivity Produced by Neutron Bombardment}, \\
      International Conference on Physics,
      London 1934. Vol. 1. Nuclear Physics, 75-77, Physical Society (London). \\

      \item{}[FNM104]
 {\it La radioattivit\`a artificiale}, \\
      Atti Soc. It. Progr. Sci., $23^{\mathrm a}$ Riunione, vol. 1, 34-39. \\

      \item{}[FNM105]
 {\it Azione di sostanze idrogenate sulla radioattivit\`a provocata da neutroni. I}, \\
      E. Fermi, E. Amaldi, B. Pontecorvo, F. Rasetti and E. Segr\`e, \\
      Ricerca Scientifica {\bf 5}(2), 282-283 (1934). \\ 

      \item{}[FNM100]
 {\it Artificial Radioactivity Produced by Neutron Bombardment}, \\
      Nature (London) {\bf 134}, 668 (1934). \\ 

      \item{}[FNM106]
 {\it Effetto di sostanze idrogenate sulla radioattivit\`a provocata da neutroni. II}, \\
      E. Fermi, E. Amaldi, B. Pontecorvo and F. Rasetti, \\
      Ricerca Scientifica {\bf 5}(2), 380-381 (1934). \\ 

      \item{}[FNM89]
 {\it Radioattivit\`a provocata da bombardamento di neutroni. VII}, \\
      E. Amaldi, O. D'Agostino, E. Fermi, B. Pontecorvo, F. Rasetti and \\ E.
      Segr\`e, \\
      Ricerca Scientifica, {\bf 5}(1), 467-470, (1934).
      \\ 

\noi \underline{\bf 1935} \\

      \item{}[FNM90]
 {\it Radioattivit\`a provocata da bombardamento di neutroni. VIII}, \\
      E. Amaldi, O. D'Agostino, E. Fermi, B. Pontecorvo, F. Rasetti and \\ E.
      Segr\`e, \\
      Ricerca Scientifica, {\bf 6}(1), 123-125,
      (1935). \\ 

      \item{}[FNM109]
 {\it On the Velocity Distribution Law for the Slow Neutrons}, \\
      Zeeman Verhandelingen, p. 128-130, Martinus Nijhoff, the Hague, 1935.
      \\ 

      \item{}[FNM107]
 {\it Artificial Radioactivity Produced by Neutron Bombardment. \ldd Part II}, \\
      E. Amaldi, O. D'Agostino, E. Fermi, B. Pontecorvo, F. Rasetti and \\ E.
      Segr\`e, \\
      Proc. Roy. Soc. (London) Series A {\bf 149}, 522-558 (1935).
      \\ 

      \item{}[FNM108]
 {\it Ricerche sui neutroni lenti}, \\
      E. Fermi and F. Rasetti, \\
      Nuovo Cimento {\bf 12}, 201-210 (1935). \\ 

      \item{}[FNM91]
 {\it Radioattivit\`a provocata da bombardamento di neutroni. IX}, \\
      E. Amaldi, O. D'Agostino, E. Fermi, B. Pontecorvo and E.
      Segr\`e, \\
      Ricerca Scientifica, {\bf 6}(1), 435-437, (1935).
      \\ 

      \item{}[FNM92]
 {\it Radioattivit\`a provocata da bombardamento di neutroni. X}, \\
      E. Amaldi, O. D'Agostino, E. Fermi, B. Pontecorvo and E.
      Segr\`e, \\
      Ricerca Scientifica, {\bf 6}(1), 581-584, (1935).
      \\ 

      \item{}[FNM110]
 {\it On the Recombination of Neutrons and Protons}, \\
      Phys. Rev. {\bf 48}, 570 (1935). \\ 

      \item{}[USP1]
 {\it Process for the production of radioactive substances}, \\
      E. Fermi, E. Amaldi, B. Pontecorvo, F. Rasetti and E. Segr\`e, \\
      Patent filed October 3, 1935 (Patent No. 2,206,634; July 2, 1940). \\

      \item{}[FNM112]
 {\it Sull'assorbimento dei neutroni lenti. I}, \\
      E. Amaldi and E. Fermi, \\
      Ricerca Scientifica {\bf 6}(2), 344-347 (1935). \\ 

      \item{}[FNM113]
 {\it Sull'assorbimento dei neutroni lenti. II}, \\
      E. Fermi and E. Amaldi, \\
      Ricerca Scientifica {\bf 6}(2), 443-447 (1935). \\ 

      \item{}[FNM111]
 {\it Recenti risultati dalla radioattivit\`a artificiale}, \\
      Ricerca Scientifica {\bf 6}(2), 399-402 (1935); \\ 
      Atti. Soc. It. Progr. Sci., $24^{\mathrm a}$ Riunione, vol. 3, 116-120. \\

\noi \underline{\bf 1936} \\

      \item{}[FNM114]
 {\it Sull'assorbimento dei neutroni lenti. III}, \\
      E. Amaldi and E. Fermi, \\
      Ricerca Scientifica {\bf 7}(1), 56-59 (1936). \\ 

      \item{}[FNM115]
 {\it Sul cammino libero medio dei neutroni nella paraffina}, \\
      E. Amaldi and E. Fermi, \\
      Ricerca Scientifica {\bf 7}(1), 223-225 (1936). \\ 

      \item{}[FNM116]
 {\it Sui gruppi di neutroni lenti}, \\
      E. Amaldi and E. Fermi, \\
      Ricerca Scientifica {\bf 7}(1), 310-315 (1936). \\ 

      \item{}[FNM117]
 {\it Sulle propriet\`a di diffusione dei neutroni lenti}, \\
      E. Amaldi and E. Fermi, \\
      Ricerca Scientifica {\bf 7}(1), 393-395 (1936). \\ 

      \item{}[FNM118a]
 {\it Sopra l'assorbimento e la diffusione dei neutroni lenti}, \\
      E. Amaldi and E. Fermi, \\
      Ricerca Scientifica {\bf 7}(1), 454-503 (1936). \\ 

      \item{}[FNM119]
 {\it Sul moto dei neutroni nelle sostanze idrogenate}, \\
      Ricerca Scientifica {\bf 7}(2), 13-52 (1936). \\ 

      \item{}[FNM118b]
 {\it On the Absorption and the Diffusion of Slow Neutrons}, \\
      E. Amaldi and E. Fermi, \\
      Phys. Rev. {\bf 50}, 899-928 (1936). \\ 

      \item{}[FNM83]
 {\it Statistica, meccanica.}, \\
      Enciclopedia Italiana di Scienze, Lettere ed Arti, Istituto G. Treccani,
      Roma, vol. 32, 518-523, (1936). \\

\noi \underline{\bf 1937} \\

      \item{}[FNM120]
 {\it Un Maestro: Orso Mario Corbino}, \\
      Nuova Antologia {\bf 72}, 313-316 (1937). \\

      \item{}[FNM121]
 {\it Un generatore artificiale di neutroni}, \\
      E. Amaldi, E. Fermi and F. Rasetti, \\
      Ricerca Scientifica {\bf 8}(2), 40-43 (1937). \\ 

      \item{}[FNM122]
 {\it Neutroni lenti e livelli energetici nucleari}, \\
      Nuovo Cimento {\bf 15}, 41-42 (1938). \\ 

      \item{}[FNM123]
 {\it (Tribute to Lord Rutherford)}, \\
      Nature (London) {\bf 140}, 1052 (1937). \\ 

\noi \underline{\bf 1938} \\

      \item{}[FNM125]
 {\it On the Albedo of Slow Neutrons}, \\
      E. Fermi, E. Amaldi and G.C. Wick, \\
      Phys. Rev. {\bf 53}, 493 (1938). \\ 

      \item{}[FNM124]
 {\it Azione del boro sui neutroni caratteristici dello iodio}, \\
      E. Fermi and F. Rasetti, \\
      Ricerca Scientifica {\bf 9}(2), 472-473 (1938). \\ 

      \item{}[FNM126]
 {\it Prospettive di applicazioni della radioattivit\`a artificiale}, \\
      Rendiconti dell'Istituto di Sanit\`a Pubblica, vol. 1, 421-432 (1938). \\

      \item{}[FNM127]
 {\it Guglielmo Marconi e la propagazione delle onde elettromagnetiche nell'alta atmosfera}, \\
      Soc. It. Progr. Sci., Collectanea Marconiana, Roma, 1-5 (1938). \\

      \item{}[FNM]
 {\it $\diamondsuit$ Intervista con Enrico Fermi, 11 Novembre 1938}, \\
      Ricerca Scientifica, {\bf 9}(2), 638-639 (1938). \\

      \item{}[FNM128]
 {\it Artificial Radioactivity Produced by Neutron Bombardment}, \\
      Les Prix Nobel en 1938. Les Conf\'erences Nobel, Stockholm p. 1-8 (1939). \\

\noi \underline{\bf 1939} \\

      \item{}[FNM129]
 {\it The Fission of Uranium}, \\
      H.L. Anderson, E.T. Booth, J.R. Dunning, E. Fermi, G.N. Glasoe
      and \\ F.G. Slack, \\
      Phys. Rev. {\bf 55}, 511-512 (1939). \\ 

      \item{}[FNM130]
 {\it Production of Neutrons in Uranium Bombarded by Neutrons}, \\
      H.L. Anderson, E. Fermi and H.B. Hanstein, \\
      Phys. Rev. {\bf 55}, 797-798 (1939). \\ 

      \item{}[FNM131]
 {\it Simple Capture of Neutrons by Uranium}, \\
      H.L. Anderson and E. Fermi, \\
      Phys. Rev. {\bf 55}, 1106-1107 (1939). \\ 

      \item{}[FNM132]
 {\it Neutron Production and Absorption in Uranium}, \\
      H.L. Anderson, E. Fermi and L. Szilard, \\
      Phys. Rev. {\bf 56}, 284-286 (1939). \\ 

      \item{}[FNM133]
 {\it The Absorption of Mesotrons in Air and in Condensed \ldd Materials}, \\
      Phys. Rev. {\bf 56}, 1242 (1939). \\ 

\noi \underline{\bf 1940} \\

      \item{}[FNM134]
 {\it The Ionization Loss of Energy in Gases and in Condensed Materials}, \\
      Phys. Rev. {\bf 57}, 485-493 (1940). \\ 

      \item{}[FNM137a]
 {\it Reactions Produced by Neutrons in Heavy Elements}, \\
      Nature {\bf 146}, 640-642 (1940);
      Science {\bf 92}, 269-271 (1940).  \\ 

      \item{}[FNM136]
 {\it Production and Absorption of Slow Neutrons by Carbon}, \\
      H.L. Anderson and E. Fermi,
      Report A-21 (September 25, 1940). \\

      \item{}[FNM137]
 {\it Branching Ratios in the Fission of Uranium (235)}, \\
      H.L. Anderson, E. Fermi and A.V. Grosse, \\
      Phys. Rev. {\bf 59}, 52-56 (1941).  \\ 

\noi \underline{\bf 1941} \\

      \item{}[FNM138]
 {\it Production of Neutrons by Uranium}, \\
      H.L. Anderson and E. Fermi, \\
      Report A-6 (January 17, 1941).  \\

      \item{}[FNM135]
 {\it Fission of Uranium by Alpha-Particles}, \\
      E. Fermi and E. Segr\`e,
      Phys. Rev. {\bf 59}, 680-681 (1941). \\ 

      \item{}[FNM139]
 {\it Capture of Resonance Neutrons by a Uranium Sphere Imbedded in Graphite}, \\
      E. Fermi, H.L. Anderson, R.R. Wilson and E.C. Creutz, \\
      Appendix A of Report A-21 to The National Defense Research Committee  \\
      by H. D. Smith, Princeton University, (June 1, 1941). \\

      \item{}[FNM140]
 {\it Standards in Slow Neutron Measurements}, \\
      H.L. Anderson and  E. Fermi, \\
      Report A-2 (June 5, 1941).  \\

      \item{}[FNM141]
 {\it Some Remarks on the Production of Energy by a Chain Reaction in \\
 Uranium}, \\
      Report A-14 (June 30, 1941).  \\

      \item{}[FNM142]
 {\it The Absorption of Thermal Neutrons by a Uranium Sphere Imbedded in Graphite}, \\
      E. Fermi and G.L. Weil, \\
      Report A-1 (July 3, 1941).  \\

      \item{}[FNM143]
 {\it Remarks on Fast Neutron Reactions}, \\
      Report A-46 (October 6, 1941).  \\

      \item{}[FNM146]
 {\it Absorption Cross Section for Rn + Be Fast Neutrons}, \\
      H.L. Anderson, E. Fermi and G.L. Weil,
      Report C-72.  \\

      \item{}[FNM145]
 {\it Fission Cross Section of Unseparated Uranium for Fast Rn + Be Neutrons}, \\
      H.L. Anderson and E. Fermi,
      Report C-83.  \\

      \item{}[FNM144]
 {\it The Effect of Chemical Binding in the Scattering and Moderation of Neutrons by Graphite}, \\
      Report C-87. \\ 

\noi \underline{\bf 1942} \\

      \item{}[FNM148]
 {\it The Absorption Cross Section of Boron for Thermal Neutrons}, \\
      H.L. Anderson and E. Fermi, \\
      Report CP-74. \\ 

      \item{}[FNM153]
 {\it A table for Calculating the Percentage of Loss Due to the \ldd Presence of Impurities in Alloy}, \\
      Report C-5 (February 10, 1942). \\

      \item{}[FNM154]
 {\it The Temperature Effect on a Chain Reacting Unit. Effect of the Change of Leakage}, \\
      Report C-8 (February 25, 1942). \\

      \item{}[FNM155]
 {\it The Use of Reflectors and Seeds in a Power Plant}, \\
      G. Breit and E. Fermi, \\
      Report C-11 (March 9, 1942). \\

      \item{}[FNM156]
 {\it Slowing Down and Diffusion of Neutrons}, \\
      Report C-29 (Notes on Lecture of March 10, 1942). \\

      \item{}[FNM149]
 {\it Neutron Production in a Lattice of Uranium and Graphite}, \\
      Theoretical Part \\
      Report CP-12 (March 17, 1942). \\

      \item{}[FNM157]
 {\it Determination of the Albedo and the Measurement of Slow Neutron Density}, \\
      Report C-31 (Notes on Lecture of March 17, 1942). \\

      \item{}[FNM158]
 {\it The Number of Neutrons Emitted by a Ra + Be Source \ldd (Source I)}, \\
      H.L. Anderson, E. Fermi, J.H. Roberts and M.D. Whitaker, \\
      Report C-21 (March 21, 1942). \\

      \item{}[FNM150]
 {\it Neutron Production in a Lattice of Uranium Oxide and Graphite (Exponential Experiment)}, \\
      H.L. Anderson, B.T. Feld, E. Fermi, G.L. Weil and W.H. Zinn,
      \\
      Report CP-20 (March 26, 1942). \\

      \item{}[FNM151]
 {\it Preliminary Report on the Exponential Experiment at Columbia University}, \\
      Report CP-26 (March, April  1942). \\

      \item{}[FNM159]
 {\it The  Determination of the Ratio Between the Absorption Cross Sections of Uranium and Carbon for Thermal
      Neutrons}, \\
      Report C-84 (May 15, 1942). \\

      \item{}[FNM152]
 {\it Effect of Atmospheric Nitrogen and of Changes of Temperature on the Reproduction Factor}, \\
      Report CP-85 (May 19, 1942). \\

      \item{}[FNM147]
 {\it Neutrons Emitted by a Ra + Be Photosource}, \\
      B.T. Feld and  E. Fermi, \\
      Report CP-89 (later issued on November 5, 1948).  \\

      \item{}[WAT1043]
 {\it $\diamondsuit$ Meeting of Engineering Council}, \\
      Present: Moore, Allison, Fermi, Leverett, Wheeler, Compton, Hilberry and
      Doan, \\
      Report CE-106 (May 28, 1942). \\

      \item{}[WAT1043]
 {\it * The Fourth Intermediate Pile}, \\
      Metallurgical Project (no explicit list of authors;
      classified by Wattenberg among the Fermi papers),\\
      Report C-102. \\

      \item{}[WAT1043]
 {\it $\diamondsuit$ Meeting of the Planning Board}, \\
      Present: Hilberry, Spedding, Allison, Wigner, Doan, Szilard, Wheeler,
      Fermi, Moore and Compton, \\
      Report CS-112, CS-185 (June 6, 1942). \\

      \item{}[WAT1043]
 {\it $\diamondsuit$ Meeting of the Engineering Council}, \\
      Present: Moore, Wilson, Seaborg, Wheeler, Leverett, Fermi, Hilberry,
      Compton, Spedding and Allison, \\
      Report CS-131 (June 11, 1942). \\

      \item{}[WAT1043]
 {\it $\diamondsuit$ Meeting of the Engineering Council}, \\
      Present: Moore, Leverett, Fermi, Wheeler, Seaborg, Doan, Wilson and
      Spedding, \\
      Report CS-135 (June 18, 1942) \\

      \item{}[FNM164]
 {\it Status of Research Problems in Experimental Nuclear Physics}, \\
      Report C-133 for Week Ending (June 20, 1942). \\

      \item{}[WAT1043]
 {\it $\diamondsuit$ Meeting of the Engineering Council}, \\
      Present: Moore, Leverett, Doan, Szilard, Allison, Teller, Seaborg, Wheeler, Fermi
      and Wilson, \\
      Report CS-147 (June 25, 1942) \\

      \item{}[FNM160]
 {\it The Absorption of Graphite for Thermal Neutrons}, \\
      Report C-154 (Notes on Lecture of June 30, 1942). \\

      \item{}[WAT1043]
 {\it $\diamondsuit$ Meeting of the Engineering Council}, \\
      Present: Fermi, Allison, Seaborg, Whitaker, Doan, Wilson, Moore,
      \ldd Leverett, Wheeler, Szilard,
      Compton, Spedding, Hilberry and Wollan, \\
      Report CS-163 (July 2, 1942). \\

      \item{}[FNM161]
 {\it Longitudinal Diffusion in Cylindrical Channels}, \\
      E. Fermi and A. M. Weinberg, \\
      Report C-170 (July 7, 1942). \\

      \item{}[WAT1043]
 {\it $\diamondsuit$ Meeting of the Engineering Council}, \\
      Present: Moore, Leverett, Fermi, Wigner, Allison, Wollan, Wheeler, Seaborg, Spedding, Szilard, Steams
      and Wilson, \\
      Report CS-174 (July 9, 1942). \\

      \item{}[WAT1043]
 {\it $\diamondsuit$ Meeting of the Technical Council}, \\
      Present: Fermi, Compton, Allison, Moore, Szilard, Wigner and
      Wheeler, \\
      Report CS-184 (July 14, 1942). \\

      \item{}[FNM162]
 {\it The Number of Neutrons Emitted by Uranium per Thermal Neutron Absorbed}, \\
      Report C-190 (July 16, 1942). \\

      \item{}[WAT1043]
 {\it $\diamondsuit$ Meeting of the Engineering Council}, \\
      Present: Moore, Leverett, Lewis, Fermi, Wollan, Hilberry, Whitaker, Wilson, Wheeler, Allison, Wigner,
      Seaborg and Doan, \\
      Report CE-194 (July 21, 1942). \\

      \item{}[WAT1043]
 {\it $\diamondsuit$ Meeting of the Technical Council}, \\
      Present: Fermi, Szilard, Wigner, Compton, Whitaker, Allison, Moore, Wheeler and
      Doan, \\
      Report CS-202 (July 25, 1942). \\

      \item{}[WAT1043]
 {\it $\diamondsuit$ Meeting of the Planning Board}, \\
      Present: Spedding, Fermi, Szilard, Wigner, Doan, Moore, Wheeler,
      Compton and Hilberry, \\
      Report CS-213 (August 1, 1942). \\

      \item{}[WAT1043]
 {\it $\diamondsuit$ Meeting of the Engineering Council}, \\
      Present: Moore, Leverett, Steinbach, Fermi, Spedding, Wheeler, Wigner,
      Seaborg and Wilson, \\
      Report CE-229 (August 8, 1942). \\

      \item{}[FNM166]
 {\it Status of Research Problems of the Physics Division}, \\
      Report CP-235 for Month Ending August 15, 1942. \\

      \item{}[FNM167]
 {\it Exponential Pile No. II}, \\
      Report CA-247 for Week Ending August 29, 1942. \\

      \item{}[WAT1043]
 {\it $\diamondsuit$ Meeting of the Technical Council}, \\
      Present: Nichols, Hilberry, Spedding, Doan, Fermi, Steinbach, Grafton, Boyd, Moore and
      Wheeler, \\
      Report CS-251 (September 4, 1942). \\

      \item{}[FNM163]
 {\it Effect of Temperature Changes on the Reproduction Factor}, \\
      R.F. Christy, E. Fermi and M. D. Whitaker, \\
      Report CP-254 (September 14, 1942). \\

      \item{}[FNM168]
 {\it Status of Research Problems of the Physics Division}, \\
      Report CP-257 for Month Ending September 15, 1942. \\

      \item{}[WAT1006]
 {\it $\diamondsuit$ Discussion of Helium Cooled Power Plant}, \\
      Present: Leverett, Cooper, Moore, Wigner, Steinbach, Fermi,
      Szilard and Wheeler, \\
      Report CS-267 (September 16, 1942). \\

      \item{}[WAT1043]
 {\it $\diamondsuit$ Meeting of the Technical Council}, \\
      Present: Compton, Wheeler, Moore, Allison, Szilard, Fermi and Spedding, \\
      Report CS-274 (September 18, 1942). \\

      \item{}[FNM169]
 {\it Purpose of the Experiment at the Argonne Forest. Meaning of the Reproduction Factor ``k''}, \\
      Report CP-283 (Notes on Lecture of September 23, 1942). \\

      \item{}[WAT1043]
 {\it $\diamondsuit$ Meeting of the Technical Council}, \\
      Present: Allison, Fermi, Moore, Wigner, Compton, Wheeler and
      \\ Oppenheimer, \\
      Report CS-281 (September 29, 1942). \\

      \item{}[FNM170]
 {\it The Critical Size-Measurement of ``k'' in the Exponential Pile}, \\
      Report CP-289 (Notes on Lecture of September 30, 1942). \\

      \item{}[WAT1043]
 {\it $\diamondsuit$ Meeting of the Technical Council}, \\
      Present: Allison, Fermi, Wigner, Compton, Whitaker, Moore, Cooper, Szilard,
      Manley, McMillan, Wheeler and Doan, \\
      Report CS-284 (October 1, 1942). \\

      \item{}[WAT1043]
 {\it $\diamondsuit$ Meeting of the Technical Council}, \\
      Present: Allison, Fermi, Szilard, Moore, Wigner, Whitaker, Wheeler, Steinbach,
      Compton and Groves, \\
      Report CS-286 (October 5, 1942). \\

      \item{}[FNM176]
 {\it Methods of Cooling Chain Reacting Piles}, \\
      Memo \#10 (October 5, 1942). \\

      \item{}[FNM171]
 {\it Problem of Time Dependance of the Reaction Rate: Effect of Delayed Neutrons Emission}, \\
      Report CP-291 (Notes on Lecture of October 7, 1942). \\

      \item{}[WAT1043]
 {\it $\diamondsuit$ Meeting of the Technical Council}, \\
      Present: Allison, Wigner, Compton, Whitaker, Szilard, Wheeler, Fermi, Moore,
      Cooper, Steinbach and Kirkpatrick, \\
      Report CS-290 (October 7, 1942). \\

      \item{}[WAT1043]
 {\it $\diamondsuit$ Meeting of the Technical Council}, \\
      Present: Allison, Szilard, Wigner, Moore, Wheeler and
      Fermi, \\
      Report CS-294 (October 12, 1942). \\

      \item{}[FNM175]
 {\it The projected Experiment at Argonne Forest and the Reproduction Factor in Metal Piles}, \\
      Report CP-297 for Month Ending October 15, 1942. \\

      \item{}[FNM172]
 {\it A simplified Control. Optimum Distribution of Materials in the Pile}, \\
      Report CP-314 (Notes on Lecture of October 20, 1942). \\

      \item{}[WAT1008]
 {\it $\diamondsuit$ Conference on Lattice Spacing}, \\
      Present: Steinbach, Leverett, Fermi, Wigner and Wheeler, \\
      Memo \#15 (October 21, 1942). \\

      \item{}[FNM177]
 {\it The Effect of Bismuth on the Reproduction Factor}, \\
      Report CA-320, Bulletin for Week Ending October 31, 1942 \\

      \item{}[FNM173]
 {\it Design of the Graphite-Uranium Lattice: Experimental Determination of ft from the Cd Ratio}, \\
      Report CP-337 (Notes on Lecture of October 27 and November 3, 1942). \\

      \item{}[WAT1043]
 {\it Report of the Committee for the Examination of the Moore-Leverett
      Design of a He-Cooled Plant}, \\
      E. Fermi, S.K. Allison, C. Cooper and E.P. Wigner, \\
      Report CE-324 (1942). \\

      \item{}[FNM174]
 {\it Calculation of the Reproduction Factor}, \\
      Report CP-358 (Notes on Lecture of November 10, 1942). \\

      \item{}[FNM178]
 {\it The Experimental Chain Reacting Pile and Reproduction Factor in Some Exponential Piles}, \\
      Report CP-341 for Month Ending November 15, 1942. \\

      \item{}[WAT1043]
 {\it $\diamondsuit$ Meeting of the Technical Council}, \\
      Present: Allison, Compton, Fermi, Moore, Spedding, Szilard
      and Wigner, \\
      Report CS-356 (November 19, 1942). \\

      \item{}[FNM179]
 {\it Feasibility of a Chain Reaction}, \\
      Report CP-383 (November 26, 1942). \\

      \item{}[FNM180]
 {\it Work Carried out by the Physics Division}, \\
      Report CP-387 for Month Ending December 15, 1942. \\

\noi \underline{\bf 1943} \\

      \item{}[FNM182]
 {\it Summary of Experimental Research Activities}, \\
      Report CP-416 for Month Ending January 15, 1943. \\

       \item{}[FNM183]
 {\it Summary of Experimental Research Activities}, \\
      Report CP-455 for Month Ending February 6, 1943. \\

      \item{}[FNM184]
 {\it The Utilization of Heavy Hydrogen in Nuclear Chain Reactions}, \\
      Memorandum of Conference between Prof. E. Fermi and Prof. H. C. Urey
      on March 6, 7, and 8, 1943, Report A-544.  \\

      \item{}[FNM185]
 {\it The Slowing Down of Neutrons in Heavy Water}, \\
      Report CP-530 (March 19, 1943). \\

      \item{}[FNM186]
 {\it Summary of Experimental Research Activities}, \\
      Report CP-570 for Month Ending April 17, 1943. \\

      \item{}[FNM187]
 {\it Summary of Experimental Research Activities}, \\
      Report CP-641 for Month Ending May 10, 1943. \\

      \item{}[FNM188]
 {\it Standardization of the Argonne Pile}, \\
      H.L. Anderson, E. Fermi, J. Marshall and L. Woods, \\
      Report CP-641 for Month Ending May 10, 1943. \\

      \item{}[FNM189]
 {\it Tests on a Shield for the Pile at Site W}, \\
      E. Fermi and W. H. Zinn, \\
      Report CP-684 for Month Ending May 25, 1943. \\

      \item{}[FNM190]
 {\it Summary of Experimental Research Activities}, \\
      Report CP-718 for Month Ending June 12, 1943. \\

      \item{}[FNM192]
 {\it Summary of Experimental Research Activities}, \\
      Report CP-781 for Month Ending July 10, 1943. \\

      \item{}[FNM193]
 {\it Range of Indium Resonance Neutrons from a Source of Fission Neutrons}, \\
      E. Fermi and G. L. Weil, \\
      Report CP-871 for Month Ending August 14, 1943. \\

      \item{}[WAT1023]
 {\it * Report for Month Ending September 25, 1943}, \\
      edited by  A.H. Compton, S.K. Allison and E. Fermi, \\
      Report CP-964. \\

      \item{}[FNM194]
 {\it Summary of Experimental Research Activities}, \\
      Report CP-1016 for Month Ending October 23, 1943. \\

      \item{}[FNM195]
 {\it Summary of Experimental Research Activities}, \\
      Report CP-1088 for Month Ending November 23, 1943. \\

      \item{}[FNM196]
 {\it The Range of Delayed Neutrons}, \\
      E. Fermi and G. Thomas, \\
      Report CP-1088 for Month Ending November 23, 1943. \\

      \item{}[FNM197]
 {\it Slowing Down of Fission Neutrons in Graphite}, \\
      E. Fermi, J. Marshall and L. Marshall, \\
      Report CP-1084 (November 25, 1943). \\

      \item{}[FNM199]
 {\it Summary of Experimental Research Activities}, \\
      Report CP-1175 for Month Ending December 25, 1943. \\

      \item{}[WAT1027]
 {\it * Report for Month Ending December 25, 1943 - Part II}, \\
      edited by  A.H. Compton, S.K. Allison and E. Fermi, \\
      Report CN-1190 \\

      \item{}[FNM198]
 {\it Fission Cross-Section and $\nu$-Value for 25}, \\
      E. Fermi, J. Marshall and L. Marshall, \\
      Report CP-1186 (December 31, 1943). \\

\noi \underline{\bf 1944} \\

      \item{}[WAT149]
 {\it Measurement of the Cross Section of Boron for Thermal \ldd Neutrons}, \\
      E. Bragdon, E. Fermi, J. Marshall and L. Marshall, \\
      Report CP-1098 (January 11, 1944). \\

      \item{}[FNM201]
 {\it Summary of Experimental Research Activities}, \\
      Report CP-1255 for Month Ending January 24, 1944. \\

      \item{}[FNM202]
 {\it Summary of Experimental Research Activities}, \\
      Report CP-1389 for Month Ending February 24, 1944. \\

      \item{}[FNM203]
 {\it Report of Fermi's Activities with The Marshall Group}, \\
      Report CP-1389 for Month Ending February 24, 1944. \\

      \item{}[FNM204]
 {\it Summary of Experimental Activities}, \\
      Report CP-1531 for Month Ending March 25, 1944. \\

      \item{}[FNM205]
 {\it Range of Fission Neutrons in Water}, \\
      H.L. Anderson, E. Fermi and D.D. Nagle, \\
      Report CP-1531 for Month Ending March 25, 1944. \\

      \item{}[FNM206]
 {\it Evidence for the Formation of 26}, \\
      E. Fermi and L. Marshall, \\
      Report CP-1531 for Month Ending March 25, 1944. \\

      \item{}[FNM207]
 {\it Summary of Experimental Research Activities}, \\
      Report CP-1592 for Month Ending April 24, 1944. \\

      \item{}[FNM208]
 {\it Absorption of 49}, \\
      Report CP-1592 for Month Ending April 24, 1944. \\

      \item{}[FNM209]
 {\it Comparison of the Ranges in Graphite of Fission Neutrons from 49 and 25}, \\
      E. Fermi, A. Heskett and D.E. Nagle, \\
      Report CP-1592 for Month Ending April 24, 1944. \\

      \item{}[FNM211]
 {\it $\diamondsuit$ Notes on Meeting of April 26, 1944}, \\
      Present: Fermi, Allison, Szilard, Wigner, Weinberg, Seitz,
      Morrison, Cooper, Vernon, Tolman, Watson and Ohlinger, \\
      Report N-1729, Eck-209. \\

      \item{}[WAT]
 {\it $\diamondsuit$ Notes on Meeting of April 28, 1944}, \\
      Present: Fermi, Allison, Wigner, Smyth, Szilard, Morrison, Watson,
      Feld, Hogness, Young, Weinberg, Creutz, Cooper, Vernon and Ohlinger, \\
      Report N-1729, Eck-209. \\

      \item{}[USP2]
 {\it Test exponential pile}, \\
      Patent filed May 4, 1944 (Patent No. 2,780,595; February 5, 1957). \\

      \item{}[FNM212]
 {\it Report on Recent Values of Constants of 25 and 49}, \\
      Report CK-1788 (May 19, 1944). \\

      \item{}[FNM213]
 {\it Summary of Experimental Research Activities}, \\
      Report CP-1729 for Month Ending May 25, 1944. \\

      \item{}[FNM214]
 {\it Dissociation Pressure of Water due to Fission}, \\
      H.L. Anderson and E. Fermi, \\
      Report CP-1729 for Month Ending May 25, 1944. \\

      \item{}[FNM215]
 {\it Summary of Experimental Research Activities}, \\
      Report CK-1761 for Month Ending May 25, 1944. \\

      \item{}[FNM216]
 {\it Summary of Experimental Research Activities}, \\
      Report CK-1827 for Month Ending June 25, 1944. \\

      \item{}[FNM217]
 {\it Collimation of Neutron Beam from Thermal Column of CP-3 and the Index of Refraction for Thermal Neutrons}, \\
      E. Fermi and W.H. Zinn, \\
      Report CK-1965 for Month Ending July 29, 1944. \\

      \item{}[FNM218]
 {\it Methods for Analysis of Helium Circulating in the 105 Unit}, \\
      Document HW 3-492 (August 7, 1944). \\

      \item{}[WAT1037]
 {\it * Report for Month Ending August 26, 1944}, \\
      edited by A.H. Compton, E. Fermi and W.H. Zinn, \\
      Report CP-2081. \\

      \item{}[FNM219]
 {\it Boron Absorption of Fission Activation}, \\
      H.L. Anderson and E. Fermi, \\
      Report CF-2161 for Month Ending September 23, 1944. \\

      \item{}[WAT1039]
 {\it * Report for Month Ending October 28, 1944}, \\
      edited by A.H. Compton, E. Fermi and W.H. Zinn, \\
      Report CP-2301 \\

      \item{}[WAT1040]
 {\it * Report for Month Ending November 25, 1944}, \\
      edited by A.H. Compton, E. Fermi and W.H. Zinn, \\
      Report CP-2436 \\

      \item{}[USP3]
 {\it Neutronic reactor}, \\
      E. Fermi and L. Szilard, \\
      Patent filed December 19, 1944 (Patent No. 2,708,656; May, 17 1955). \\

\noi \underline{\bf 1945} \\

      \item{}[USP4]
 {\it Chain reacting system}, \\
      E. Fermi M.C. Leverett, \\
      Patent filed February 16, 1945 (Patent No. 2,837,447; June 3, 1958). \\

      \item{}[USP5]
 {\it Neutronic reactor}, \\
      Patent filed May 12, 1945 (Patent No. 2,931,762; April 5, 1960). \\

      \item{}[USP6]
 {\it Air cooled neutronic reactor}, \\
      E. Fermi and L. Szilard, \\
      Patent filed May 29, 1945 (Patent No. 2,836,554; May 27, 1958). \\

      \item{}[FNM221]
 {\it $\diamondsuit$ Relation of Breeding to Nuclear Properties}, \\
      Present: Fermi {\it et al.}, \\
      Report CF-3199 
      (June 19-20, 1945). \\

      \item{}[USP7]
 {\it Testing material in a neutronic reactor}, \\
      E. Fermi and H.L. Anderson, \\
      Patent filed August 28, 1945 (Patent No. 2,768,134; October. 23, 1956). \\

      \item{}[USP8]
 {\it Neutron velocity selector}, \\
      Patent filed September 18, 1945 (Patent No. 2,524,379; October 3, 1950). \\

      \item{}[USP9]
 {\it Neutronic reactor}, \\
      E. Fermi and L. Szilard, \\
      Patent filed October 11, 1945 (Patent No. 2,807,581; September 25, 1957). \\

      \item{}[USP10]
 {\it Neutronic reactor}, \\
      E. Fermi, W.H. Zinn and H.L. Anderson, \\
      Patent filed October 11, 1945 (Patent No. 2,852,461; September 16, 1958). \\

      \item{}[USP11]
 {\it Neutronic reactor}, \\
      E. Fermi and W.H. Zinn, \\
      Patent filed November 2, 1945 (Patent No. 2,714,577; August 2, 1955). \\

      \item{}[USP12]
 {\it Method of testing thermal neutron fissionable material for purity}, \\
      E. Fermi and H.L. Anderson, \\
      Patent filed November 21, 1945 (Patent No. 2,969,307; January 24, 1961). \\

      \item{}[USP13]
 {\it Method of sustaining a neutronic chain reacting system}, \\
      E. Fermi and M.C. Leverett, \\
      Patent filed November 28, 1945 (Patent No. 2,813,070; November 12, 1957). \\

\noi \underline{\bf 1946} \\

      \item{}[USP14]
 {\it Neutronic reactor shield}, \\
      E. Fermi and W.H. Zinn, \\
      Patent filed January 16, 1946 (Patent No. 2,807,727; September 24, 1957). \\

      \item{}[FNM222]
 {\it * Neutron Physics. A Course of Lectures by E. Fermi}, \\
      Notes by I. Halpern. Part I, Document LADC-225 (February 5, 1946);
      Part II (Declassification in 1962). \\

      \item{}[CHAD]
 {\it * Neutron Physics. A Course of Lectures by E. Fermi}, \\
      Notes by A.P. French (June 23, 1947). \\

      \item{}[FNM223]
 {\it The Development of the First Chain Reacting Pile}, \\
      Proc. Amer. Philosophical Soc. {\bf 90}, 20-24 (1946). \\

      \item{}[FNM224]
 {\it Atomic Energy for Power}, \\
      The George Westinghouse Centennial Forum.
      {\it Science and Civilization - The Future of Atomic Energy};
      also MDDC-1 (May 27, 1946). \\

      \item{}[FNM225]
 {\it Elementary Theory of the Chain-Reacting Pile}, \\
      Proceedings of the American Physical Society, Meeting at Chicago, June
      20-22, 1946 (see Document MDDC-74 and the announcement in Phys. Rev.
      {\bf 70}, 99 (1946)); \\
      Science {\bf 105}, 27-32 (1947). \\ 

      \item{}[FNM220]
 {\it Reflection of Neutrons on Mirrors}, \\
      E. Fermi and W.H. Zinn, \\
      International Conference on Fundamental Particles and
      Low Temperature Physics, Cambridge, July 1946.
      Physical Society (Cambridge), p.92 (1947). \\

      \item{}[FNM227]
 {\it Phase of Neutron Scattering}, \\
      E. Fermi and L. Marshall, \\
      International Conference on Fundamental Particles and
      Low Temperature Physics, Cambridge, July 1946.
      Physical Society (Cambridge), pp.94-97 (1947). \\

      \item{}[FNM191]
 {\it Production of Low Energy Neutrons by Filtering Through Graphite}, \\
      H.L. Anderson, E. Fermi and L. Marshall, \\
      Phys. Rev. {\bf 270} 815-817 (1946). \\ 

\noi \underline{\bf 1947} \\

      \item{}[FNM226]
 {\it The Transmission of Slow Neutrons through Microcrystalline Materials}, \\
      E. Fermi, W.J. Sturm and R.G. Sachs, \\
      Phys. Rev. {\bf 71}, 589-594 (1947). \\ 

      \item{}[FNM232]
 {\it The Decay of Negative Mesotrons in Matter}, \\
      E. Fermi, E. Teller and V. Weisskopf, \\
      Phys. Rev. {\bf 71}, 314-315 (1947). \\ 

      \item{}[FNM228]
 {\it Inference Phenomena of Slow Neutrons}, \\
      E. Fermi and L. Marshall,
      Phys. Rev. {\bf 71}, 666-667 (1947). \\ 

      \item{}[FNM210]
 {\it Method for Measuring Neutron-Absorption Cross Section by Effect on
      the Reactivity of a Chain-Reacting Pile}, \\
      H.L. Anderson, E. Fermi, A. Wattenberg, G.L. Weil and W.H.
      Zinn, \\
      Phys. Rev. {\bf 72}, 16-23 (1947). \\ 

      \item{}[FNM231]
 {\it Further Experiments with Slow Neutrons}, \\
      E. Fermi and L. Marshall, \\
      Reports CF-3574 (July 26, 1946), CP-3750 (January 17, 1947)
      and CP-3801 (April 14, 1947). \\

      \item{}[FNM200]
 {\it A thermal Neutron Velocity Selector and its Application to the Measurement of the Cross-Section of Boron}, \\
      E. Fermi, J. Marshall and L. Marshall, \\
      Phys. Rev. {\bf 72} 193-196 (1947). \\ 

      \item{}[FNM229]
 {\it Phase of Scattering of Thermal Neutrons by Aluminium and Strontium}, \\
      E. Fermi and L. Marshall, \\
      Phys. Rev. {\bf 71}, 915 (1947). \\ 

      \item{}[FNM230]
 {\it Spin Dependence of Scattering of Slow Neutrons by} Be, Al {\it and} Bi, \\
      E. Fermi and L. Marshall, \\
      Phys. Rev. {\bf 72}, 408-410 (1947). \\ 

      \item{}[FNM233]
 {\it The Capture of Negative Mesotrons in Matter}, \\
      E. Fermi and E. Teller, \\
      Phys. Rev. {\bf 72}, 399-408 (1947). \\ 

      \item{}[FNM234]
 {\it On the Interaction Between Neutrons and Electrons}, \\
      E. Fermi and L. Marshall, \\
      Phys. Rev. {\bf 75}, 1139-1146 (1947). \\ 

\noi \underline{\bf 1948} \\

      \item{}[FNM236]
 {\it Note on Census-Taking in Monte-Carlo Calculations}, \\
      E. Fermi and R.D. Richtmyer, \\
      Document LAMS-805 (July 11, 1948). \\

      \item{}[FNM235]
 {\it Spin Dependence of Slow Neutrons Scattering by Deuterons}, \\
      E. Fermi and L. Marshall, \\
      Phys. Rev. {\bf 75}, 578-580 (1949). \\ 

\noi \underline{\bf 1949} \\

      \item{}[FNM237]
 {\it On the Origin of the Cosmic Radiation}, \\
      Phys. Rev. {\bf 75}, 1169-1174 (1949). \\ 

      \item{}[FNM238]
 {\it An Hypothesis on the Origin of the Cosmic Radiation}, \\
      Nuovo Cimento {\bf 6}, Suppl. 317-323 (1949). \\

      \item{}[FNM239]
 {\it Are Mesons Elementary Particles?}, \\
      E. Fermi and C.N. Yang, \\
      Phys. Rev. {\bf 76}, 1739-1743 (1949). \\ 

      \item{}[FNM240a]
 {\it La visita di} Enrico Fermi {\it al Consiglio Nazionale delle Ricerche}, \\
      Ricerca Scientifica {\bf 19}, 1113-1118 (1949). \\ 

      \item{}[FNM240]
 {\it Conferenze di Fisica Atomica}, \\
      (Fondazione Donegani). Accademia Nazionale dei Lincei (1950).
      \\ 

\noi \underline{\bf 1950} \\

      \item{}[FNM241]
 {\it High Energy Nuclear Events}, \\
      Prog. Theor. Phys. {\bf 5}, 570-583 (1950). \\

      \item{}[FNM242]
 {\it Angular Distribution of the Pions Produced in High Energy Nuclear Collision}, \\
      Phys. Rev. {\bf 81}, 683-687 (1951). \\ 

\noi \underline{\bf 1951} \\

      \item{}[FNM243]
 {\it Lecture on Taylor Instability}, \\
      given during the Fall of 1951 at Los Alamos Scientific Laboratory, \\
      issued on November 1955. \\

      \item{}[FNM244]
 {\it Taylor Instability of an Incompressible Liquid}, \\
      Part 1 of Document AECU-2979 (September 4, 1951) \\

      \item{}[FNM246]
 {\it Fundamental Particles}, \\
      International Conference on Nuclear Physics and the Physics of Fundamental
      Particles,
      The University of Chicago (September 17 to 22, 1951). \\

\noi \underline{\bf 1952} \\

      \item{}[FNM251]
 {\it Letter to Feynman} (January 18, 1952). \\

      \item{}[FNM248]
 {\it Total Cross Sections of Negative Pions in Hydrogen}, \\
      H.L. Anderson, E. Fermi, E.A. Long, R. Martin and D.E.
      Nagle, \\
      Phys. Rev. {\bf 85}, 934-935 (1952). \\ 

      \item{}[FNM249]
 {\it Ordinary and Exchange Scattering of Negative Pions in Hydrogen}, \\
      E. Fermi, H.L. Anderson, A. Lundby, D.E. Nagle and G.B.
      Yodh, \\
      Phys. Rev. {\bf 85}, 935-936 (1952). \\ 

      \item{}[FNM250]
 {\it Total Cross Sections of Positive Pions in Hydrogen}, \\
      H.L. Anderson, E. Fermi, E.A. Long, and D.E. Nagle, \\
      Phys. Rev. {\bf 85}, 936 (1952). \\ 

      \item{}[FNM249a]
 {\it Scattering of Negative Pions by Hydrogen}, \\
      A. Lundby, E. Fermi, H.L. Anderson, D.E. Nagle and G. Yodh,
      \\
      Phys. Rev. {\bf 86}, 603 (1952). \\ 

      \item{}[]
 {\it Total Collision Cross Section of Negative Pions on Protons}, \\
      D.E. Nagle, H.L. Anderson, E. Fermi, E.A. Long and R.L. Martin,
      \\
      Phys. Rev. {\bf 86}, 603 (1952). \\ 

      \item{}[FNM252]
 {\it Deuterium Total Cross Sections for Positive and Negative Pions}, \\
      H.L. Anderson, E. Fermi, D.E. Nagle, and G.B. Yodh, \\
      Phys. Rev. {\bf 86}, 413 (1952). \\ 

      \item{}[FNM247]
 {\it The Nucleus}, \\
      Physics Today {\bf 5}, 6-9 (March 1952). \\

      \item{}[FNM253]
 {\it Angular Distribution of Pions Scattered by Hydrogen}, \\
      H.L. Anderson, E. Fermi, D.E. Nagle and G.B.Yodh, \\
      Phys. Rev. {\bf 86}, 793-794 (1952). \\ 

      \item{}[FNM254]
 {\it Scattering and Capture of Pions by Hydrogen}, \\
      H.L. Anderson and E. Fermi, \\
      Phys. Rev. {\bf 86}, 794 (1952). \\ 

      \item{}[FNM181]
 {\it Experimental Production of a Divergent Chain Reaction}, \\
      Am. J. Phys. {\bf 20}, 536-558 (1952). \\ 

      \item{}[FNM256]
 {\it Numerical Solution of a Minimum Problem}, \\
      E. Fermi and N. Metropolis, \\
      Document LA-1492, (November 19, 1952). \\

      \item
 {\it Fermi's own story}, \\
      Chicago Sun-Times, November 23, 1952. \\

      \item{}[USP15]
 {\it Method of operating a neutronic reactor}, \\
      E. Fermi and L. Szilard, \\
      Patent filed December 1, 1952 (Patent No. 2,798,847; July 9, 1957). \\

      \item{}[FNM255]
 {\it Report on Pion Scattering}, \\
      Proceeding of the Third Annual Rochester Conference
      (December 18-20, 1952). \\

\noi \underline{\bf 1953} \\

      \item{}[FNM257]
 {\it Angular Distribution of Pions Scattered by Hydrogen}, \\
      H.L. Anderson, E. Fermi, R. Martin and D.E. Nagle, \\
      Phys. Rev. {\bf 91}, 155-168 (1953). \\ 

      \item{}[FNM261]
 {\it Magnetic Fields in Spiral Arms}, \\
      S. Chandrasekhar and E. Fermi, \\
      Astrophysical Journal {\bf 118}, 113-115 (1953). \\ 

      \item{}[FNM262]
 {\it Problems of Gravitational Stability in the Presence of a Magnetic Field}, \\
      S. Chandrasekhar and E. Fermi, \\
      Astrophysical Journal {\bf 118}, 116-141 (1953). \\ 

      \item{}[FNM258a]
 {\it Scattering of 169 and 192 MeV Pions by Hydrogen}, \\
      R.L. Martin, E. Fermi and D.E. Nagle, \\
      Phys. Rev. {\bf 91}, 467 (1953). \\ 

      \item{}[FNM258]
 {\it Nucleon Polarization in Pion-Proton Scattering}, \\
      Phys. Rev. {\bf 91}, 947-948 (1953). \\ 

      \item{}[FNM259]
 {\it Scattering of Negative Pions by Hydrogen}, \\
      E. Fermi, M. Glicksman, R. Martin and D.E. Nagle, \\
      Phys. Rev. {\bf 92}, 161-163 (1953). \\ 

      \item{}[FNM264]
 {\it Multiple Production of Pions in Nucleon-Nucleon Collision at Cosmotron Energies}, \\
      Phys. Rev. {\bf 92}, 452-453 (1953); \\
      Errata Corrige Phys. Rev. {\bf 93}, 1954, pp. 1434-1435. \\ 

      \item{}[FNM245]
 {\it Taylor Instability at the Boundary of Two Incompressible \ldd Liquids}, \\
      E. Fermi and J. Von Neumann, \\
      Part 2 of Document AECU-2979 (August 19, 1953). \\

\noi \underline{\bf 1954} \\

      \item{}[FNM267]
 {\it Polarization of High Energy Protons Scattered by Nuclei}, \\
      Nuovo Cimento {\bf 11}, 407-411 (1954). \\ 

      \item{}[FNM268]
 {\it Polarization in the elastic Scattering of High Energy Protons by Nuclei}, \\
      Private Communication (March 24, 1954). \\

      \item{}[FNM260]
 {\it Phase Shift Analysis of the Scattering of Negative Pions by Hydrogen}, \\
      E. Fermi, N. Metropolis and E.F. Alei, \\
      Phys. Rev. {\bf 95}, 1581-1585 (1954). \\ 

      \item{}[FNM263]
 {\it Multiple Production of Pions in Pion-Nucleon Collision}, \\
      Academia Brasileira de Ciencias {\bf 26}, 61-63 (1954). \\

      \item{}[FNM265]
 {\it Galactic Magnetic Fields and the Origin of Cosmic Radiation}, \\
      Astrophysical Journal {\bf 119}, 1-6 (1954). \\

\noi \underline{\bf 1955} (posthumous) \\

      \item{}[FNM270]
 {\it Lectures on Pions and Nucleons}, \\
      Nuovo Cimento {\bf 2}, Suppl., 17-95 (1955). \\ 

      \item{}[FNM266]
 {\it Studies of Nonlinear Problems. I}, \\
      E. Fermi, J. Pasta and S. Ulam, \\
      Document LA-1940 (May 1955). \\

      \item{}[FNM269]
 {\it Physics at Columbia University - The Genesis of the Nuclear Energy Project}, \\
      Physics Today {\bf 8}, 12-16 (November 1955). \\

\end{enumerate}

\newpage

\begin{figure}
\begin{center}
\vspace{1.5truecm} %
\epsfysize=7cm 
\epsffile{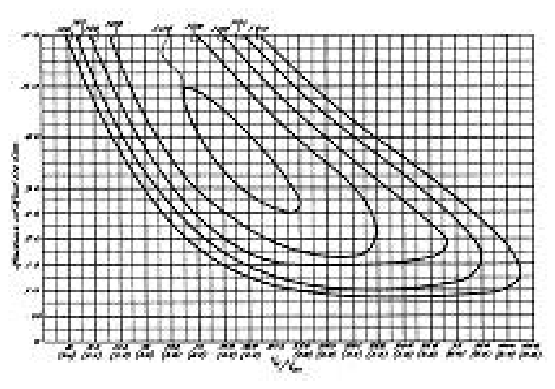}
\\
\begin{tabular}{cc}
\epsfysize=5cm 
\epsffile{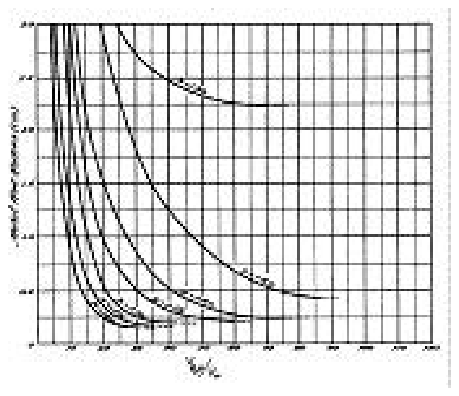} &
\epsfysize=5cm 
\epsffile{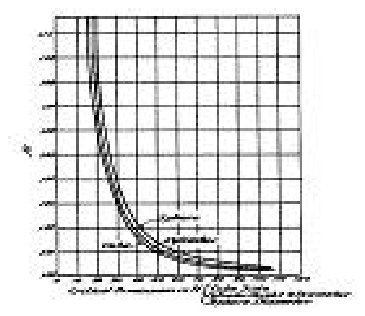}
\end{tabular}
\caption{Graphs taken from the patent USP3 testifying for the
accurate numerical determinations performed by Fermi and
collaborators on the physics of nuclear piles. The first two
graphs show contour lines representing various reproduction
constants $k$ for systems employing uranium oxide (in the form of
cylindrical rods) and graphite or systems employing uranium metal
rods immersed in heavy water, respectively. The third graph,
instead, shows the change in critical size in uranium-graphite
reactors with change in $k$ for different geometries.}
\label{fig1}
\end{center}
\end{figure}

\newpage

\begin{figure}
\begin{center}
\begin{tabular}{c}
\epsfysize=9cm 
\epsffile{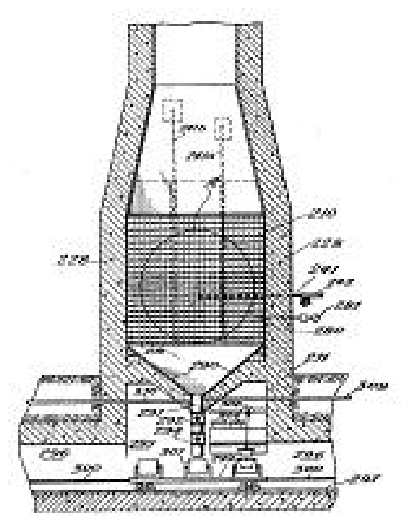} \\
\epsfysize=9cm 
\epsffile{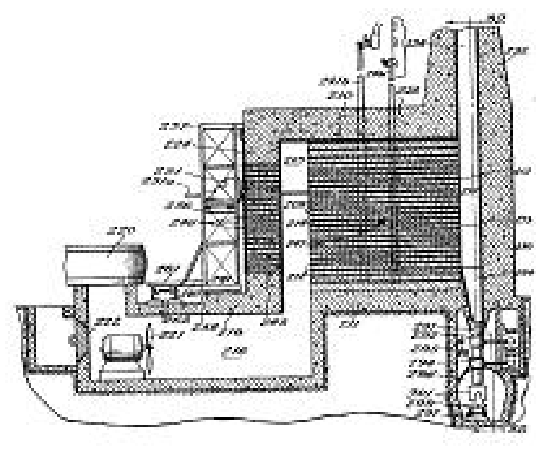}
\end{tabular}
\caption{Longitudinal and cross sectional views of an air cooled
chain reacting system, taken from the patent USP3.} \label{fig2}
\end{center}
\end{figure}

\newpage

\begin{figure}
\begin{center}
\begin{tabular}{c}
\epsfysize=9cm 
\epsffile{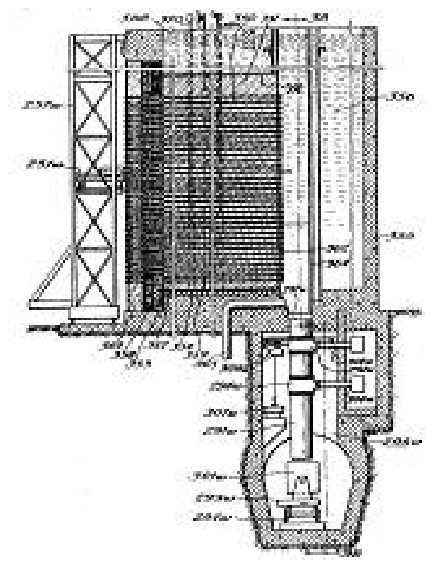} \\
\epsfysize=9cm 
\epsffile{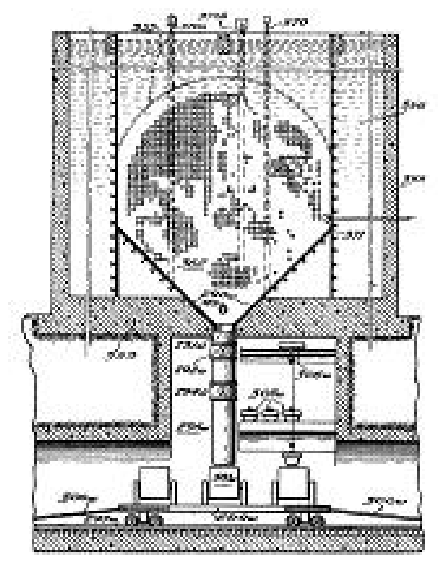}
\end{tabular}
\caption{Vertical sectional views of a liquid cooled reactor,
taken from the patent USP3.} \label{fig3}
\end{center}
\end{figure}

\newpage

\begin{figure}
\begin{center}
\begin{tabular}{c}
\epsfysize=15cm 
\epsffile{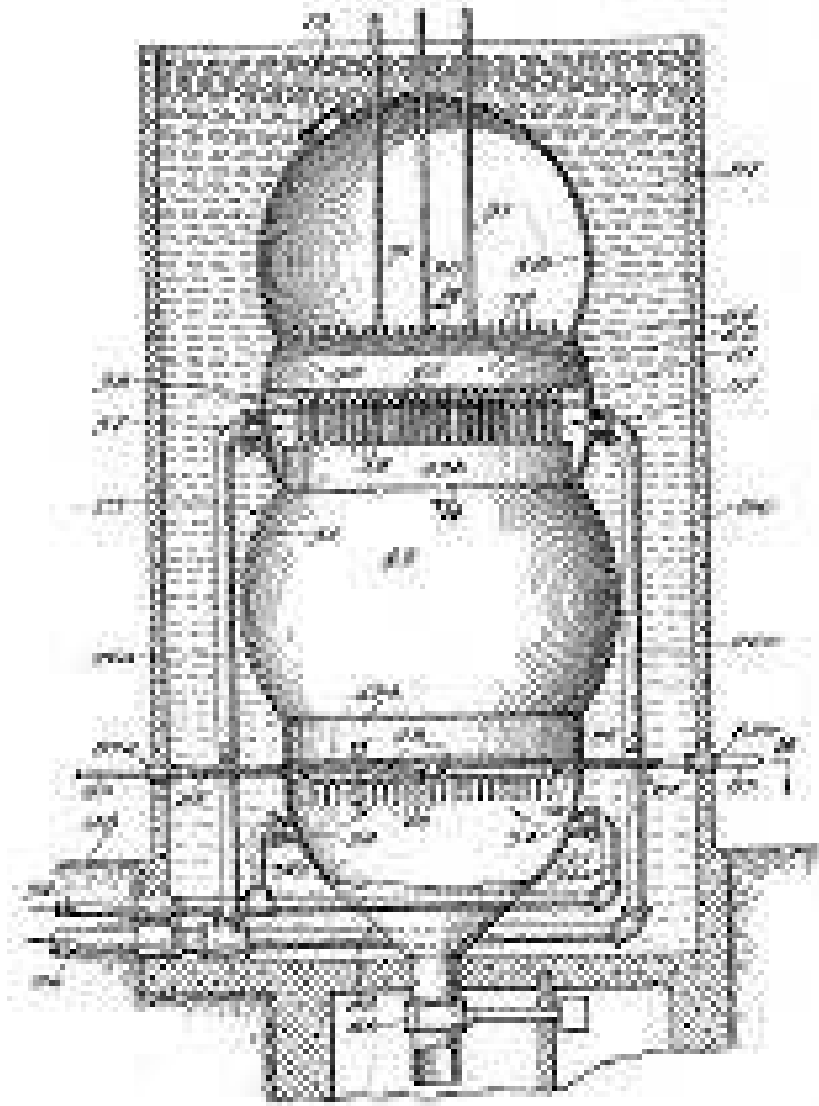}
\end{tabular}
\caption{A vertical sectional view, taken from the patent USP4,
showing a nuclear fission power plant in which the heat equivalent
of 100000 kW is removed by circulation of 400000 pounds of helium
per hour.} \label{fig4}
\end{center}
\end{figure}

\newpage

\begin{figure}
\begin{center}
\begin{tabular}{c}
\epsfysize=10cm 
\epsffile{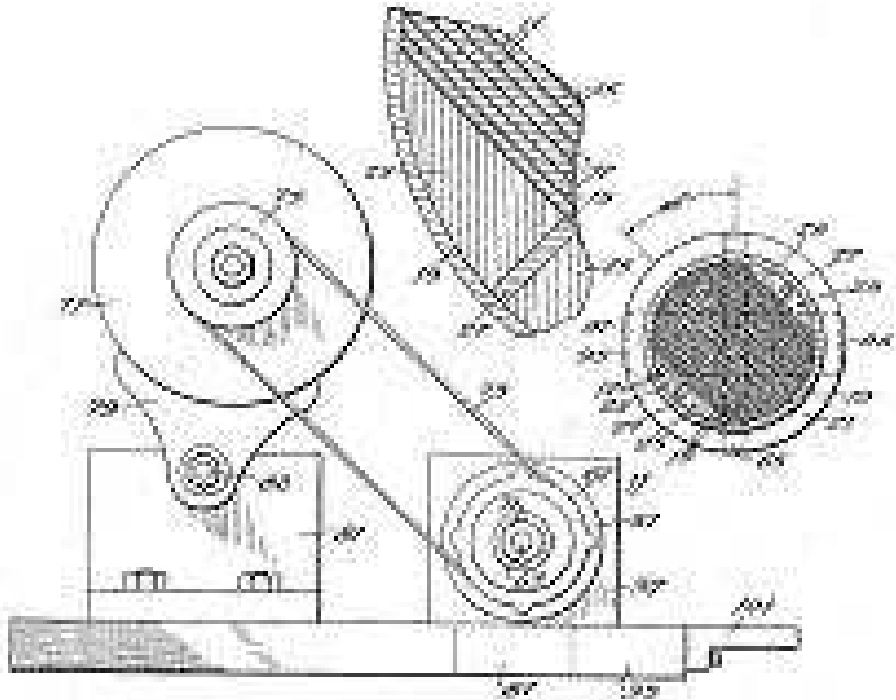}
\end{tabular}
\caption{A drawing of the rotatable shutter unit and the driving
motor of a velocity selector constructed according to what
described in the patent USP8.} \label{fig5}
\end{center}
\end{figure}

\newpage

\begin{figure}
\begin{center}
\begin{tabular}{c}
\epsfysize=8cm 
\epsffile{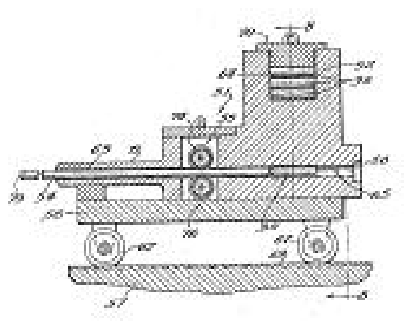} \\
\epsfysize=8cm 
\epsffile{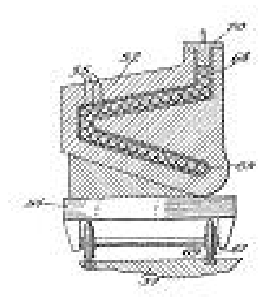}
\end{tabular}
\caption{The loading device for the air cooled uranium-graphite
reactor considered in the patent USP13.} \label{fig6}
\end{center}
\end{figure}

\newpage

\begin{figure}
\begin{center}
\begin{tabular}{c}
\epsfysize=8cm 
\epsffile{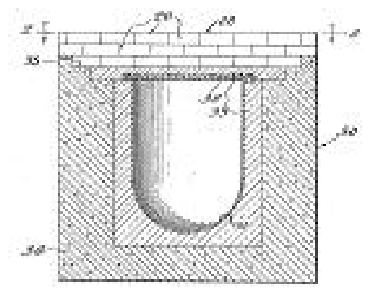} \\
\epsfysize=8cm 
\epsffile{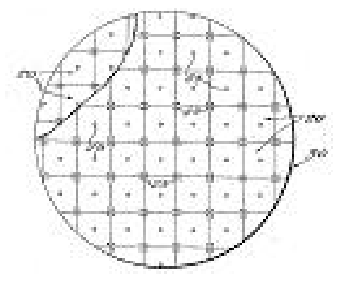}
\end{tabular}
\caption{Drawings showing the center of a nuclear reactor equipped
with a shield constructed according to what described in the
patent USP14. In particular, the second figure shows the top
construction of the shield mentioned.} \label{fig7}
\end{center}
\end{figure}

\newpage

\end{document}